\documentclass{aa}

\usepackage{graphicx}
\usepackage{txfonts}
\usepackage{xcolor}
\usepackage{placeins}
\usepackage{multirow}
\usepackage{float}
\usepackage{longtable}
\newcommand{\ocen}{\ensuremath{\omega\,\mathrm{Cen}\,}}

\begin{document}

   \title{Identifying substructure associations in the Milky Way halo using chemo-kinematic tagging}

   \subtitle{}

   \author{K. Youakim \thanks{kristopher.youakim@astro.su.se}
          \inst{1}
          \and
          K. Lind 
          \inst{1}
          }

   \institute{$^{1}$Department of Astronomy, Stockholm University, AlbaNova University Centre, Roslagstullsbacken 21, 106 91 Stockholm, Sweden\\
              \email{kristopher.youakim@astro.su.se}
             }

 
  \abstract
   {The Milky Way halo has been built-up over cosmic time through the accretion and dissolution of star clusters and dwarf galaxies as well as through their complex interactions with the Galactic disc. Traces of these accreted structures persist to the present day in the chemical and kinematic properties of stars and their orbits and allow for the disentangling of the accretion history of the Galaxy through observations of Milky Way stars.}
   {We utilised 6D phase-space information in combination with [Fe/H] measurements to facilitate a clustering analysis of stars using their kinematics and chemistry simultaneously, a technique known as chemo-kinematic tagging. We aim to associate large halo substructure groups with Milky Way halo globular clusters, stellar streams, and satellite galaxies in order to investigate the common origins of these groups of structures.}
   {We implemented t-distributed stochastic neighbour embedding (t-SNE) to perform dimensionality reduction and identify stars from clusters and streams that are co-localised in the kinematic and chemical parameter space. We used the orbital parameters E, J$_r$, J$_z$, L$_z$, r$_{apo}$, r$_{peri}$, and eccentricity as well as [Fe/H] as input into the algorithm, and we performed a clustering analysis for a sample of 5347 stars from 229 individual Milky Way substructures.}
   {Most notably, we recovered several large-scale structures that have been reported in the literature, including GSE, Thamnos, Sequoia, I'itoi, LMS-1/Wukong, Sagittarius, Kraken/Koala, the splashed disc, and a candidate structure recently found in the literature. We assigned globular cluster populations to each of these accreted structures and find that 44\% of Milky Way globular clusters are consistent with having an accreted origin. Of the remaining in situ clusters, we find that they can be separated into four distinct groups: bulge, pre- and post- disc, and the splashed disc. In addition, we identified many small-scale structures, including several stream-progenitor associations. Most notably amongst these is a connection between the Orphan-Chenab stream and the Grus II ultra-faint dwarf galaxy, which supports previous findings that these two objects were brought into the Galaxy in the same accretion event.}
   {}


   \maketitle
%

\section{Introduction}

Our Galaxy has undergone many mergers and accretion events over its lifetime, and these events are an integral part of the formation process that created the Milky Way we see today. Present day observations of Milky Way stars can reveal information about their origins, as these stars retain chemical and kinematic signatures of their formation environments over long periods of cosmic time \citep[e.g.][]{Helmi_1999, Freeman_2002, Bland-Hawthorn_2010}. Therefore, with this information it is possible to, for example, determine whether stars were formed inside our Galaxy or if they were formed in another galaxy that was later accreted. In some cases, stars can even be traced back to their common birth clusters \citep[e.g.][]{Meza_2005, Haywood_2018, Myeong_2018b, Naidu_2020}. Recent advances in the precision and scale of spectroscopic and astrometric measurements of these Milky Way stars have increasingly allowed for the disentanglement of this complex formation history. Based on this idea, several accreted structures were identified in the Galaxy by their spatial coherence and similarity of stellar properties, for example, the Sagittarius dwarf galaxy and stream \citep{Ibata_1994}, the Helmi streams \citep[][]{Helmi_1999, Koppelman_2019a}, and the Orphan stream \citep[][]{Grillmair_2006, Belokurov_2007}, amongst others. 

Since the availability of the European Space Agency's Gaia mission \citep{Gaia2016} third data release \cite[Gaia DR3;][]{Gaia2021, Lindegren2021}, which has provided astrometric measurements for over 1.8 billion halo stars, several more accreted halo structures have been identified. In particular, these are structures that are less visible due to occlusion from other Galactic components and because they are fainter and more phase-mixed within the stellar halo. Some examples include the Gaia-Sausage-Enceladus \citep[GSE;][]{Belokurov_2018, Helmi_2018, Haywood_2018}, Sequoia \citep{Myeong_2019}, Thamnos \citep{Koppelman_2019b}, and LMS-1/Wukong \citep{Yuan_2020, Naidu_2020, Malhan_2021}. In addition to these accreted structures, a recent work by \citet{Belokurov_2020} has described a peculiar population of metal-rich stars on highly eccentric orbits that they dubbed the splashed disc, or simply `the Splash', and which the authors suggest is likely the result of a perturbation of Milky Way disc stars by an ancient major merger.  

Gaia has also revolutionised the study of stellar streams, long elongated populations of stars that disperse along the orbital trajectory of in-falling progenitors during their interaction with the gravitational potential of the Milky Way. Recent work using the STREAMFINDER algorithm \citep{Malhan_2018a, Malhan_2018b} has resulted in a litany of new stellar stream discoveries, many of which have been followed-up spectroscopically and compiled in a catalogue \citep{Ibata_2021, Ibata_2022_cat, Ibata_2024}. The S5 survey has also undertaken a targeted survey of stellar streams in the southern hemisphere, and the spectroscopic follow-up of several streams has been included in their first data release, DR1 \citep{Li_2019}. Using the latest available data, \citet{Mateu_2023} painstakingly compiled a list of as many stellar streams as possible and computed stream tracks to trace each stream's proper motion, radial velocity, and sky position, and published these in a comprehensive library called galstreams. Taken together, these efforts have provided a large homogeneous dataset with which to investigate associations between stellar streams and other structures in the Galaxy. 

Accreted structures in the Milky Way halo have been associated with globular cluster (GC) and dwarf galaxy populations based on similarities of their orbital properties, suggesting that the associated clusters were accreted into the Milky Way halo in a common merger event \citep[e.g.][]{Belokurov_2018, Helmi_2018, Myeong_2018a, Myeong_2018b, Malhan_2019}. Individual stellar streams have also been associated with their progenitors based on similarities in their orbits and chemistry \citep[e.g.][]{Ibata_2019, Malhan_2019, Bonaca_2021}.

Furthermore, it has been shown that there are two distinct populations of GCs in our Galaxy: one of accreted clusters, located in the halo and with a distinctly lower metallicity distribution, and one that formed in situ, which is more centrally located in the Galaxy and has a characteristically more metal-rich metallicity distribution \citep[e.g.][]{Zinn_1993}. These two populations can be distinguished as two sequences in the age-metallicity relation \citep[e.g.][]{Forbes_2010, Leaman_2013}, and they show respective disc- and halo- like kinematics and spatial distributions \citep[e.g.][]{Dinescu_1997}. More recently with improved kinematics, it has been shown that this separation between in situ and accreted GCs can be made using a selection in the E-L$_z$ space \citep{Belokurov_2023, Belokurov_2024} and that groups of GCs can also be connected to specific accretion events \citep[e.g.][]{Massari_2019, Forbes_2020, Malhan_2022a, Sun_2023}.
It has even been suggested that the entire stellar halo may be made up of an accreted substructure \citep{Naidu_2020} and that a substantial portion of the phase-mixed halo stars may be attributable to disrupted GCs \citep{Martell_2016, Gnedin_1997}.

Recently, much effort has been directed towards attempting to trace halo stars back to a common progenitor based solely on their chemistry, a technique known as chemical tagging \citep{Freeman_2002, Bland-Hawthorn_2010}. In order for chemical tagging to be viable, stars born in the same cluster should exhibit homogeneous abundances \citep{Hawkins_2020}, and unique chemical signatures of stars from a given cluster must be distinguishable from other clusters and the background signature of the overwhelmingly more numerous halo and disc stars \citep{Price-Jones_2019, Cheng_2021}. In addition, abundances must be derived with adequate precision such that measurement uncertainties and systematic broadening do not erase the distinguishing abundance signature of the cluster. Indeed, this results in a picture where chemical tagging may be successful on certain progenitors with more distinct or unusual abundance patterns, whereas recovering clusters that have generic halo-like abundances is significantly more challenging, if not entirely unfeasible. This limitation was investigated by \citet{Casamiquela_2021}, who showed that by using high-precision differential abundances of stars in open clusters, they were able to identify over 40\% of member stars for only about one third of open clusters in their sample despite having 16 high-precision elemental abundances as input for their clustering algorithm. They concluded that this was because the overlap in the chemical parameter space was too large for most clusters, unless they had a somewhat unusual abundance signature with respect to field stars.

It has also been shown that clusters and substructures can be associated with a common progenitor based on their kinematic signatures. For example, \citet{Myeong_2018a} showed an association between the GSE and eight old halo GCs in energy and action space. In another study, \citet{Myeong_2018b} also suggest a tentative association between five newly discovered retrograde halo substructures and the retrograde atypical globular cluster Omega Centauri (\ocen). Such studies leverage the fact that stars belonging to structures with distinct orbital properties are relatively easy to identify compared to those with orbits closer to the bulk disc and halo populations. Therefore, it is no surprise that most kinematic substructures and associations that we know of consist of stars that are on retrograde orbits or have highly eccentric and energetic orbits. However, a recent work by \citet{Sestito_2020} showed that by using [Fe/H] measurements, they were able to find a group of low-metallicity stars that are confined to prograde disc-like orbits and do not venture out of the disc plane. They propose that these may be an ancient accreted population that has interacted in a complex way with the Galactic disc and would have been indistinguishable from the disc if only orbital parameters were considered.

By combining both chemistry and kinematics, the dimension of the discovery space increases and therefore so do the chances of having a unique vector in the parameter space with which to differentiate stars belonging to substructures and field stars. This is the basis for chemo-kinematic tagging, a method in which kinematic and chemical quantities are re-parameterised such that they can be simultaneously used to identify stars clustered in both of these parameter spaces via dimensionality reduction techniques. This technique has recently been used to identify tidally stripped members of \ocen, even at large distances from the cluster body \citep{Youakim_2023}. 

In a recent work, \citet{Malhan_2022a} performed a comprehensive analysis of stellar structure in the Milky Way halo using the latest kinematic data of known stellar streams and stream candidates from the STREAMFINDER survey \citep{Ibata_2021} and catalogues of GCs and satellite galaxies from the literature. They accomplished this by computing the action-angle coordinates (J$_r$, J$_{\phi}$, J$_z$) and energy for these objects using 6D kinematic information and subsequently searched for groups in the kinematic action plus energy phase-space using a clustering algorithm called ENLINK. They found six main kinematic groups in the halo, which they suggested were each the result of a previous Galactic merger. Five of these had been previously described in the literature, and one was proposed to be a new merger called Pontus, which they subsequently characterised in a follow-up study \citep{Malhan_2022b}.

In this paper, we take this analysis further by implementing chemo-kinematic tagging as a means to connect halo substructures, adding the independent chemical parameter [Fe/H] as an input into our clustering algorithm. We also include several other conserved orbital parameters, including apocentre distance, pericentre distance, and eccentricity, which despite being correlated with the orthogonal action coordinates are useful in modifying the relative weighting of the input parameters and improving clustering results. We also include a larger number of stream candidates from the updated STREAMFINDER catalogue \citep{Ibata_2024} and the S5 survey's first data release \citep{Li_2019}, and we use the t-distributed stochastic neighbour embedding (t-SNE) dimensionality reduction algorithm. Several previous works have used t-SNE for Milky Way studies predominantly pertaining to tagging and classifying stellar populations \citep[e.g.][]{Traven_2017, Kos_2018, Anders_2018, Traven_2020, Hughes_2022, Santiago_2022, Youakim_2023, Ortigoza-Urdaneta_2023, Cantelli_2024}. However, this work marks the first time it is used on such a broad dataset to associate substructures with specific accretion events. 

In Section \ref{sec:data}, we describe the dataset and data processing, in Section \ref{sec:methods}, we describe the dimensionality reduction method and input parameters, and in Section \ref{sec:results} we present the results, including the validation of clustered objects found in the clustering analysis. In Section \ref{sec:discussion} we discuss the identified associations, including large groups, subgroups, and small-scale stream-progenitor associations, and finally we provide our conclusions in Section \ref{sec:conclusions}.

\renewcommand{\arraystretch}{1.1}
\begin{table*}
\caption{Summary of the compiled data used in this work.}
\label{tab:compiled_data}
\centering
\begin{tabular}{|c|c|c|c|}
\hline
Substructure       & Sample                                                   & \# Objects  & Source \\ \hline \hline
Globular clusters&          -                                                & 147         & \begin{tabular}[c]{@{}l@{}}\citealt{Vasiliev_2019}  (Kinematics) \\ \citealt{Harris_1996} ({[}Fe/H{]}) \end{tabular} \\ \hline
Satellite galaxies &           -                                               & 30          & \citealt{McConnachie_2020a, McConnachie_2020b}                                                                                  \\ \hline
Halo structures& GSE                                                      & 557*        & \citealt{Kushniruk_2025}                                                                                                                          \\ \hline
                   & Splash                                                   & 2200*       & \citealt{Kushniruk_2025}                                                                                                                          \\ \hline
                   & Thamnos                                                  & 111*        & \citealt{Kushniruk_2025}                                                                                                                          \\ \hline
Stellar streams    & STREAMFINDER                                             & 43 (1805*)  & \citealt{Ibata_2024, Ibata_2021}                                                                                                \\ \hline
                   & S5 survey DR1 & 12 (603*)   & \citealt{Li_2019}                                                                                                                \\ \hline
                   & Wukong                                                   & 13*         & \citealt{Limberg_2024}                                                                                                           \\ \hline
                   & LMS-1                                                    & 48*         & \citealt{Malhan_2021}                                                                                                           \\ \hline
                   & C-19                                                     & 10*         & \begin{tabular}[c]{@{}l@{}}\citealt{Yuan_2022} \\ \citealt{Martin_2022} \end{tabular}     \\ \hline \hline
Total              &            -                                              & 238 (5347*) &   -                                                                                                                                              \\ \hline 
\end{tabular}
\tablefoot{The first column shows the substructure group, the second column describes the sub sample of the group, the third column describes the number of objects in the group (GCs, galaxies or streams), with the * symbol indicating the number of individual stars. The final column shows the literature source from which the data was taken.}
\end{table*}

\section{Data}
\label{sec:data}
We assembled a catalogue of substructures from several different sources in the literature, which included the most up-to-date data on Galactic GCs, satellite galaxies, stellar streams and kinematic structures. We included all of the data on these objects which had available 6D phase-space and [Fe/H] information. 
In this section, we provide more details on the data that were used as well as the processing involved to homogenize the dataset and prepare it for input into the t-SNE algorithm. 

\subsection{Dataset}
\label{subsec:data_set}
Table \ref{tab:compiled_data} summarises the total compiled dataset used in this work and includes the source of each individual input table taken from the literature. In total, we included 147 GCs, 30 satellite galaxies, 3 kinematic halo structures, and 49 distinct stellar streams, for a total of 229 objects (Table \ref{tab:compiled_data} shows 238 total objects since there were seven stellar streams in common between the S5 and STREAMFINDER samples, and LMS-1 and C-19 were both also included in the STREAMFINDER sample). 

\subsubsection{Halo substructures: GSE, Splash, and Thamnos}

In order to identify the regions of the latent space where stars from various halo structures are expected to be located, we used a sample of stars selected in \citet{Kushniruk_2025}. Using the most recent data release of the GALAH survey \citep[GALAH DR4;][]{Buder_2025}, they selected GSE, Splash, and Thamnos stars using a wavelet analysis to identify over-dense regions in the commonly used $\sqrt{\mathrm{J}_r}$ - L$_z$ parameter space and then refined the selection with a cut to select halo stars using the [Mg/Cu] versus [Na/Fe] abundance plane. 

\subsubsection{Globular clusters}

As input for Milky Way GCs, we used the catalogue of \citet{Vasiliev_2019}, who provide positions, proper motions (from Gaia DR2), radial velocities, and distances for over 150 GCs. From these data we computed the orbital parameters and integrals of motion used in the t-SNE analysis. We also supplemented the kinematics with [Fe/H] values of each GC from the \citet{Harris_1996} catalogue of GC parameters. The final sample consisted of 147 GCs with 6D + [Fe/H] data. 

\subsubsection{Stream sample}
For the stream sample, we used the STREAMFINDER catalogue of spectroscopically followed-up stream candidates from \citet{Ibata_2024}, which includes $\sim 25\,000$ candidates from 86 stellar streams (some of which are still unconfirmed). We selected all stars which had available radial velocities in the catalogue, which were a combination of STREAMFINDER spectroscopic follow-up and literature radial velocities from publicly available spectroscopic surveys. We assigned metallicities to each of the stream candidates based on stream metallicities published in the literature (see Table 2 of \cite{Malhan_2022a} for the compilation used as a reference), such that every star belonging to the same stellar stream was assigned the same metallicity. 

We assigned distances to each stream candidate using the stellar stream tracks from the galstreams catalogue of \cite{Mateu_2023}, which is a compilation of parameters made by fitting tracks to available data for almost 100 stellar streams in the literature. We computed the nearest neighbour from the stream track to each stream star from our sample and assigned it the corresponding distance value, resulting in a distance gradient amongst our observed stream sample that matched that expected for a given stellar stream. This effectively removed the highly uncertain distance parameter from the orbit determinations and put all of the weight on the proper motions and radial velocities, which are astrometrically and spectroscopically measured quantities, respectively. This approach is similar to what was used for the STREAMFINDER catalogue, which determined distances self-consistently during the orbit calculations \citep{Ibata_2021}. Indeed, our approach yielded distance values that were very similar to those provided in the STREAMFINDER catalogue.

We also included stream candidates identified in the S5 survey \citep{Li_2019}, using radial velocities and [Fe/H] measurements selected from the first data release. For this sample we also updated the distances using galstreams tracks as described above. Deriving distances this way resulted in a larger sample of stream stars which could be included in our analysis compared to using only the stars with distance estimations in \citet{Ibata_2024} and \citet{Li_2019}, and had the added advantage that all distances for the whole stream sample were derived in a uniform way. Furthermore, distances provided in DR1 of the S5 Survey were photo-geometrically derived using the StarHorse bayesian inference code \citep{Queiroz_2018, Anders_2022}, which relies heavily on Milky Way priors and is therefore not suitable to provide precise individual distances to stream stars.

We selected stream stars from the S5 DR1 catalogue with priority = 9, logg50 < 4.1, and feh50 < -1. This the strictest selection to select stream candidates based on a tight selection in projected proper motion space in the reference frame of the stream, as well as a removal of main sequence and metal-rich stars to remove contamination from nearby disc stars, and was shown to effectively select stream stars in velocity space in \citet{Li_2019}. We also applied the qso\_flag\_wise = 0 and good\_star = 1 flags as advised by the authors of the catalogue to ensure that the objects were stars and that the photometry was well behaved. For our purposes, we were more interested in purity rather than completeness of the stream samples, and therefore a strict cut was preferred. We also implemented a cut on the standard deviation of the radial velocity to < 5 km/s to reduce the uncertainty of the orbit determinations.
Finally, we supplemented the sample with detailed spectroscopic follow-up samples from the literature for Wukong \citep[][from the H3 survey]{Limberg_2024}, LMS-1 \citep{Malhan_2021}, and C-19 \citep{Martin_2022, Yuan_2022}.

The final sample consisted of 1805 stream candidate stars from 43 streams from STREAMFINDER, 603 stars from 12 streams from the S5 survey, (five unique streams not already included in the STREAMFINDER sample) and the additional Wukong, LMS-1, and C-19 samples. Therefore, in the full dataset we included data from a total of 49 distinct stellar streams. 

\subsubsection{Satellite galaxies}
We used the comprehensive compilation of Local Group dwarf galaxies and their parameters from \cite{McConnachie_2020a, McConnachie_2020b}. We made a cut on the sample to only include galaxies for which the measured apocentre was larger than the measured distance (r$_{apo}$ > dist), and the measured distance was less than 250 kpc from the Milky Way Galactic centre. This resulted in a sample of 30 satellite galaxies.

\subsection{Data processing and computation of orbital parameters}

We used galpy \citep{Bovy_2015} to compute orbits and derive the orbital parameters of each object in our dataset. For the clustering analysis, we chose to use conserved quantities of the orbit, namely actions, energy, eccentricity, and Galactic apo- and peri-centre radii (J$_r$, J$_{\phi}$, J$_z$, E, e, r$_{apo}$, r$_{peri}$), as these quantities are largely conserved over Galactic timescales in a slowly evolving potential. 

Although it is true that the entirety of the orbits are described by the three action-angle coordinates J$_r$, J$_{\phi}$, J$_z$, the orbital energy is also a commonly used parameter for selecting and identifying halo substructures in the literature in the E-L$_z$ plane. Indeed, adding E does not add any additional independent physical information to the action-angle coordinates, but including it can help improve the clustering in the latent space by providing a beneficial weighting of the input parameters. This was previously shown to be the case in \citet{Malhan_2022a}, where they used E in addition to J$_r$, J$_{\phi}$, J$_z$ and found an improvement in the results of their clustering algorithm. We tested this empirically and indeed the inclusion of E, as well as r$_{apo}$, r$_{peri}$ and eccentricity improved the coherence of structures in the latent space, and made selecting groups easier. In practice, including these extra, redundant parameters resulted in more detailed separation between groups, and tighter clusters of individual groups, particularly on small-scales. Therefore, rather than apply individual weights to the input parameters, which requires subjective fine-tuning and optimization, we chose instead to include these extra orbital parameters to provide a natural weighting.

For the orbit calculations, we used the four component MWPotential2014 described in \citet{Bovy_2015}. (For a more in-depth discussion on this as well as the detailed properties of the Galactic components used, see Section 2.4 and Table 1 in \citet{Youakim_2023}.)
    
\section{Methods}
\label{sec:methods}

For this work, we used an unsupervised manifold learning algorithm called t-SNE to perform a clustering analysis of Milky Way halo substructures. This algorithm was originally developed by \citet{Hinton2002} as a general purpose dimensionality reduction tool for visualisation of high-dimensional data. It was later modified to use a t-distribution to initialise parameter weights in the high-dimensional parameter space \citep{vanderMaaten2008} and subsequently updated to implement a variation of a Barnes-Hut algorithm to substantially reduce the computation time and allow for the embedding of millions of objects \citep{vanderMaaten2013}. In this work, we used the MulticoreTSNE python package \citep{Ulyanov_2016}, which is a version of the algorithm optimised to run on parallelised computing systems.  
The general method for the dimensionality reduction follows a similar procedure to \citet{Youakim_2023}, and we refer to that paper for the full details.  

In short, the dataset was prepared for input into the t-SNE algorithm with a standardization step, transforming the distribution for each parameter to have a median of zero and a standard deviation of one, such that the scaling of each parameter was uniform and equally weighted. In addition, we removed all objects with a Galactocentric apocentre distance r$_{apo} > 250$ kpc, to remove objects with nonphysical orbital parameters from the orbit computation. We chose not to remove outliers in E, J$_r$, L$_{z}$ and J$_z$ as we did not want to exclude high energy dwarf galaxies from the analysis.

The two hyper-parameters relevant for the t-SNE analysis were n-jobs -- the number of cores used in parallel while running the algorithm -- and perplexity -- which controlled the width of the initialised probability distributions used to map from the high-dimensional space to the latent space, and can roughly be summarised as the number of data points expected in a given cluster. Given that we were interested in a range of scales from large kinematic halo structures to small scale stream-progenitor associations, we performed the selection of groups considering three different scales: large structures corresponding to GSE-like mergers, medium sized structures representing sub-groups within a GSE-like merger, and small scales which consisted of at most a few objects and were mostly pairs of individual stellar streams and their progenitors. These selections were informed by considering t-SNE maps computed at several different values for perplexity, ranging from 5 < p < 3000, which are shown in Figure \ref{fig_app:perp_grid} in the Appendix. At the smallest value of perplexity the latent space was uniformly populated and small clusters were clearly separated, but larger groups were not. On the largest scales, large groups were clearly visible but group boundaries were not well defined, and small sub-groups were not easily identifiable.

\section{Results}
\label{sec:results}

Figure \ref{fig:tsne_plot} displays the entire t-SNE latent space with all input data and with a perplexity value of 50. For clarity, we chose to show only the latent space at this intermediate perplexity value, which separated clusters on both large and small scales. All objects and stars are coloured according to their identification in the literature. 

\subsection{Selection of groups and subgroups}
\label{subsec:results_selection}

From the t-SNE latent space analysis, we identified 9 distinct groups (bottom left panel of Figure \ref{fig:tsne_plot}), which we further separated into 16 subgroups (bottom middle panel). These groupings comprised of known GCs, satellite galaxies, and stellar streams. Additionally, we investigated 20 small groupings of stellar streams and their potential progenitors (bottom right panel). 

The initial selection of groups in the t-SNE latent space was performed visually on the t-SNE coordinates in the latent space projection, identifying regions that appeared kinematically and chemically distinct from neighbouring structures. We used the labels from known Milky Way structures, GCs, satellite galaxies and streams to guide our selection, as our goal was to make associations between known Milky Way substructures, not to perform blind tagging of groups.

We also used t-SNE maps computed with different perplexity values to inform our selections at different scales, which we show in the Appendix as Figure \ref{fig_app:perp_grid}. For example, for the main group selections we predominantly used the p~=~300 and p~=~500 maps, which clearly show the main larger groups. For the subgroup selection, we used the p~=~30, p~=~50 and p~=~100 maps, which show medium scale clustering, and for the small groups and stream progenitor associations, we used the smallest scale maps of p~=~5, p~=~10, p~=~30 and p~=~50 to inform our selections. 

For these selections, we took into account several different scales of clustering, the types of objects in a given cluster (e.g. Stars, GCs, satellite galaxies, streams), and the populations of previously tagged stars. Therefore, due to the complexity, use of an automated clustering algorithm for selecting groups was not feasible, and we made these selections by hand.

\renewcommand{\arraystretch}{1.1}
\begin{table}
\centering
\caption[]{Selected groups and their corresponding structures.}
\label{tab:small_associations}
\begin{tabular}{|c|c|}
\hline
Group & Name \\ \hline
G-1 & Splash \\ \hline
G-2 & GSE \\ \hline
G-3 & Thamnos \\ \hline
G-4 & Sequoia/I'itoi \\ \hline
G-5 & Accreted Structures \\ \hline
G-6 & Bulge GCs \\ \hline
G-7 & Post-disc GCs \\ \hline
G-8 & Pre-disc GCs \\ \hline
G-9 & \begin{tabular}[c]{@{}c@{}}High-energy\\ Satellite Galaxies \end{tabular}  \\ \hline
\end{tabular}
\end{table}

Table \ref{tab:small_associations} summarises the main groups identified with this selection, along with their correspondence to previously identified structures from the literature. A longer version of this table is available as Table \ref{tab:large_associations} in the Appendix, which provides the full complement of halo structures, GCs, streams, and satellite galaxies belonging to each of these groups.

\subsection{Validation of groups and subgroups}
\label{subsec:results_validation}

\begin{figure*}
   \centering
     \includegraphics[width=\textwidth]{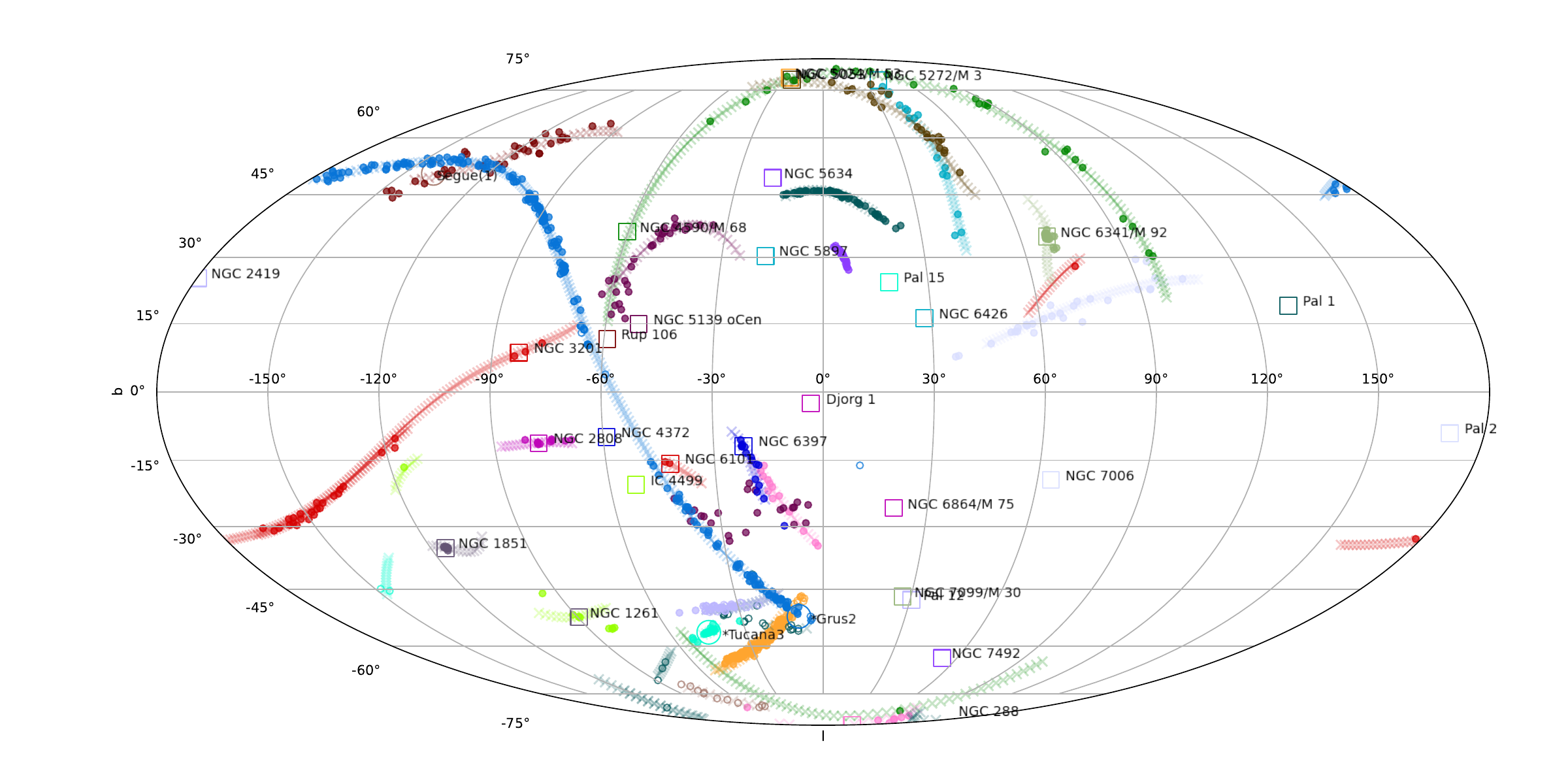}
     \includegraphics[width=\textwidth]{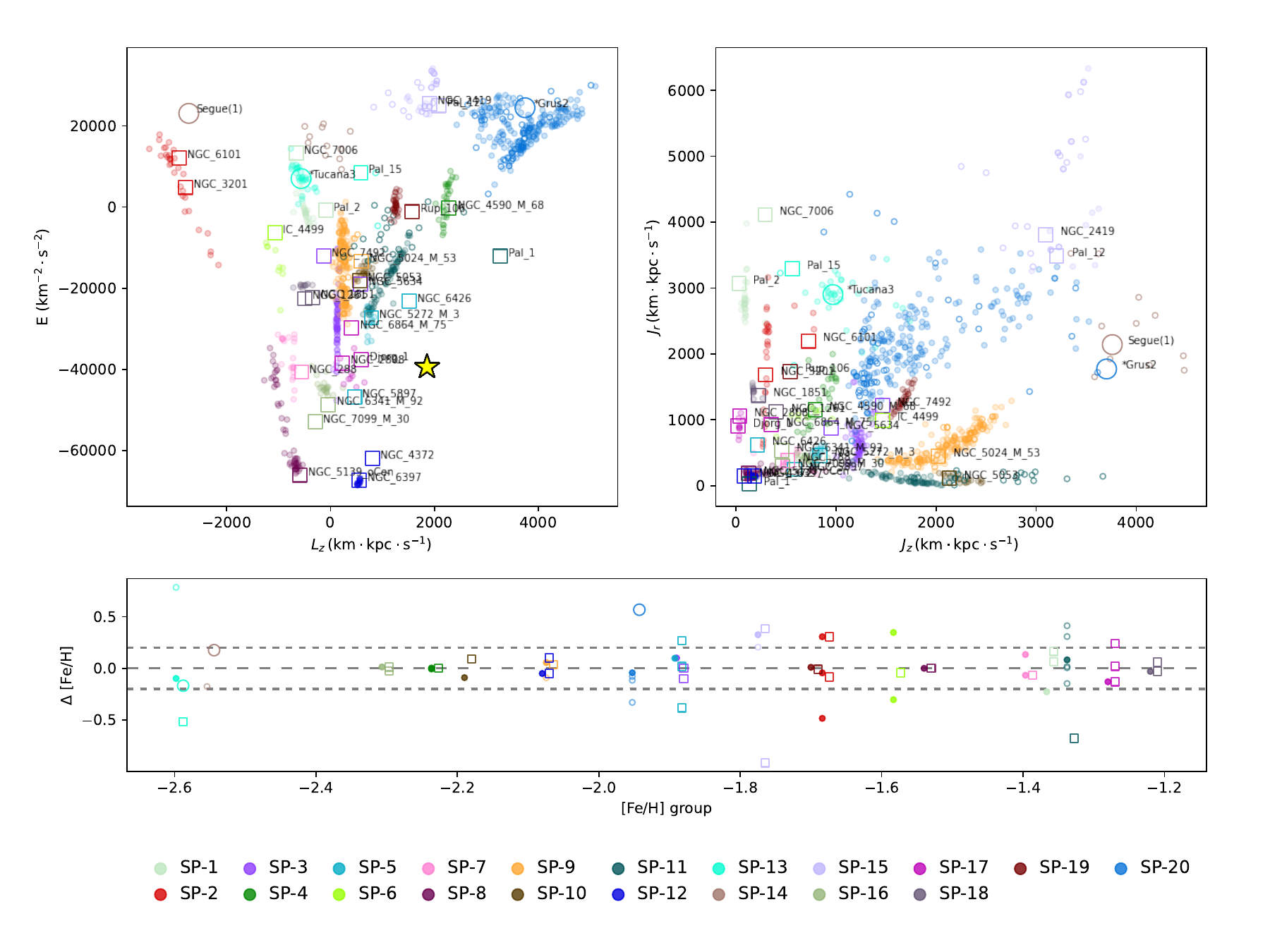}
   \caption{Top: Streams and associated progenitors plotted in Galactic coordinates. Middle: Kinematic parameter spaces E-L$_z$ and J$_r$-J$_z$ are shown in the left and right panels, respectively. Bottom: Differential [Fe/H] of each structure vs the mean [Fe/H] for all streams and progenitors in the group. Filled and open points are stream stars from the STREAMFINDER sample and the S5 sample, respectively. Large open square symbols are for GCs, and large open circle symbols are for Local Group dwarf galaxies.}
   \label{fig:stream_progenitor_plot}
    \end{figure*}

To validate these groupings, we examined their distributions across multiple parameter spaces, including E-L$_z$, J$_r$-J$_z$, and [Fe/H] distributions. Figures \ref{fig:group_validation}, \ref{fig:sub_group_validation} and \ref{fig:stream_progenitor_plot} demonstrate the coherence of these structures in phase space for the larger groups, subgroups, and stream-progenitor associations, respectively. The bottom panels of these Figures show the mean [Fe/H] for all objects in a given group on the x-axis, with each point on the y-axis showing the difference $\Delta$[Fe/H] = [Fe/H]$_\textrm{group}$ - [Fe/H]$_\textrm{object}$ for each GC (open squares), satellite galaxy (open circles), STREAMFINDER stream (filled points) and S5 stream (unfilled points), where each stellar stream is a single point representing the average [Fe/H] of the stream. We applied a horizontal offset of 0.01 dex between the streams and clusters within the groups in order to better visualise the distributions. We opted to depict the metallicity distributions this way as the large number of groups and their broad and highly variable metallicity distributions made histograms messy and difficult to interpret.

\section{Discussion}
\label{sec:discussion}

Within the nine identified groups, we recognised several previously described accreted halo structures, including GSE (G-2), Thamnos (G-3), and Sequoia/Arjuna/I'itoi (G-4) as well as the three distinct populations of GCs that we identified (see Section \ref{subsection:GC_populations}) as bulge (G-6), post-disc formation (G-7), and pre-disc formation (G-8) and a group of satellite galaxies (G-9). We also identified a population of structures linked to a heated disc population (Splash: G-1) and a substantial group of accreted structures (G-5) containing multiple subgroups, including the Helmi streams (sg-10), LMS-1/Wukong (sg-12), Sagittarius (sg-14), and a new potential structure (sg-7) previously discovered in \citet{Malhan_2022a}. In this section, we discuss these identified groups in further detail and put them into context within existing classifications from the literature.

   \begin{figure*}
   \centering
     \includegraphics[width=\textwidth]{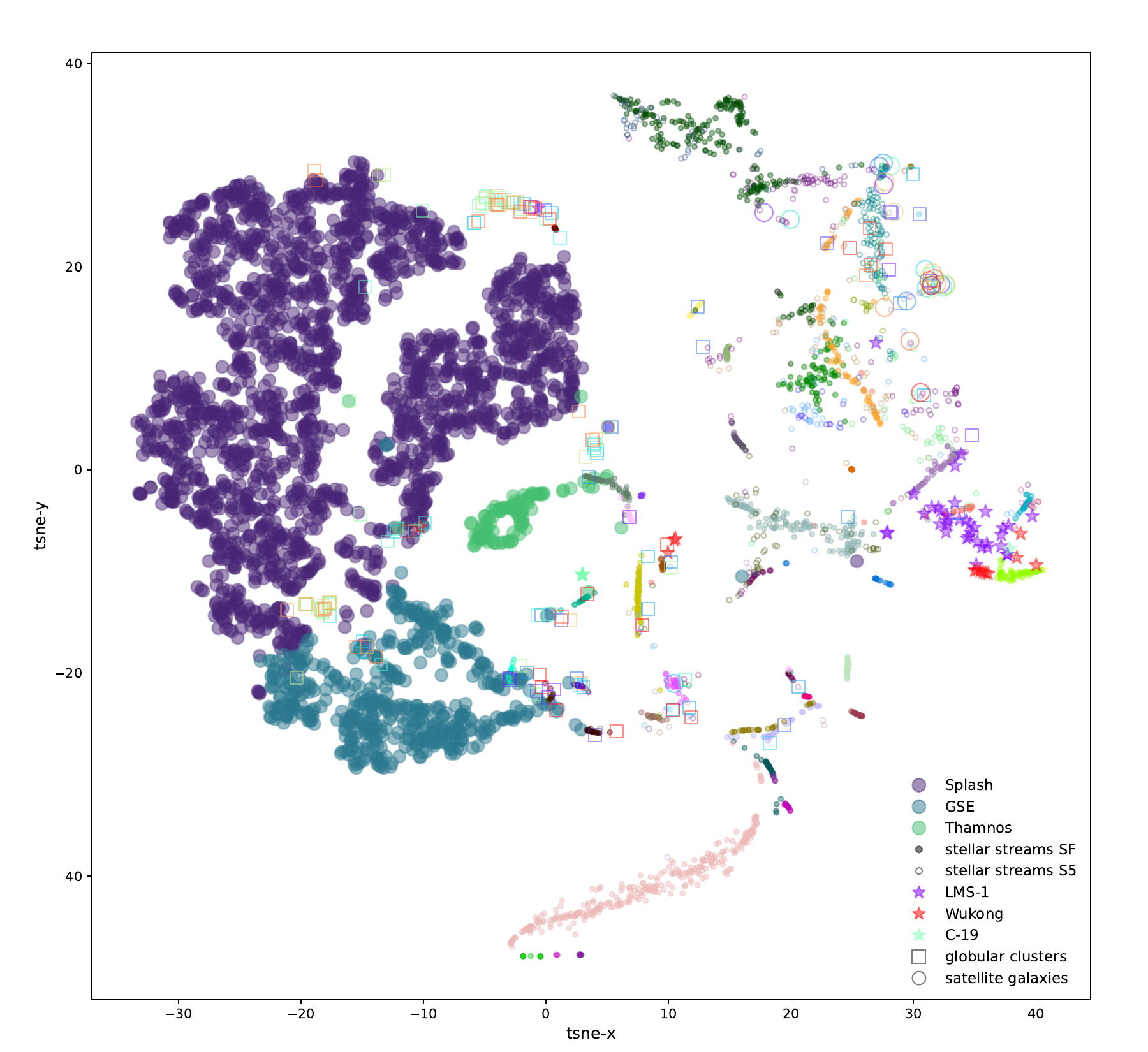}
     \vspace{0.1cm}
     \includegraphics[width=\textwidth]{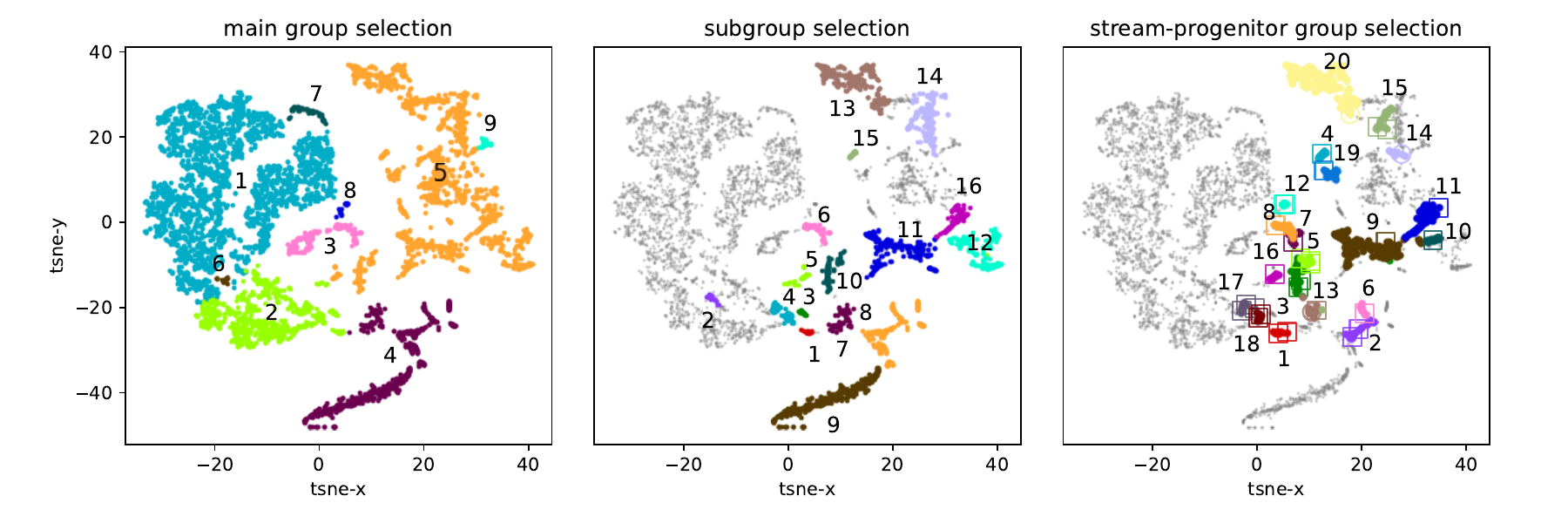} 
   \caption{Top: t-SNE latent space at a perplexity of p = 50. Stream stars from STREAMFINDER are shown as filled coloured points, and stream stars from S5 as open coloured points, GCs are open coloured squares, satellite galaxies are open coloured circles, halo substructures are filled coloured circles, and extra stream stars from the literature are star symbols, coloured according to the legend. Bottom: t-SNE latent space coloured by the selected groups, with the left panel showing the selection of the main groups, the middle panel showing the selection of the subgroups, and the right panel showing the selection of the stream-cluster progenitors. Each group, subgroup and stream-progenitor pair is labelled by its group number which corresponds to the groups listed in Tables \ref{tab:small_associations}, \ref{tab:selected_stream_progenitor}, and \ref{tab:large_associations}.}
   \label{fig:tsne_plot}
    \end{figure*}

   \begin{figure*}
   \centering
     \includegraphics[width=\textwidth]{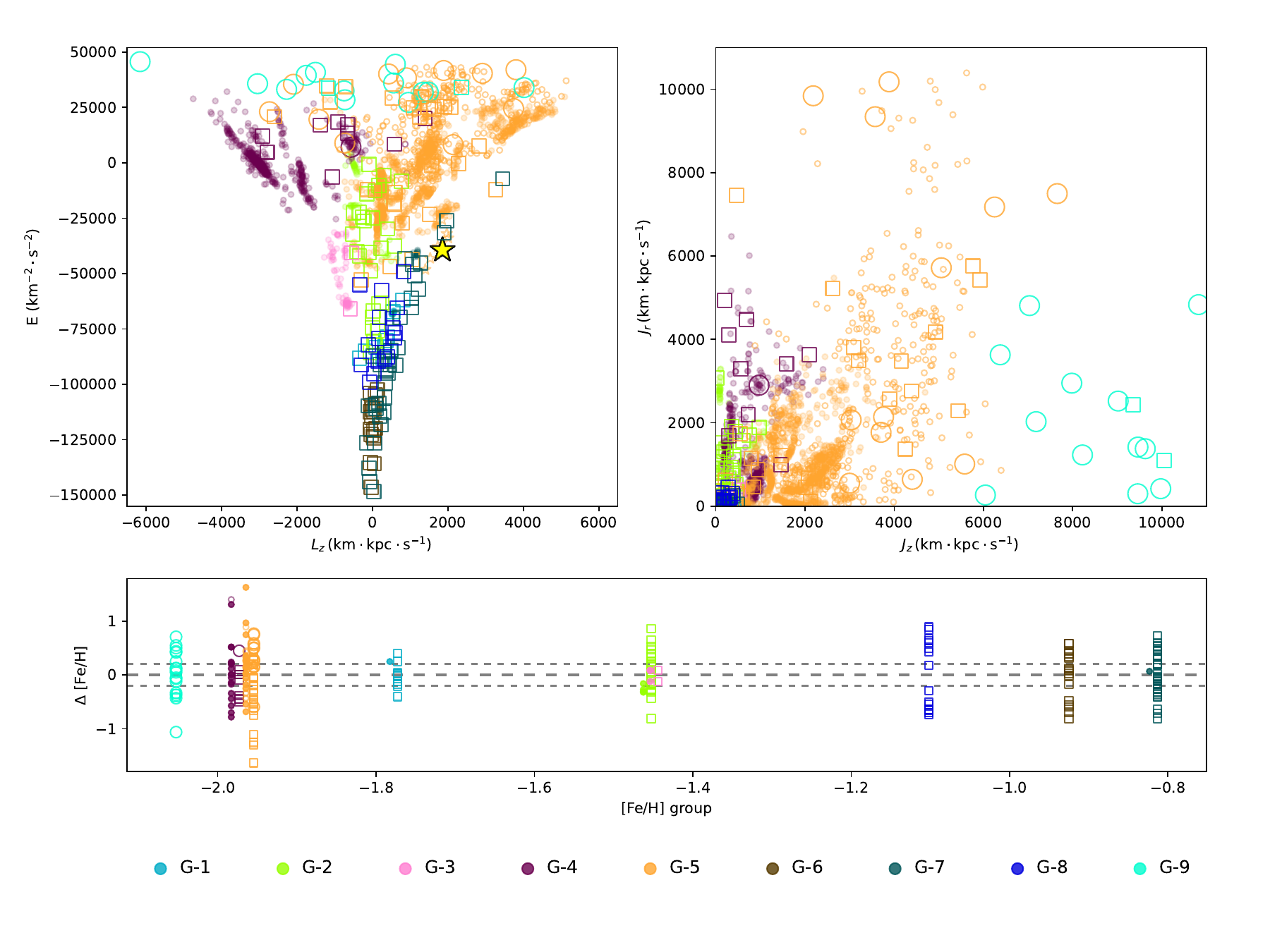}
   \caption{Validation plots for the nine identified groups. Top: Kinematic parameter spaces of E-L$_z$ (left panel) and J$_r$-J$_z$ (right panel). Bottom: Differential [Fe/H] of each structure vs the mean [Fe/H] for all structures in the group (see Section \ref{subsec:results_validation} for a more detailed explanation.) The grey dashed line shows the mean value for each group, and the faint dashed lines show a 0.2 dex dispersion to guide the eye. A horizontal offset of 0.01 dex has been applied between streams and clusters within the groups in order to better visualise the distributions. Structures are coloured consistently with the selection shown in the bottom left panel of Figure \ref{fig:tsne_plot} and described in Table \ref{tab:small_associations}. Filled and open points are stream stars from the STREAMFINDER sample and the S5 sample, respectively. Large open square symbols are GCs and large open circle symbols are Local Group dwarf galaxies.}
   \label{fig:group_validation}
    \end{figure*}

           \begin{figure*}
   \centering
     \includegraphics[width=\textwidth]{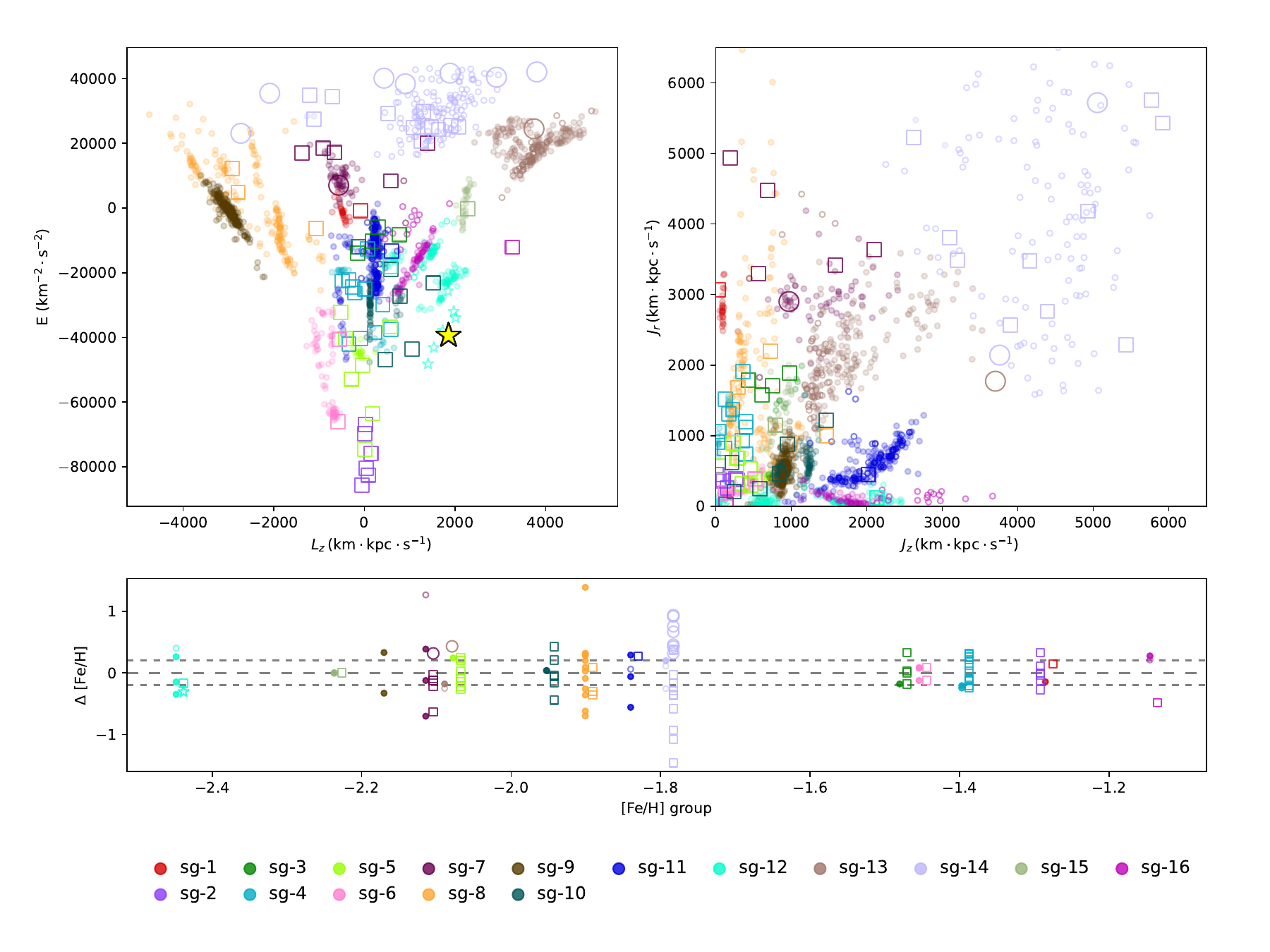}
   \caption{Validation plots for the 16 identified subgroups. Structures are coloured consistently with the selection shown in the bottom-left middle panel of Figure \ref{fig:tsne_plot} and described in Table \ref{tab:large_associations}. Filled and open points are stream stars from the STREAMFINDER sample and the S5 sample, respectively. Large open square symbols are GCs, and large open circle symbols are Local Group dwarf galaxies.}
   \label{fig:sub_group_validation}
    \end{figure*}

\subsection{Main groups}

\subsubsection{Group G-1 (Splash)}
\label{subsubsec:splash}

This group of GCs had chemistry and kinematics that were consistent with tagged stars in the splashed disc population. Therefore, it likely contains both in situ and accreted GCs that are on heated, disc-like orbits. We discuss the properties of this group of GCs further and place them into context with the other groups of Milky Way GCs in Section \ref{subsection:GC_populations}. 

From the purple filled circles in Figure \ref{fig:tsne_plot}, it appears that there could be two subgroups of Splash stars. However, looking at Figure \ref{fig_app:perp_grid} reveals that these subgroups are not present at any other clustering scale. Indeed, if the perplexity value is less than the size of a given group, then the t-SNE projection is likely to artificially separate the larger group into multiple smaller groups. Therefore, we do not attribute any physical significance to these two subgroups.

\subsubsection{Group G-2 (Gaia Sausage-Enceladus)}

Based on the sample of GSE stars selected using GALAH DR4 data from \citet{Kushniruk_2025}, we were able to define the region of the t-SNE latent space in which we expected to find substructures associated with GSE (large blue circles in the top panel of Figure \ref{fig:tsne_plot}). We made this selection using a t-SNE map optimised for large scale clustering, with a perplexity of p = 300. In this larger scale map, the three subgroups of GCs that are separate from the rest of the GSE stars in the p = 50 map (sg-1, sg-3, and sg-5 in the bottom middle panel of Figure \ref{fig:tsne_plot}), are fully connected to the rest of the GSE group. We therefore include these subgroups as part of GSE in Table \ref{tab:large_associations}. In this section we discuss the global properties of this GSE population, and leave the details of investigating these different subgroups to Section \ref{subsubsec:subgroups_GSE}. 

Several previous works have investigated the substructures associated with the GSE merger. For example, \citet{Myeong_2018a} associated 10 GCs to the GSE merger using dynamical arguments. More recent works have associated far more GCs to GSE, with \citet{Massari_2019} associating 32 GCs, selected in action space and corroborated using an age-metallicity relation, and \citet{Forbes_2020} associating 28 GCs. In contrast, \citet{Malhan_2022a} associate fewer GCs with GSE, (16-18 GCs and streams) and suggest that the rest of the GCs belong to another structure, called Pontus \citet{Malhan_2022b}. Based on cross-referencing the GC populations with the literature, we identify subgroup sg-5 as Pontus, and subgroup sg-2 as the Kraken (see Section \ref{subsubsec:subgroups_GSE} for more details). 

In total, we found 29 GCs and 6 stellar streams to be associated with group G-5, GSE. At face value, this number agrees with the higher estimates from the literature \citep{Massari_2019, Forbes_2020}. However, if we exclude the GCs from subgroups sg-2 and sg-5, which likely were brought into the Galaxy through other progenitors, these numbers are reduced to 16 GCs and 4 stellar streams, which is in better agreement with the lower estimates \citep{Myeong_2018a, Malhan_2022a}. Therefore, the misclassification of Pontus and Kraken/Koala GCs as GSE GCs appears to account for the discrepancy between the estimates in the literature. 

\subsubsection{Group G-3 (Thamnos)}
\label{subsubsec:thamnos}

We find two GCs and two stellar streams associated with the Thamnos structure. Notably, \ocen and its associated stellar stream Fimbulthul are part of this group, as well as NGC 288 and its associated stellar stream. We discuss the plausibility of these stream-progenitor pairs to belong to Thamnos along with a more in depth discussion of the associated progenitor of \ocen in Section \ref{subsection:omega_cen}. 

\subsubsection{Group G-4 (Sequoia/Arjuna/I'itoi)}
\label{subsubsec:sequoia}

The Sequoia structure was first discovered as a retrograde moving group in \citet{Myeong_2019}. The authors of that work suggested an association with 6 GCs, FSR 1758, NGC 3201, \ocen (NGC 5139), NGC 6101, NGC 5635, and NGC 6388, as well as a tentative association with the GD-1 stream. In comparison, we found this structure to be associated with NGC 3201, IC 4499, and NGC 6101, but not to the other retrograde GCs. This is consistent with the findings of \citet{Malhan_2022a}, who only associated NGC 6101 and NGC 3201 to this structure.

\citet{Bonaca_2021} associated the GD-1, Gj\"oll, Leiptr, Phlegethon, Wambelong, and Ylgr stellar streams with the Sequoia/Arjuna/I'itoi merger event. Looking at sg-8 and sg-9 of group G-4 in Table \ref{tab:large_associations}, we support the finding that all of these streams are indeed associated with the Sequoia/Arjuna/I'itoi structure (with the exception of Wambelong which was not included in our sample). 
In addition, we add the streams Gaia-1, Gaia-6, Gaia-9, Gaia-11, Gaia-12, NGC 1261, NGC 6101 and Kshir to those belonging to the Sequoia/Arjuna/I'itoi structure. In their analysis, \citet{Malhan_2022a} also found that GD-1, Kshir, and Gaia-9 were associated with this structure.

\subsubsection{Group G-5 (Accreted structures)}
\label{subsubsec:accreted_structures}

The t-SNE latent space in Figure \ref{fig:tsne_plot} shows a large group of stellar streams and GCs, as well as some accreted satellite galaxies. This group corresponds to high-energy, large radial and vertical motion, and unbound objects. We therefore refer to this group as the accreted structures, which corresponds to group G-5 in Tables \ref{tab:small_associations} and \ref{tab:large_associations}. We do not expect that all objects in this group are physically related to a common origin, but we do expect that some smaller groups share a host progenitor galaxy. Figure \ref{fig:tsne_plot} clearly shows a large number of substructures within this larger group. We identified these subgroups through visual inspection of the latent space, paying attention to medium sized structures in the mid-level perplexity maps of p = 30 and p = 50. These selections are shown in the bottom middle panel of Figure \ref{fig:tsne_plot}, and these selected subgroups are summarised in Table \ref{tab:large_associations} and discussed further in Section \ref{subsec:sub_groups}. 

We then performed another visual selection in the latent space, this time focusing on individual objects or small groups of a few objects, paying close attention to instances where streams clearly overlapped with GCs or satellite galaxies. These smallest selected groups are visible in the bottom right panel of Figure \ref{fig:tsne_plot}, and are  discussed further in Section \ref{subsec:stream_cluster_progenitors}. 

\subsubsection{Groups G-6, G-7, and G-8 (bulge GCs and disc GCs)}
\label{subsubsec:disc_bulge_halo_gcs}

Groups G-6, G-7, and G-8 are all compact groups of GCs in the latent space. We assessed whether these represent distinct populations of GCs in Section \ref{subsection:GC_populations}.

\subsubsection{Group G-9 (satellite dwarf galaxies)}
\label{subsubsec:satellite_dgs}

Group G-9 is a tight cluster of satellite Galaxies which were not associated with any other structures. Figure \ref{fig:group_validation} reveals that these are objects with very high energy (E > 25\,000 km$^{-2}$ s$^{-2}$) and very high vertical action ($J_z$ > 6\,000 kpc km s$^{-1}$). This suggests that these satellite galaxies are orbiting in the outskirts of the Milky Way and thus are not likely to be associated with any detectable streams or other structures and are therefore not considered further in the current work.

\subsection{Subgroups}
\label{subsec:sub_groups}

\subsubsection{Subgroups sg-1, sg-3, and sg-4 (subgroups of GSE)}
\label{subsubsec:subgroups_GSE}

In Table \ref{tab:large_associations}, we identify five possible substructures within the GSE population (group G-2), namely sg-1, sg-2, sg-3, sg-4 and sg-5. Upon a more detailed analysis during the validation of our selected groups, we determined that subgroups sg-2 and sg-5 did not belong to the GSE, but rather belong to separate accretions events previously described in the literature, namely Kraken/Koala \citep{Kruijssen_2020, Forbes_2020} and Pontus \citep{Malhan_2022b}, respectively (see Sections \ref{subsubsec:subgroups_kraken} and \ref{subsubsec:subgroups_pontus}). This left three remaining possible subgroups of the GSE: sg-1, sg-3 and sg-4.

Figures \ref{fig:group_validation} and \ref{fig:sub_group_validation} show the orbital properties and [Fe/H] distributions of the whole GSE group, and its subgroups, respectively, but they are difficult to see due to the occlusion from all of the other groups and subgroups also shown in these Figures. Therefore, we also included Figure \ref{fig:gse_validation_plot} in Appendix \ref{app:gc_analyis} to more clearly analyze these different subgroups. Each square represents one GC and they are colour-coded by the subgroup to which they belong. The bottom panel shows the [Fe/H] distributions of each group as a kernel density estimation.

In Figure \ref{fig:gse_validation_plot}, subgroups sg-1, sg-3 and sg-4 are clustered together, but do show a slight gradient in orbital energy and vertical action, J$_r$. Subgroup sg-1 also stands out as having the highest orbital energy, and significantly higher J$_r$ than the other subgroups, but only consists of a single GC, Pal 2. Furthermore, we note that in the top panel of Figure \ref{fig:gse_validation_plot}, 3 out of 4 (75\%) of GCs in sg-3 are in the northern hemisphere, and 9 out of 11 (82\%) GCs in sg-4 are in the southern hemisphere. 

A recent work by \citet{Sun_2023} used kinematics, metallicities and ages to classify Milky Way GCs into different accreted groups and separated the GSE into four separate groups, GSE, GSE-a, GSE-b and GSE-c. After analyzing the detailed properties of these groups they suggested that the three subgroups were likely unrelated to GSE, but noted that further validation is required to confirm this. Compared to our groupings, the main GSE group of \citet{Sun_2023} shares many GCs with our subgroup sg-4, including NGC 362, NGC 1261, NGC 1851, NGC 1904/M 79, NGC 5286, IC 1257, NGC 6981/M 72, and NGC 7089/M 2. None of their other subgroups correspond to either our sg-1 or sg-3 subgroups based on their assigned GCs.

It is important to mention that interpreting present-day Milky Way phase-space is challenging, even with high-precision data. Using N-body simulations, \cite{Koppelman_2020} showed that a GSE-like merger with a Milky Way Galaxy can result in a complicated and highly fragmented phase-space structure of the mixed debris. One consequence of this is that sub-populations of substructures with differing dynamical properties of the in-falling satellite may present themselves as distinct progenitors in purely dynamical data. This reinforces the need for using chemistry when disentangling complex structures in phase-space, which is the aim of the current analysis. Nevertheless, confirming these distinct populations will require a more comprehensive chemical analysis with additional elements, along with sophisticated modelling of such complex merger events.

Therefore, based on these observations alone, we suggest that this is insufficient evidence to conclude that these subgroups represent real progenitors that accreted into the GSE galaxy before it merged with the Milky Way. However, we do remark that our results, as well as those from \citet{Sun_2023}, provide compelling evidence to suggest that there may be several sub-populations within the GSE, warranting further, more detailed investigation in future works.

\subsubsection{Subgroup sg-2 (Kraken/Koala low orbital energy group)}
\label{subsubsec:subgroups_kraken}

Given that they are centrally concentrated in the Galaxy and located at distinctly lower orbital energies than the rest of the GSE GCs, sg-2 is likely a group of low-energy GCs that may not belong to GSE, despite their initial classification. In \citet{Massari_2019}, they found a group of 25 GCs which they failed to label based on their selection criteria, and so dubbed them the unassociated low-energy GCs (L-E). By comparison, our subgroup sg-2 only has six GCs, but we note that all six of them are part of the \citet{Massari_2019} L-E sample. The progenitor for this group of low-energy GCs was later re-named by \citet{Kruijssen_2020} as the Kraken, and predicted to have brought in 13 GCs into the Milky Way with an initial mass of 10$^{8.28}\,\mathrm{M}_{\odot}$. In another work, \citet{Forbes_2020} fit an age-metallicity relation to these low energy GCs and suggest that 21 GCs belong to this progenitor, which they call the Koala. Of the six GCs in our group sg-2, four of them are in the Koala structure. \citet{Forbes_2020} classify NGC 6284 as belonging to GSE, and classify NGC 6121 as an in situ GC due to its age and metallicity, which they point out is consistent with the classification in \citet{Horta_2020} which was based on the cluster's high alpha-element abundance. Furthermore, in their detailed chemical analysis of more than three thousand individual stars associated with GCs, \citet{Horta_2020} claim that the low energy structure from \citet{Massari_2019} is broadly consistent with an in situ origin, with some individual GCs likely being accreted. This picture is consistent with what we find in the current work, namely far fewer GCs in this group than were found in previous works, with the rest of the GCs from the L-E/Kraken/Koala groups being classified as in situ in our analysis and distributed somewhat evenly amongst groups G-1, G-6 and G-8. Therefore, the number of GCs belonging to the Kraken/Koala structure presented in \citet{Kruijssen_2020} and \citet{Forbes_2020} is likely overestimated due to contamination from in situ GCs, and therefore the estimated masses for these structures in those works are likely overestimated.

\subsubsection{Subgroup sg-5 (Pontus)}
\label{subsubsec:subgroups_pontus}

As previously mentioned in Section \ref{subsubsec:subgroups_GSE}, the subgroup sg-5 has a distinctly more metal-poor metallicity distribution than the rest of the selected GSE subgroups, as is shown by the lime green curve in the bottom panel of Figure \ref{fig:gse_validation_plot}. This subgroup has slightly lower orbital energies (E $< -25\,000 \,\mathrm{km}^{-2}\,\mathrm{s}^{-2}$) and radial actions (J$_r$ < 1\,000 km kpc s$^{-1}$) than the other subgroups of GSE clusters, but is very hard to differentiate based on its orbital properties alone.

In a recent work, \citet{Malhan_2022a} used action-angle coordinates plus energy as the inputs into a clustering algorithm and identified a new structure which they named Pontus. They report seven GCs associated with this progenitor: NGC 288, NGC 5286, NGC 7099/M 30, NGC 6205/M 13, NGC 6341/M 92, NGC 6779/M 56, and NGC 362. In a subsequent work, they performed a more detailed chemical and dynamical analysis of the stellar population of Pontus, and supported the initial finding that it is an independent accretion event separate from GSE \citep{Malhan_2022b}.

Our selected subgroup sg-5 shares several GCs in common with Pontus, including NGC 6341/M 92, NGC 7099/M 30, and NGC 6779/M 56. We did not find an association with NGC 288, NGC 5286, NGC 6205/M 13, or NGC 362, but additionally found an association with NGC 2298, NGC 6287, NGC 4833, and ESO 280/SC 06, as well as the C-19 stream. All of these clusters are metal-poor, falling in the range of -2.3 < [Fe/H] < -1.8, with the exception of the C-19 stream which is the most metal-poor stellar stream in the Milky Way with [Fe/H]~$\sim -3.4$. Therefore, despite the differences is specific membership of some GCs, we identify subgroup sg-5 as the Pontus structure from \citet{Malhan_2022a}, and report a somewhat different set of associated GCs.

\subsubsection{Subgroup sg-6 (\ocen/Fimbulthul, C-7 and NGC 288)}
\label{subsubsec:subgroup_ocen}

The associations between NGC 288, \ocen and their respective stellar streams have been described in detail in the literature \citep[e.g.][]{Shipp_2018, Ibata_2019}. Our medium-scale selection in t-SNE latent space selects these four objects plus the C-7 stellar stream \citep{Ibata_2021} in the same subgroup. The light pink points and open squares in Figure \ref{fig:sub_group_validation} shows that this group of objects all occupy the same region in E-L$_z$ space, on retrograde orbits at low orbital energies. Furthermore, they all share very similar metallicities with all of them falling in the narrow range of -1.30 > [Fe/H] < -1.55, which agrees with the association that these structures were accreted with the Thamnos progenitor, which has a broad metallicity distribution with a peak at [Fe/H] $\approx$ -1.5).

However, the sky distribution in the top panel of Figure \ref{fig:sub_group_validation} shows that the NGC 288 GC and stream are located at the souther Galactic cap, while \ocen/Fimbulthul and C-7 are close to the Galactic plane, with C-7 overlapping with the Fimbulthul stars just south of the disc. The Galactic distribution of Thamnos stars are restricted to b < |50|, making it highly unlikely that NGC 288 could be part of the debris brought in with Thamnos, barring some unlikely, complex past dynamical interactions that have drastically altered its orbit to a polar one. Therefore, we conclude that \ocen/Fimbulthul and C-7 were accreted with the same host as Thamnos, but NGC 288 was likely brought into the Galaxy in a different accretion event.

\subsubsection{Subgroup sg-7 (Candidate merger)}
\label{subsubsec:sg7}
During our selection, we initially decided to include this subgroup as part of group G-4, the Sequoia/Arjuna/I'itoi structure, based on its proximity in higher perplexity maps. However, in all the t-SNE maps, it makes a rather tight cluster in the latent space, warranting further investigation that perhaps this subgroup may be related to a separate accretion event. Indeed, this group is clearly separate from the rest of the Sequoia/Arjuna/I'itoi stars in the E-L$_z$ space, sitting at equally high orbital energy, but distinctly less retrograde in its orbit. This group is also distinguished by larger vertical action in the J$_r$-J$_z$ panel, on the right-hand side of Figure \ref{fig:sub_group_validation}. Finally, it is at a lower mean metallicity than both other subgroups belonging to group G-4.

Interestingly, \citet{Malhan_2022a} also noted a similar structure in their clustering analysis. They remark that it was not detected by their automatic algorithm, but rather that they noticed it when manually inspecting the clustering data. Their group contains NGC 5466 (GC and stream), NGC 7492, Gaia-10 and Tucana III. By comparison, sg-7 contains all of the same structures with the exception of NGC 7492 (sg-10), and with the addition of NGC 5694, Pal 15, NGC 6934, Pal 13, and the Tuc-III stream. NGC 6934 and Pal 15 are on prograde orbits and are outliers from the rest of the group in the E-L$_z$ panel of Figure \ref{fig:sub_group_validation}, and NGC 5694 and NGC 6934 are also an outliers in J$_r$-J$_z$ with J$_r$ > 4000 kpc km s$^{-1}$.

If we ignore the above outliers in orbital properties, the rest of the streams and clusters have metallicities in the narrow range -2.1 < [Fe/H] < -1.9, with the exception of Tuc-III with [Fe/H] $\sim$ -2.5 and Gaia-10 with [Fe/H] $\sim$ -1.4. Therefore, we deem this to be a likely group accreted with a common progenitor. This is in agreement with the conclusions of \citet{Malhan_2022a}, who classify this as a candidate merger, after analysis of the orbits and CMDs of these objects.

\subsubsection{Subgroups sg-8 and sg-9 (subgroups of Sequoia/Arjuna/I'itoi)}
\label{subsubsec:subgroups_sg8_sg9}

Now that we have established that subgroup sg-7 is not part of Sequoia/Arjuna/I'itoi, we have two remaining subgroups, sg-8 and sg-9.

\citet{Naidu_2020} demonstrated that the retrograde halo can be separated into three structures, Arjuna, Sequoia and I'itoi, which can be differentiated by their metallicities distributions. More specifically, they claim that Arjuna stars are the most metal-rich, occupying the regime of [Fe/H] > -1.5, Sequoia stars are found with -2 < [Fe/H] < -1.5, and I'itoi populating the metal-poor regime at [Fe/H] < -2.

The bottom panel of Figure \ref{fig:sub_group_validation} shows that subgroup sg-9 is the most metal-poor subgroup of group G-5, with [Fe/H] < -2.1. Therefore, we propose that GD-1 ([Fe/H] $\sim$ -2.5) and Kshir (sg-9) belong to the I'itoi structure, while subgroup sg-8 represents the Sequoia accretion event. However, looking into more detail at the [Fe/H] distribution in the bottom panel of Figure \ref{fig:sub_group_validation}, the orange points of sg-8 show a broad metallicity distribution, with some outliers. When we investigated this further, we found that the three of the most metal-poor streams in sg-8, Leiptr ([Fe/H] $\sim$ -2.17, Ylgr ([Fe/H] $\sim$ -2.09), and Gaia-12 ([Fe/H] $\sim$ -3.28) all fall in the region of the latent space between sg-8 and sg-9, at [tsne-x, tsne-y] = [20, 30] in Figure \ref{fig:tsne_plot}. Therefore, we propose that these metal-poor streams were initially selected in the wrong subgroup, and are actually part of the I'itoi structure rather than Sequoia.

\subsubsection{Subgroups sg-10 and sg-11 (Helmi streams)}
\label{subsubsec:subgroups_helmi_streams}

It has been suggested that the Helmi streams are an important donor to the Milky Way halo, contributing $\sim$15\% of its mass in field stars and 10\% of its GCs \citep{Koppelman_2019a}. Furthermore, \citet{Koppelman_2019a} suggest through dynamical arguments that the Helmi streams are associated with seven halo GCs: NGC 4590, NGC 5024, NGC 5053, NGC 5272, NGC 5634, NGC 5904, and NGC 6981. 

Given that subgroups sg-10 and sg-11 contain NGC 5024, NGC 5272/M 3, and NGC 5634, it is likely that these groups are related to the Helmi streams. However, we note that more than 50\% of GCs associated with the Helmi streams in \citet{Koppelman_2019a} and \citet{Forbes_2020} are assigned to different structures in our analysis. Furthermore, sg-10 and sg-11 have different J$_z$ distributions in the right panel of Figure \ref{fig:sub_group_validation}, which may indicate that sg-11 is connected to a different accretion event separate from the Helmi streams.

As for the remaining GCs expected to be associated with the Helmi streams, NGC 4590 and its associated stellar stream Fj\"orm are in sg-15, located far away in the latent space from sg-10 and not associated with any larger structure. NGC 5904 and NGC 6981 are associated with GSE, although they are located spatially near to sg-10 in the latent space. NGC 5053 is part of subgroup sg-12 which is adjacent to sg-10 and sg-11 in the latent space but clearly a separate structure, and is associated with the LMS-1/Wukong structure.

\subsubsection{Subgroup sg-12 (LMS-1/Wukong)}
\label{subsubsec:lms1_wukong}

Due to the inclusion of the tagged samples of know members from Wukong \citep{Limberg_2024} and LMS-1 \citep{Malhan_2021}, we identified subgroup sg-12 as the LMS-1/Wukong structure. In addition to LMS-1 and Wukong stars, we also associated the metal-poor Phoenix and Kwando streams, with [Fe/H] $\sim$ -2.7 and -2.3, respectively. We also found the GC NGC 5053 to be associated with this group, which is corroborated by previous results which have reported a connection between this GC and the LMS-1 stream based on their similar orbital properties \citep{Yuan_2020, Malhan_2021}.

\citet{Malhan_2022a} report three other GCs and five other stellar streams associated with this group, namely NGC 5272/M 3, NGC 5024/M 53, Pal 5, C-19, Indus, Sylgr and Jhelum. In our current analysis, the stream-progenitor pair SP-9 of Jhelum and NGC 5024/M 53 is adjacent to subgroup sg-12 in the latent space and therefore could be connected to the LMS-1/Wukong structure. However, in our current analysis, NGC 5272/M 3 is associated with subgroup sg-10, the Helmi streams, which is in agreement with previous findings, although those also associate NGC 5024/M 53 and NGC 5053 with the Helmi streams \citep{Massari_2019, Koppelman_2019a}. This emphasises the challenge with associating GCs to specific accretion events, and highlights the different results that occur when different selection criteria are implemented.

Finally, we note that \citet{Malhan_2022a} associate the metal-poor streams Sylgr, Phoenix and C-19 to LMS-1/Wukong and therefore describe it as the most metal-poor accretion event in the Milky Way's history. However, they do not associate the metal-poor stream Kwando, which we add to the metal-poor streams associated with LMS-1/Wukong based on our analysis.

\subsubsection{Subgroup sg-14 (Sagittarius)}
\label{subsubsec:sagittarius_dwarf}

This subgroup contains twelve GCs, three stellar streams and seven satellite galaxies. The GCs that we associate with this group are very similar to previous the GCs associated with the Sagittarius progenitor in previous studies \citep{Massari_2019, Forbes_2020, Malhan_2022a}. Interestingly, \citet{Malhan_2022a} only associate one stream, Elqui, to the Sagittarius structure. By comparison, we do not associate the Elqui stream, but instead the Sagittarius stream, Indus stream, and NGC 5466 stream. These streams are entirely different than those associated with Sagittarius based on their orbital characteristics by \citet{Bonaca_2021}, namely Aliqu-Uma, ATLAS, Fj\"orm, Slidr and Turranburra. This discrepancy highlights the inherent uncertainties of associating streams to progenitor galaxies, and reinforces the need for higher precision measurements of streams and their progenitors to improve orbital calculations.

Looking at Table \ref{tab:large_associations}, we also associated several satellite galaxies to the Sagittarius structure during the selection of subgroups in the t-SNE latent space. However, given that sg-14 is the structure with the highest orbital energy of all of the identified structures (see lilac symbols in Figure \ref{fig:sub_group_validation}), it is highly probable that the high energy satellite galaxies only clustered in the same region of the latent space coincidentally. Therefore, we do not consider these satellite galaxies to be physically associated with the Sagittarius structure.

\subsection{Stream-progenitor associations}
\label{subsec:stream_cluster_progenitors}

We also identified small-scale structures representing stream-progenitor pairs that connect stellar streams with their original GCs or dwarf galaxies. These small scale structures provide crucial insight into the accretion and disruption processes shaping the Milky Way's halo. We also allowed for the possibility of small associations of multiple streams and GCs, or a single stream and multiple GCs. These associations, shown in the bottom right panel of Figure \ref{fig:tsne_plot}, are summarised in Table \ref{tab:selected_stream_progenitor}, with identifiers designated as SP-X (Stream-Progenitor-X).

Figure \ref{fig:stream_progenitor_plot} shows the sky distribution of these stream-progenitor systems in Galactic coordinates, as well as the E versus L$_z$ and J$_r$ versus J$_z$ planes, and the [Fe/H] distributions of the stellar streams and the associated progenitor GCs or satellite galaxies. Typically, if streams and progenitors share a common origin, they should be spatially coherent along the stream's orbital trajectory, share similar orbital properties, and have similar metallicity distributions. Of course, we expect that the identified groups should be somewhat clustered in these parameter spaces, since these were the input parameters used to generate the t-SNE latent space from which the groups were identified. However, here we investigate in detail these parameter spaces in order to validate these associations, and discuss the plausibility of these identified groups having a shared Galactic origin.

We first selected all small groups in the t-SNE latent space where stellar streams and GCs or satellite galaxies appeared to clustered in the latent space. We made this selection by hand using the smallest scale t-SNE maps, with perplexity of p~=~5, p~=~10, and p~=~30, and intentionally selected all plausible associations in this step so as to favour completeness over purity. We then removed all streams from each group where the fraction of the stream stars was less than 30\%, which excluded streams for which only a few stars were randomly scattered into the selected group. The selection of these objects is shown in the bottom right panel of Figure \ref{fig:tsne_plot}, and the selected groups are summarised in Table \ref{tab:selected_stream_progenitor}.  

We recovered several previously identified pairs of streams and progenitors from the literature. These include: \ocen - Fimbulthul, Gj\"oll - NGC 3201, NGC 6101 (GC) - NGC 6101 (stream), Fj\"orm - NGC 4590/M68, Svol - NGC 5272/M 3, NGC 288 (GC) - NGC 288 (stream), LMS-1 - NGC 5053, NGC 6397 (GC) - NGC 6397 (stream), Tucana III (DG) - Tuc-III (stream), NGC 6341/M 92 (GC) and M 92 (stream), NGC 2808 (GC) - NGC 2808 (stream), and NGC 1851 (GC) - NGC 1851 (stream). These are indicated in Table \ref{tab:selected_stream_progenitor} in the discovery column with a listing of the sources in the literature where the association was first presented. 

In this table, we also included a qualifier for how confident the detection of a given stream was. In order for a stream-progenitor pair to have been given a value of 'strong' in the detection column, they must have satisfied all three of the following criteria: overlapping in E-L$_z$ and J$_r$-J$_z$ space, metallicity values within 0.2 dex for stream/GC pairs or 0.5 dex for stream/satellite galaxy pairs, and clear proximity on the sky of stream tracks and progenitor positions. Associations that satisfied two of the three criteria were labelled 'tentative', associations that satisfied only one of the criteria were labelled as 'weak', and objects which were found not to be associated after further analysis were given a label of 'none'.

The top panel of Figure \ref{fig:stream_progenitor_plot} shows the sky positions in Galactic coordinates of each of the selected stream-progenitor associations, colour-coded based on the selected SP-X groups in the bottom right panel of Figure \ref{fig:tsne_plot}. We also plot the galstreams tracks underneath the stream stars so as to show the expected orbital trajectory for each stream, which allowed for the confirmation of the spatial connection in some instances when the progenitor was located far away from the stream stars, which was the case for Gj\"oll - NGC 3201, Fj\"orm - NGC 4590/M68, and Svol - NGC 5272/M 3.

In addition to the stream-progenitor pairs previously described in the literature, we also present several new stream-progenitor associations in this work.

\subsubsection{Hrid - Pal 2 and NGC 7006}

The SP-1 group consists of the Hrid stream \citep{Ibata_2021} and the two GCs Pal 2 and NGC 7006. The three objects form a very tight group in the latent space, and also trace the same regions in the E-$L_z$ and J$_r$-J$_z$ diagrams in the middle panels of Figure \ref{fig:stream_progenitor_plot}. However, Hrid has a spectroscopically measured metallicity of -1.13, whereas the GCs Pal 2 and NGC 7006 are more metal poor with [Fe/H] = -1.42 and -1.52, respectively. Although the [Fe/H] measurements are only discrepant by 0.3-0.4 dex, we would expect them to be closer if either Pal 2 or NGC 7006 were the progenitor of the Hrid stream, given that they are GCs and have narrow spreads in [Fe/H]. However, the [Fe/H] measurements for the Hrid stream is based on a small number of stars and therefore may be refined to a slightly lower metallicity in with future follow-up efforts. The spatial distribution of these three objects in Figure \ref{fig:stream_progenitor_plot} shows that they are all located nearby on the sky, but that neither Pal 2 nor NGC 7006 lies on the stream track of the Hrid stream available from current data. Therefore, the current data are suggestive, but not sufficient to support the that either of Pal 2 or NGC 7006 are the progenitor for the Hrid stream. We therefore give this association a classification of `weak' in Table \ref{tab:selected_stream_progenitor}. It is, however, likely that these three objects were brought into the Galaxy with a common progenitor.

\subsubsection{NGC 6101 - Gj\"oll - NGC 3201}

Both the NGC 6101 GC - stream \citep{Ibata_2021} and the Gj\"oll - NGC 3201 \citep{Bianchini_2019, Riley_2020, Hansen_2020a} pairs have been previously individually described in the literature, but not previously associated as a larger complex. In our current analysis, all four of these objects were selected together as part of subgroup sg-10 and therefore may be connected to a common progenitor. In the top panel of Figure \ref{fig:stream_progenitor_plot}, the two stream-progenitor pairs do not obviously trace similar orbits in galactic coordinates, but do have somewhat similar metallicities, with [Fe/H] = -1.98 for NGC 6101 \citep{Ibata_2021}, -1.63 for Gj\"oll \citep{Ibata_2021} and -1.59 for NGC 3201 \citep{Vasiliev_2019}. Therefore, we propose that these two GCs and streams came into the Galaxy with the same progenitor that formed the retrograde Sequoia/Arjuna/I'itoi structure.

\subsubsection{Ophiucus - NGC 5634, NGC 7492}

The Ophiucus stream \citep{Bernard_2014} is a short, thin stream in the Northern Galactic hemisphere. \citet{Sesar_2015} performed spectroscopic follow-up and identified 14 members of the stream, from which they inferred that the stream originated from an old, metal-poor globular cluster progenitor with an age of 11.7 Gyr and [Fe/H] = -1.95. However, they ultimately do not detect any overdensity along the stream and conclude that the stream is all that is left of the progenitor.

In our current analysis, we find two GCs that may potentially be the progenitor of the Ophiucus stream, namely NGC 5634, and NGC 7492. Looking at SP-3 in Figure \ref{fig:stream_progenitor_plot}, both of these GCs have a plausible connection to Ophiucus based on their positions in E-L$_z$ and J$_r$-J$_z$, and also their [Fe/H] values. Ophiucus has [Fe/H] = -1.98 \citep{Ibata_2021}, which is very much consistent with the values of [Fe/H] = -1.88 and -1.78 for NGC 5634 and NGC 7492, respectively. Even the Galactic coordinates of both GCs could be aligned with the orbital path of Ophiucus, although they are both located well beyond the streams visible extent on the sky. 

Generally, it is expected that the velocity dispersion of the progenitor should be less than or equal to that of the tidal tails of the stream. Therefore, NGC 5634 can be ruled out on the basis of its high velocity dispersion of $5.3 \, \mathrm{km} \,\mathrm{s}^{-1}$, compared to that of Ophiucus which is $1.6 \pm 0.4 \, \mathrm{km} \,\mathrm{s}^{-1}$ \citep{Caldwell_2020}. NGC 7492 also has a velocity dispersion of $1.6 \, \mathrm{km} \,\mathrm{s}^{-1}$, but is also located much farther away on the sky from the Ophiucus stream stars in the Southern hemisphere, reducing its likelihood of being connected to Ophiucus. 

\citet{Lane_2020} conducted detailed N-body simulations based on the current observed properties of the Ophiucus stream, and suggest that the progenitor was likely a faint, weakly bound GC with a mass of $2 \times 10^3 M_{\odot}$ or less, a half-mass radius in the range 60–100 pc, and began disrupting just 360 Myr ago. Of course, this assumes that most of the streams material is contained within its present day location, and that no progenitor is currently visible. However, this further reduces the possibility of either NGC 5634 or NGC 7492 being the progenitor. We therefore give both of these a detection significance of weak in Table \ref{tab:selected_stream_progenitor}.

\subsubsection{Jhelum - NGC 5024}

The t-SNE projection in Figure \ref{fig:tsne_plot} shows the Jhelum stream and NGC 5024/M 53 clearly overlapping in the latent space. Further investigation of the orange symbols in Figure \ref{fig:stream_progenitor_plot} shows that these objects are overlapping in the E-L$_z$ and J$_r$-J$_z$ plots and that they have a remarkably similar metallicity, both with [Fe/H] $\sim$ -2.1. The top panel of Figure \ref{fig:stream_progenitor_plot} shows that Jhelum is located in the southern hemisphere, while NGC 5024/M 53 is located at the north pole. Given that it appears both objects are on polar orbits, this does not rule out a connection between them although they are located far away from one another on the sky.

In a previous work, \citet{Malhan_2021} proposed a connection between LMS-1, NGC 5053, Indus and NGC 5024. They specifically pointed out that since Jhelum was unrelated to LMS-1, it effectively implied that it was also unrelated to Indus and NGC 5024. However, this result is in contention with a recent analysis that indicated a common origin of the Indus and Jhelum streams \citep{Bonaca_2021}, and several previous works that have suggested that Indus and Jhelum have very similar orbital properties and thus may represent different orbital wraps of the same stream \citep[e.g.][]{Shipp_2018, Bonaca_2019}.

Therefore, due to their very similar orbits and metallicities, but lack of a direct spatial connection between Jhelum and NGC 5024/M 53, we propose that NGC 5024/M 53 is the progenitor of the Jhelum stream, and classify this as a tentative association in Table \ref{tab:selected_stream_progenitor}.

\subsubsection{Elqui - Segue I}

Elqui is a metal-poor and has a broad metallicity distribution, with [Fe/H] $\sim$ -2.22 $\pm$ 0.27 \citep{Li_2022}. The spectroscopic sample used in this work has a range of -3.1 < [Fe/H] < -1.5. Therefore, due to its broad metallicity distribution, Elqui is likely to be the stellar stream of a dwarf galaxy progenitor \citep{Ji_2020, Li_2022}. Segue I is a dwarf galaxy with a measured metallicity of [Fe/H] $\sim$ -2.72, a r$_{peri}$ = 21 kpc, r$_{apo}$ = 82 kpc, and e = 0.6 according to our orbital analysis. Elqui has 8 kpc < r$_{peri}$ < 20 kpc, 50 kpc < r$_{apo}$ < 85 kpc, and 0.5 < e 0.7, making their orbital properties very similar.

Figure \ref{fig:stream_progenitor_plot} shows that Elqui and Segue I have similar radial and vertical actions, but fairly discrepant angular momenta. Elqui is on a near polar, slightly retrograde orbit while Segue I is on a highly retrograde orbit. Furthermore, Elqui is currently located at the southern pole of the Galaxy, while Segue I is in the northern hemisphere. Taken together, despite the orbital properties of Elqui and Segue I being somewhat similar, and their compatible [Fe/H] distributions, the likelihood of Segue I being the progenitor of Elqui is weak given the difference in angular momenta of their orbits.

\subsubsection{Indus - NGC 2419}

In the current analysis, we find a grouping SP-15 which includes Indus, NGC 2419, and Pal 12. The top two panels of Figure \ref{fig:stream_progenitor_plot} show that these three objects are located on a prograde orbit at very high orbital energy (E $> 20\,000 \,\mathrm{km}^{-2}\,\mathrm{s}^{-2}$), and large values of vertical and radial action.

Based on sky position, Pal 12 is favoured as the potential progenitor for Indus, which is located at [l, b] = [-30$^{\circ}$, -50$^{\circ}$], while NGC 2419 is very far away at [l, b] = [-180$^{\circ}$, 25$^{\circ}$]. However, based on the [Fe/H] value of Indus of [Fe/H] $\sim$ -2.1, NGC 2419 is favoured with [Fe/H] $\sim$ -2.15 while Pal 12 is ruled because it is much more metal-rich at with [Fe/H] $\sim$ -0.85. Therefore, we exclude Pal 12 as a possible progenitor for Indus, and suggest that NGC 2419 is an unlikely progenitor, but likely was accreted in the same accretion event as Indus.

\subsubsection{Slidr - Rup 106}

In the bottom two panels of Figure \ref{fig:stream_progenitor_plot}, the Slidr stream and Rup 106 GC both have similar energy and L$_z$, but very different vertical actions, J$_z$. The bottom panel shows a negligible metallicity difference, with Slidr having [Fe/H] $\sim$ -1.7 \citep{Ibata_2024} and Rup 106 [Fe/H] $\sim$ -1.68, which is likely why they clustered so tightly in the latent space. However, their sky positions do not support a likely stream progenitor connection. Therefore, due to their spatial separation and differences in $J_z$, it is not likely that Rup 106 is the progenitor of the Slidr stream, although these two objects may have come into the Milky Way with a common host.

\subsubsection{Orphan-Chenab - Grus II}

The Orphan stream \citep{Grillmair_2006, Belokurov_2007} is a very long cold stream in the Milky Way halo that has been extensively studied. Nevertheless, no convincing progenitor has yet been found for this stream.

A recent study by \cite{Koposov_2019} used a sample of RR Lyrae stars to map out the stream and made some estimates of its progenitor properties. They estimate that the progenitor mass is most likely to be $\sim 4 \times 10^6 M_\odot$, with a luminosity of $L_V = 3.8 \times 10^5 L_\odot$. In another recent work, \citet{Hawkins_2023} performed a detailed chemical analysis of members of the Orphan stream and concluded that based on the [Mg/Al] abundances, as well as the metallicity spread, the Orphan stream is not likely to have a GC progenitor but rather a dwarf spheroidal galaxy, and they provide a mass estimate of this progenitor of $\sim 10^6 M_\odot$.

Several candidates have previously been investigated as potential progenitors for the Orphan stream, including Segue I  and UMa II \citep{Fellhauer_2007, Newberg_2010}, but neither of these have resulted in a convincing connection to the stream. More recently, \citet{Koposov_2019} proposed a connection between Grus II and the Orphan stream when they noticed a clear coincidence in sky coordinates, proper motion, and velocity space. They remarked that the distance of Grus II is about 10 kpc farther than the debris from the Orphan stream at the same location. Though they conclude that a connection is likely, they still suggest that further chemical analysis would be required to confirm this.

Our chemo-dynamical analysis with t-SNE also showed a clear connection between the Orphan stream and the Grus II ultra-faint dwarf galaxy. Figure \ref{fig:stream_progenitor_plot} shows that both the Orphan stars and Grus II share very high E and L$_z$ values in the top right portion of the plot. In the J$_r$-J$_z$ plane, however, Grus II has a significantly higher value of vertical action than the rest of the Orphan stream stars. Grus II has [Fe/H] = -2.5 \citep{Simon_2020}, which makes it more metal-poor than the Orphan stream with [Fe/H] = -1.85. This fact alone is not problematic, as the more metal-poor Grus II could still lie in the metal-poor tail of the metallicity distribution of the Orphan stream. However, the fact that Orphan stream is more metal-rich than its potential progenitor cluster does pose an issue for a connection between the two objects, as pointed out by \citet{Prudil_2021}. They reasoned that as Grus II interacted with the Milky Way its outer envelope would have been shed to produce the Orphan stream. Given that these stars are more metal-rich than the remaining core of the UFD, it follows that for Grus II to be the progenitor of the Orphan stream, it would need to have an inverse metallicity gradient. Although inverse metallicity gradients have been observed in high-red shift galaxies \citep{Grossi_2020}, they have not yet been observed in any galaxies in the Local Group.

In a recent work, \citet{Hansen_2020b} performed a high resolution analysis of three stars in Grus II, and found very low metallicities and enhanced [Al/Fe] for these stars. Although, with just three stars with detailed abundance measurements, a comprehensive chemical comparison to the Orphan stream sample is not currently possible. However, if future follow-up of Grus II stars show evidence of an inverse metallicity gradient in the UFD, this would highly favour the scenario that Grus II is the progenitor of the Orphan stream.

\citet{Koposov_2023} performed an in-depth analysis of the stream properties for the Orphan-Chenab stream, and conducted backward integration orbital analysis of all Milky Way satellite galaxies and GCs. They found that the Orphan-stream was consistent with several close passages with Grus II ($\lesssim$ 0.1 kpc, and at low relative velocities of $\lesssim$ 50 km s$^{-1}$) and therefore suggested that Grus II could have been bound to the same original host as the Orphan-Chenab stream.

Taken with these previous investigations, the current results strongly support a connection between the OC stream and the Grus II UFD and thus also support the scenario that they share the same original host galaxy. Nonetheless, these results are insufficient to claim that Grus II is the progenitor of the OC stream. However, future work to better understand the metallicity gradient in Grus II could offer new insights into this scenario.

\begin{table*}
\centering
\caption{Stream-progenitor groups from the smallest scale selection in the t-SNE latent space.}
\label{tab:selected_stream_progenitor}
\begin{tabular}{|c|l|l|l|l|}
\hline
Group & \multicolumn{1}{c|}{Stellar stream(s)} & \multicolumn{1}{c|}{Progenitor(s)} & Detection & Discovery \\ \hline
SP-1 & Hrid & Pal 2, NGC 7006 & Weak & New \\ \hline
SP-2-1 & \begin{tabular}[c]{@{}l@{}}NGC 6101,\\ Gaia-11\end{tabular} & NGC 6101 & Strong & \citealt{Ibata_2021} \\ \hline
\multicolumn{1}{|l|}{SP-2-2} & Gj\"oll & NGC 3201 & Strong & \begin{tabular}[c]{@{}l@{}} \citealt{Bianchini_2019}, \\ \citealt{Riley_2020}, \\\citealt{Hansen_2020a}\end{tabular} \\ \hline
SP-3 & Ophiuchus & NGC 5634, NGC 7492 & Weak & New \\ \hline
SP-4 & Fj\"orm & NGC 4590/M 68 & Tentative & \citealt{Palau_2019} \\ \hline
SP-5 & Svol & NGC 5272/M 3 & Strong & \begin{tabular}[c]{@{}l@{}}\citealt{Bonaca_2021}, \\ \citealt{Yang_2023}\end{tabular} \\ \hline
SP-6 & \begin{tabular}[c]{@{}l@{}}NGC 1261,\\ NGC 2298\end{tabular} & IC 4499 & None & None \\ \hline
SP-7 & NGC 288 & NGC 288 & Strong & \citealt{Shipp_2018} \\ \hline
SP-8 & Fimbulthul & NGC 5139/\ocen & Strong & \citealt{Ibata_2019} \\ \hline
SP-9 & Jhelum & NGC 5024/M 53 & Tentative & New \\ \hline
SP-10 & LMS-1 & NGC 5053 & Strong & \begin{tabular}[c]{@{}l@{}}\citealt{Yuan_2020}, \\ \citealt{Malhan_2021}\end{tabular} \\ \hline
SP-11 & Pal-5, Ravi & Pal 1 & None & None \\ \hline
SP-12 & NGC 6397 & NGC 6397 & Strong & \citealt{Ibata_2021} \\ \hline
SP-13 & Tuc-III, & Tucana III & Strong & \begin{tabular}[c]{@{}l@{}}\citealt{Drlica_2015}, \\ \citealt{Shipp_2018}\end{tabular} \\ \hline
SP-14 & Elqui & Segue I & Weak & New \\ \hline
SP-15 & Indus & NGC 2419 & Weak & New \\ \hline
SP-16 & M 92 & NGC 6341/M 92 & Strong & \begin{tabular}[c]{@{}l@{}}\citealt{Sollima_2020}, \\ \citealt{Thomas_2020} \end{tabular}\\ \hline
SP-17 & NGC 2808 & NGC 2808 & Strong & \begin{tabular}[c]{@{}l@{}}\citealt{Carballo_2018}, \\ \citealt{Kundu_2021}, \\ \citealt{Ibata_2021}\end{tabular} \\ \hline
SP-18 & NGC 1851 & NGC 1851 & Strong & \begin{tabular}[c]{@{}l@{}}\citealt{Carballo_2018}, \\ \citealt{Shipp_2018}, \\ \citealt{Ibata_2021}\end{tabular} \\ \hline
SP-19 & Slidr & Rup 106 & Weak & New \\ \hline
SP-20 & Orphan & Grus II & Strong & \citealt{Koposov_2019} \\ \hline
\end{tabular}
\tablefoot{The detection column summarises the results of the validation of the pairs with a value of strong, tentative, weak, or none, as described in Section \ref{subsec:stream_cluster_progenitors}. The discovery column lists the source where the pairing was first described in the literature, or has a value of 'new' if this is a new association.}
\end{table*}

\vspace{0.5cm}
\subsection{Globular cluster populations}
\label{subsection:GC_populations}
One of the most interesting outcomes of our clustering analysis is the identification of eight seemingly distinct groups of GCs. Each of these groups either clearly belongs to a tagged population (for example Splash, GSE, Thamnos, Sequoia/Arjuna/I'itoi) or form tight groups in the latent space that are clearly separate from any other groups. Therefore, our selection is a robust and effective method for separating Milky Way GCs into similar groups, with only a few cases where GCs fall in between selection groups. Below we discuss how these groups of GCs compare to previous classifications in the literature.

In a recent work, \citet{Belokurov_2023} define a selection criteria in E-L$_z$ space that is calibrated using [Al/Fe] abundances, which is an empirically derived boundary drawn in this space, above which the accreted GCs are located and below which the in situ GCs are located. Looking at the lower left panel of Figure \ref{fig:group_validation}, our selection also broadly separates the in situ and accreted groups into different regions of E-L$_z$ space, which qualitatively resembles the selection made in \citet{Belokurov_2023}. In that work, they argue that this simple selection is very effective at separating the two populations, but with some small amount of contamination, for example that they label \ocen and NGC 6273 as in situ, despite both of these GCs being likely remnant nuclear star clusters of accreted galaxies. In our selection, we correctly identify \ocen as an accreted GC in group G-3, Thamnos, but also erroneously classify NGC 6273 as in situ as part of group G-8.

Broadly speaking, groups G-1 (Splash), G-6, G-7, and G-8 make up the in situ GCs, and groups G-2 (GSE), G-3 (Thamnos), G-4 (Sequoia/Arjuna/I'itoi), and G-5 (accreted structures) make up the accreted GCs. This is consistent with our previous classification of these groups in Table \ref{tab:small_associations}, and is apparent from the orbital parameters and [Fe/H] distributions shown in Figure \ref{fig:group_validation}. More specifically, all GCs in groups G-1, G-6, G-7 and G-8 lie at highly bound orbital energies, are on disc-like, prograde orbits with low radial and vertical actions, and have higher metallicities than the other groups, with all of them having mean [Fe/H] > -1.3, with the exception of the Splash group which has a mean [Fe/H] $\sim$ -1.78. In total, of the 147 GCs in our sample, we find 80 of them (54\%) to be consistent with an in situ origin, 64 (44\% to be accreted, 2 (1.3\%) to be unbound, and 1 (0.7\%) unclassified. By comparison, \citet{Belokurov_2024} find a somewhat lower accreted percentage of 35\%, \citet{Sun_2023} find a percentage of 38.4\%, \citet{Forbes_2020} find a percentage of 54\%, and \citet{Massari_2019} find 60\%. We note that of the 44\% of Milky Way GCs that we report to be accreted, we are assuming all of the GCs in group G-5 are accreted, even though we can only attribute some of them to known accretion events. 

For the accreted groups belonging to G-2, G-3, G-4, and G-5, we can attribute the majority of these (59 GCs or 41\% of the GC sample) to individual accretion events, which we expect to have distinct chemical and dynamical properties.
In addition, the different in situ groups also show different characteristics. For example, the bottom panel of Figure \ref{fig:group_validation} shows different values for the mean metallicity of these groups, although there is a very large spread in these distributions and groups G-6, G-7 and G-8 are all overlapping, but the Splash group G-1 is noticeably more metal-poor than the others. Furthermore, there appears to be some separation in the E-L$_z$ space, with groups G-6 and G-7 sitting lower in the Galactic potential at lower values of orbital energy than groups G-1 and G-8, although this is difficult to see due to crowding in this region of the plot. In order to investigate these GC populations beyond Figure \ref{fig:group_validation}, we also supplemented this analysis with Figure \ref{fig:all_gcs_validation_plot} in Appendix \ref{app:gc_analyis} showing the sky distribution in Galactic coordinates, and a plot of the Galactic apocentre and pericentre distances for all of the GCs in the sample.  

Group G-6, the brown squares in Figure \ref{fig:gse_validation_plot} are centrally located in the Galaxy, with all of the GCs in this group having r$_{apo} \, \lesssim$ 4 kpc, thus representing the bulge GCs. This group makes up 12\% of the GC sample, which is somewhat lower than the 23\% that \citet{Massari_2019} attribute to the bulge, where they consider all GCs with r$_{apo} \, \lesssim$ 3.5 kpc to be part of this population. We remark that it is interesting that when using a multi-dimensional classification approach, several GCs with small apocentre distances are classified into other groups. Future work should investigate if this corresponds to differences in chemistry and ages in these populations as well.

Group G-7 (dark green squares) extends beyond the Galactic centre, and consists of GCs with orbits that are largely circular, with 1 > e > 0.5. Groups G-8 and G-1 are both centrally located in the Galaxy, although they extend beyond group G-6, out to r$_{apo} \, \lesssim$ 10 kpc and have somewhat eccentric orbits with e $\simeq$ 0.5.

Using tangential velocities distributions, \citet{Belokurov_2024} have suggested that the in situ Milky Way GC population can be separated into two groups. The first, GCs at [Fe/H] < -1.3 showing low tangential velocities likely formed during turbulent pre-disc stages of Milky Way evolution, while the second, GCs at -1.3 < [Fe/H] show much higher tangential velocities, suggesting that they formed after the formation of the disc and therefore show this disc spin-up signature. This could explain the formation mechanism of our observed groups G-7 and G-8, where a later formation time of GCs in G-7 would explain their more circularised orbits and distribution that extends farther out from the Galactic centre. Therefore, we define the in situ GCs from group G-7 as the post-disc GCs, and the GCs from group G-8 as the pre-disc GCs, signifying that they were likely to have formed after and before the formation of the Milky Way disc, respectively. 

\subsection{Omega Centauri}
\label{subsection:omega_cen}

The origin of \ocen (NGC 5139), the most massive globular cluster in the Milky Way, has for many years been the subject of considerable debate. Recently, \citet{Massari_2019} and \citet{Bonaca_2021} have proposed that \ocen could be dynamically associated with the GSE merger event, or be the nuclear core of the GSE galaxy. Alternatively, \citet{Myeong_2019} and \citet{Forbes_2020} have suggested an association between \ocen and the Sequoia structure, and proposed that \ocen might represent either the most massive GC of the Sequoia progenitor, or the nuclear core of the progenitor itself.

At the time of the discovery of Sequoia, \citet{Myeong_2019} described the structure as extending down to low orbital energies consistent with those of \ocen. However, a subsequent work by \citet{Koppelman_2019b} later showed that Sequoia could be broken up into at least three different structures, Sequoia, Thamnos 1 and Thamnos 2, with the Thamnos structures occupying a lower orbital energy than Sequoia. They supported this result based on the differing abundances between the three structures and the expected mass of the Sequoia not being large enough to account for such a large distribution in E-L$_z$ space. Based on clustering in integrals of motion space, Thamnos 1 and Thamnos 2 were later proposed to be one single structure, but with a complex chemistry representing a mix of stellar populations. Ultimately, \citet{Koppelman_2019b} conclude that \ocen is below the orbital energy of Thamnos in the E-L$_z$ space and do not claim a connection between them.

Our current analysis, however, supports that the chemo-dynamic properties of \ocen and Fimbulthul align most closely with the Thamnos structure (Group G-3). This is supported by the similar, broad metallicity distributions of \ocen and Thamnos (with peak [Fe/H] $\approx$ -1.5), as well as similar orbital parameters, particularly that both Thamnos and \ocen are on retrograde orbits and have lower orbital energies than those of Sequoia stars. 

Furthermore, the proposed mass of the Thamnos structure is $M_* < 5 \times 10^6 M_{\odot}$, which is remarkably similar to the present day stellar mass of \ocen at $M_* = 4 \times 10^6 M_{\odot}$. This would imply that if this scenario is true, \ocen would have lost half of its mass during its interactions with the Milky way which is now visible as the Thamnos structure. 


\section{Conclusions}
\label{sec:conclusions}
This work represents a large-scale clustering analysis of known substructures in the Milky Way halo. We gathered a dataset consisting of 6D phase space + [Fe/H] for 147 GCs, 30 satellite Local Group galaxies, candidate stars from 49 unique stellar streams from the STREAMFINDER, and S5 spectroscopic follow-up samples supplemented with data for the LMS-1, Wukong and C-19 streams and also clean selected samples of the GSE, Splash, and Thamnos halo structures from the GALAH DR4 data (from \citealt{Kushniruk_2025}). 

We computed the orbital parameters E, J$_r$, J$_z$, L$_z$, r$_{apo}$, r$_{peri}$, and eccentricity using Galpy, which we complimented with [Fe/H] values from available spectroscopic observations. We then standardised these seven parameters and used them as input into the t-SNE dimensionality reduction algorithm. Although the orbits are fully characterised by the three action-angle parameters J$_r$, J$_z$, and L$_z$, we find that adding additional input parameters that are important for separating groups in dynamical space added relevant weighting to the inputs that empirically improved the clustering in the latent space, and therefore we chose to include them. We selected groups of these structures identified from the latent space as being related in their kinematics and chemistry, resulting in 9 main groups, 16 subgroups, and 20 stream-progenitor associations, as summarised in Table \ref{tab:large_associations} and Table \ref{tab:selected_stream_progenitor}. We then discussed these groups in the context of previously described structures from the literature and the implications of these findings for our understanding of the accretion history of the Milky Way halo. The main findings from this analysis are as follows:

   \begin{itemize}
      \item We recovered several previously described halo structures, including GSE, Thamnos, Sequoia/Arjuna/I'itoi, LMS-1/Wukong, Sagittarius, Helmi streams, Kraken/Koala, and Pontus.
      \item We also confirmed several established stream-progenitor associations, including \ocen - Fimbulthul, Gj\"oll - NGC 3201, NGC 6101 (GC) - NGC 6101 (stream), Fj\"orm - NGC 4590/M 68, Svol - NGC 5272/M 3, NGC 288 (GC) - NGC 288 (stream), LMS-1 - NGC 5053, NGC 6397 (GC) - NGC 6397 (stream), Tucana III (DG) - Tuc-III (stream), NGC 6341/M 92 (GC) - M 92 (stream), NGC 2808 (GC) - NGC 2808 (stream), and NGC 1851 (GC) - NGC 1851 (stream).
      \item We report several new tentative stream-progenitor associations: Hrid - Pal 2, Ophiucus - NGC 5634, Jhelum - NGC 5024/M 53, Elqui - Segue I, and Slidr - Rup 106.
      \item Most notably, we found a connection between the Orphan stream and the ultra-faint dwarf galaxy Grus II, which after further analysis we conclude are unlikely to be a true stream-progenitor pair but were likely to have shared a common progenitor. 
       \item We determined an accreted GC fraction of 44\%, which is in line with previous determinations. Furthermore, we identified four distinct groups of in situ GCs: bulge GCs (G-6), post-disc GCs (G-7), pre-disc GCs (G-8), and GCs associated with the splashed disc population (G-1).
      \item We found evidence of a substructure within the GSE accretion event, namely, five potential subgroups (sg-1, sg-2, sg-3, sg-4, and sg-5). Upon further investigation of orbital characteristics and metallicity distributions, we conclude that subgroups sg-2 and sg-5 are not part of the GSE but are the Kraken/Koala and Pontus structures, respectively. Of the remaining subgroups, we found insufficient evidence to show clear separate populations within the GSE clusters, but our results still suggest that GSE itself may have had internal structure before merging with the Milky Way, and this should be investigated further in future work.
      \item We found a chemo-dynamical association between \ocen and the Thamnos structure. Previous works have suggested that \ocen is dynamically associated with GSE \citep{Bonaca_2021}, while others have suggested that it could be the remnant core of the retrograde Sequoia/Arjuna/I'itoi \citep{Myeong_2019, Forbes_2020}. However, in this work, we find that the chemo-dynamics of \ocen are most consistent with it being associated with Thamnos and suggest that it may even be the remnant core of the progenitor that brought in the Thamnos stars. Future, more detailed N-body simulations and chemical analyses, for example of neutron capture elements, could shed light on this hypothesis.

   \end{itemize}

   Associations between substructures in the Galactic halo provide valuable information for the accretion history of our Galaxy and the identification of common progenitors for Milky Way stars. These findings provide valuable constraints on the properties of past Milky Way mergers, informing chemical evolution models and dynamical simulations of galaxy formation and ultimately enhancing our understanding of the formation and evolution of the Galaxy. Future work with larger samples of high-precision abundance measurements and detailed spectroscopic follow-up will help further refine these associations and resolve remaining ambiguities in classification.

\begin{acknowledgements}
      KY and KL thank the European Research Council (ERC) for providing funds under the European Union's Horizon 2020 research and innovation programme (grant agreement number 852977). KY would like to thank Fabio De Ferrari for many productive co-working days and valuable feedback which provided crucial support and motivation for the completion of this manuscript. KY would also like to thank F\aa gel\"angens catering and the Albanova cafeteria and staff, for providing delicious lunches and the daily energy required to conduct this research. This work has made use of the typesetting software \texttt{overleaf}~\footnote{\url{https://www.overleaf.com/}}, the plotting and table handling environment
\textsc{topcat} \citep{Taylor2005}, and extensively used the
\textsc{python} programming language \citep{python} for the analysis, including the following packages:
\textsc{matplotlib} \citep{Hunter2007}, \textsc{scipy} \citep{scipy}, \textsc{numpy}
\citep{Harris2020}, \textsc{pandas}
\citep{Mckinney2010}, and \textsc{astropy} \citep{astropy2022, astropy2018, astropy2013}.

\end{acknowledgements}

%
\bibliographystyle{aa} 
\bibliography{tsne_streams} 

\begin{thebibliography}{114}
\expandafter\ifx\csname natexlab\endcsname\relax\def\natexlab#1{#1}\fi

\bibitem[{{Anders} {et~al.}(2018){Anders}, {Chiappini}, {Santiago}, {Matijevi{\v{c}}}, {Queiroz}, {Steinmetz}, \& {Guiglion}}]{Anders_2018}
{Anders}, F., {Chiappini}, C., {Santiago}, B.~X., {et~al.} 2018, \aap, 619, A125

\bibitem[{{Anders} {et~al.}(2022){Anders}, {Khalatyan}, {Queiroz}, {Chiappini}, {Ard{\`e}vol}, {Casamiquela}, {Figueras}, {Jim{\'e}nez-Arranz}, {Jordi}, {Mongui{\'o}}, {Romero-G{\'o}mez}, {Altamirano}, {Antoja}, {Assaad}, {Cantat-Gaudin}, {Castro-Ginard}, {Enke}, {Girardi}, {Guiglion}, {Khan}, {Luri}, {Miglio}, {Minchev}, {Ramos}, {Santiago}, \& {Steinmetz}}]{Anders_2022}
{Anders}, F., {Khalatyan}, A., {Queiroz}, A.~B.~A., {et~al.} 2022, \aap, 658, A91

\bibitem[{{Astropy Collaboration} {et~al.}(2022){Astropy Collaboration}, {Price-Whelan}, {Lim}, {Earl}, {Starkman}, {Bradley}, {Shupe}, {Patil}, {Corrales}, {Brasseur}, {N{\"o}the}, {Donath}, {Tollerud}, {Morris}, {Ginsburg}, {Vaher}, {Weaver}, {Tocknell}, {Jamieson}, {van Kerkwijk}, {Robitaille}, {Merry}, {Bachetti}, {G{\"u}nther}, {Aldcroft}, {Alvarado-Montes}, {Archibald}, {B{\'o}di}, {Bapat}, {Barentsen}, {Baz{\'a}n}, {Biswas}, {Boquien}, {Burke}, {Cara}, {Cara}, {Conroy}, {Conseil}, {Craig}, {Cross}, {Cruz}, {D'Eugenio}, {Dencheva}, {Devillepoix}, {Dietrich}, {Eigenbrot}, {Erben}, {Ferreira}, {Foreman-Mackey}, {Fox}, {Freij}, {Garg}, {Geda}, {Glattly}, {Gondhalekar}, {Gordon}, {Grant}, {Greenfield}, {Groener}, {Guest}, {Gurovich}, {Handberg}, {Hart}, {Hatfield-Dodds}, {Homeier}, {Hosseinzadeh}, {Jenness}, {Jones}, {Joseph}, {Kalmbach}, {Karamehmetoglu}, {Ka{\l}uszy{\'n}ski}, {Kelley}, {Kern}, {Kerzendorf}, {Koch}, {Kulumani}, {Lee}, {Ly}, {Ma}, {MacBride}, {Maljaars}, {Muna}, {Murphy}, {Norman},
  {O'Steen}, {Oman}, {Pacifici}, {Pascual}, {Pascual-Granado}, {Patil}, {Perren}, {Pickering}, {Rastogi}, {Roulston}, {Ryan}, {Rykoff}, {Sabater}, {Sakurikar}, {Salgado}, {Sanghi}, {Saunders}, {Savchenko}, {Schwardt}, {Seifert-Eckert}, {Shih}, {Jain}, {Shukla}, {Sick}, {Simpson}, {Singanamalla}, {Singer}, {Singhal}, {Sinha}, {Sip{\H{o}}cz}, {Spitler}, {Stansby}, {Streicher}, {{\v{S}}umak}, {Swinbank}, {Taranu}, {Tewary}, {Tremblay}, {de Val-Borro}, {Van Kooten}, {Vasovi{\'c}}, {Verma}, {de Miranda Cardoso}, {Williams}, {Wilson}, {Winkel}, {Wood-Vasey}, {Xue}, {Yoachim}, {Zhang}, {Zonca}, \& {Astropy Project Contributors}}]{astropy2022}
{Astropy Collaboration}, {Price-Whelan}, A.~M., {Lim}, P.~L., {et~al.} 2022, \apj, 935, 167

\bibitem[{{Astropy Collaboration} {et~al.}(2018){Astropy Collaboration}, {Price-Whelan}, {Sip{\H{o}}cz}, {G{\"u}nther}, {Lim}, {Crawford}, {Conseil}, {Shupe}, {Craig}, {Dencheva}, {Ginsburg}, {VanderPlas}, {Bradley}, {P{\'e}rez-Su{\'a}rez}, {de Val-Borro}, {Aldcroft}, {Cruz}, {Robitaille}, {Tollerud}, {Ardelean}, {Babej}, {Bach}, {Bachetti}, {Bakanov}, {Bamford}, {Barentsen}, {Barmby}, {Baumbach}, {Berry}, {Biscani}, {Boquien}, {Bostroem}, {Bouma}, {Brammer}, {Bray}, {Breytenbach}, {Buddelmeijer}, {Burke}, {Calderone}, {Cano Rodr{\'\i}guez}, {Cara}, {Cardoso}, {Cheedella}, {Copin}, {Corrales}, {Crichton}, {D'Avella}, {Deil}, {Depagne}, {Dietrich}, {Donath}, {Droettboom}, {Earl}, {Erben}, {Fabbro}, {Ferreira}, {Finethy}, {Fox}, {Garrison}, {Gibbons}, {Goldstein}, {Gommers}, {Greco}, {Greenfield}, {Groener}, {Grollier}, {Hagen}, {Hirst}, {Homeier}, {Horton}, {Hosseinzadeh}, {Hu}, {Hunkeler}, {Ivezi{\'c}}, {Jain}, {Jenness}, {Kanarek}, {Kendrew}, {Kern}, {Kerzendorf}, {Khvalko}, {King}, {Kirkby}, {Kulkarni},
  {Kumar}, {Lee}, {Lenz}, {Littlefair}, {Ma}, {Macleod}, {Mastropietro}, {McCully}, {Montagnac}, {Morris}, {Mueller}, {Mumford}, {Muna}, {Murphy}, {Nelson}, {Nguyen}, {Ninan}, {N{\"o}the}, {Ogaz}, {Oh}, {Parejko}, {Parley}, {Pascual}, {Patil}, {Patil}, {Plunkett}, {Prochaska}, {Rastogi}, {Reddy Janga}, {Sabater}, {Sakurikar}, {Seifert}, {Sherbert}, {Sherwood-Taylor}, {Shih}, {Sick}, {Silbiger}, {Singanamalla}, {Singer}, {Sladen}, {Sooley}, {Sornarajah}, {Streicher}, {Teuben}, {Thomas}, {Tremblay}, {Turner}, {Terr{\'o}n}, {van Kerkwijk}, {de la Vega}, {Watkins}, {Weaver}, {Whitmore}, {Woillez}, {Zabalza}, \& {Astropy Contributors}}]{astropy2018}
{Astropy Collaboration}, {Price-Whelan}, A.~M., {Sip{\H{o}}cz}, B.~M., {et~al.} 2018, \aj, 156, 123

\bibitem[{{Astropy Collaboration} {et~al.}(2013){Astropy Collaboration}, {Robitaille}, {Tollerud}, {Greenfield}, {Droettboom}, {Bray}, {Aldcroft}, {Davis}, {Ginsburg}, {Price-Whelan}, {Kerzendorf}, {Conley}, {Crighton}, {Barbary}, {Muna}, {Ferguson}, {Grollier}, {Parikh}, {Nair}, {Unther}, {Deil}, {Woillez}, {Conseil}, {Kramer}, {Turner}, {Singer}, {Fox}, {Weaver}, {Zabalza}, {Edwards}, {Azalee Bostroem}, {Burke}, {Casey}, {Crawford}, {Dencheva}, {Ely}, {Jenness}, {Labrie}, {Lim}, {Pierfederici}, {Pontzen}, {Ptak}, {Refsdal}, {Servillat}, \& {Streicher}}]{astropy2013}
{Astropy Collaboration}, {Robitaille}, T.~P., {Tollerud}, E.~J., {et~al.} 2013, \aap, 558, A33

\bibitem[{{Belokurov} {et~al.}(2018){Belokurov}, {Erkal}, {Evans}, {Koposov}, \& {Deason}}]{Belokurov_2018}
{Belokurov}, V., {Erkal}, D., {Evans}, N.~W., {Koposov}, S.~E., \& {Deason}, A.~J. 2018, \mnras, 478, 611

\bibitem[{{Belokurov} {et~al.}(2007){Belokurov}, {Evans}, {Irwin}, {Lynden-Bell}, {Yanny}, {Vidrih}, {Gilmore}, {Seabroke}, {Zucker}, {Wilkinson}, {Hewett}, {Bramich}, {Fellhauer}, {Newberg}, {Wyse}, {Beers}, {Bell}, {Barentine}, {Brinkmann}, {Cole}, {Pan}, \& {York}}]{Belokurov_2007}
{Belokurov}, V., {Evans}, N.~W., {Irwin}, M.~J., {et~al.} 2007, \apj, 658, 337

\bibitem[{{Belokurov} \& {Kravtsov}(2023)}]{Belokurov_2023}
{Belokurov}, V. \& {Kravtsov}, A. 2023, \mnras, 525, 4456

\bibitem[{{Belokurov} \& {Kravtsov}(2024)}]{Belokurov_2024}
{Belokurov}, V. \& {Kravtsov}, A. 2024, \mnras, 528, 3198

\bibitem[{{Belokurov} {et~al.}(2020){Belokurov}, {Sanders}, {Fattahi}, {Smith}, {Deason}, {Evans}, \& {Grand}}]{Belokurov_2020}
{Belokurov}, V., {Sanders}, J.~L., {Fattahi}, A., {et~al.} 2020, \mnras, 494, 3880

\bibitem[{{Bernard} {et~al.}(2014){Bernard}, {Ferguson}, {Schlafly}, {Abbas}, {Bell}, {Deacon}, {Martin}, {Rix}, {Sesar}, {Slater}, {Penarrubia}, {Wyse}, {Burgett}, {Chambers}, {Draper}, {Hodapp}, {Kaiser}, {Kudritzki}, {Magnier}, {Metcalfe}, {Morgan}, {Price}, {Tonry}, {Wainscoat}, \& {Waters}}]{Bernard_2014}
{Bernard}, E.~J., {Ferguson}, A.~M.~N., {Schlafly}, E.~F., {et~al.} 2014, \mnras, 443, L84

\bibitem[{{Bianchini} {et~al.}(2019){Bianchini}, {Ibata}, \& {Famaey}}]{Bianchini_2019}
{Bianchini}, P., {Ibata}, R., \& {Famaey}, B. 2019, \apjl, 887, L12

\bibitem[{{Bland-Hawthorn} {et~al.}(2010){Bland-Hawthorn}, {Krumholz}, \& {Freeman}}]{Bland-Hawthorn_2010}
{Bland-Hawthorn}, J., {Krumholz}, M.~R., \& {Freeman}, K. 2010, \apj, 713, 166

\bibitem[{{Bonaca} {et~al.}(2019){Bonaca}, {Conroy}, {Price-Whelan}, \& {Hogg}}]{Bonaca_2019}
{Bonaca}, A., {Conroy}, C., {Price-Whelan}, A.~M., \& {Hogg}, D.~W. 2019, \apjl, 881, L37

\bibitem[{{Bonaca} {et~al.}(2021){Bonaca}, {Naidu}, {Conroy}, {Caldwell}, {Cargile}, {Han}, {Johnson}, {Kruijssen}, {Myeong}, {Speagle}, {Ting}, \& {Zaritsky}}]{Bonaca_2021}
{Bonaca}, A., {Naidu}, R.~P., {Conroy}, C., {et~al.} 2021, \apjl, 909, L26

\bibitem[{Bovy(2015)}]{Bovy_2015}
Bovy, J. 2015, The Astrophysical Journal Supplement Series, 216, 29

\bibitem[{{Buder} {et~al.}(2025){Buder}, {Kos}, {Wang}, {McKenzie}, {Howell}, {Martell}, {Hayden}, {Zucker}, {Nordlander}, {Montet}, {Traven}, {Bland-Hawthorn}, {de Silva}, {Freeman}, {Lewis}, {Lind}, {Sharma}, {Simpson}, {Stello}, {Zwitter}, {Amarsi}, {Armstrong}, {Banks}, {Beavis}, {Beeson}, {Chen}, {Ciuc{\u{a}}}, {da Costa}, {de Grijs}, {Martin}, {Nataf}, {Ness}, {Rains}, {Scarr}, {Vogrin{\v{c}}i{\v{c}}}, {Wang}, {Wittenmyer}, {Xie}, \& {The Galah Collaboration}}]{Buder_2025}
{Buder}, S., {Kos}, J., {Wang}, X.~E., {et~al.} 2025, \pasa, 42, e051

\bibitem[{{Caldwell} {et~al.}(2020){Caldwell}, {Bonaca}, {Price-Whelan}, {Sesar}, \& {Walker}}]{Caldwell_2020}
{Caldwell}, N., {Bonaca}, A., {Price-Whelan}, A.~M., {Sesar}, B., \& {Walker}, M.~G. 2020, \aj, 159, 287

\bibitem[{{Cantelli} \& {Teixeira}(2024)}]{Cantelli_2024}
{Cantelli}, E. \& {Teixeira}, R. 2024, \mnras, 530, 2648

\bibitem[{{Carballo-Bello} {et~al.}(2018){Carballo-Bello}, {Mart{\'\i}nez-Delgado}, {Navarrete}, {Catelan}, {Mu{\~n}oz}, {Antoja}, \& {Sollima}}]{Carballo_2018}
{Carballo-Bello}, J.~A., {Mart{\'\i}nez-Delgado}, D., {Navarrete}, C., {et~al.} 2018, \mnras, 474, 683

\bibitem[{{Casamiquela} {et~al.}(2021){Casamiquela}, {Castro-Ginard}, {Anders}, \& {Soubiran}}]{Casamiquela_2021}
{Casamiquela}, L., {Castro-Ginard}, A., {Anders}, F., \& {Soubiran}, C. 2021, \aap, 654, A151

\bibitem[{{Cheng} {et~al.}(2021){Cheng}, {Price-Jones}, \& {Bovy}}]{Cheng_2021}
{Cheng}, C.~M., {Price-Jones}, N., \& {Bovy}, J. 2021, \mnras, 506, 5573

\bibitem[{{Dinescu} {et~al.}(1997){Dinescu}, {Girard}, {van Altena}, {Mendez}, \& {Lopez}}]{Dinescu_1997}
{Dinescu}, D.~I., {Girard}, T.~M., {van Altena}, W.~F., {Mendez}, R.~A., \& {Lopez}, C.~E. 1997, \aj, 114, 1014

\bibitem[{{Drlica-Wagner} {et~al.}(2015){Drlica-Wagner}, {Bechtol}, {Rykoff}, {Luque}, {Queiroz}, {Mao}, {Wechsler}, {Simon}, {Santiago}, {Yanny}, {Balbinot}, {Dodelson}, {Fausti Neto}, {James}, {Li}, {Maia}, {Marshall}, {Pieres}, {Stringer}, {Walker}, {Abbott}, {Abdalla}, {Allam}, {Benoit-L{\'e}vy}, {Bernstein}, {Bertin}, {Brooks}, {Buckley-Geer}, {Burke}, {Carnero Rosell}, {Carrasco Kind}, {Carretero}, {Crocce}, {da Costa}, {Desai}, {Diehl}, {Dietrich}, {Doel}, {Eifler}, {Evrard}, {Finley}, {Flaugher}, {Fosalba}, {Frieman}, {Gaztanaga}, {Gerdes}, {Gruen}, {Gruendl}, {Gutierrez}, {Honscheid}, {Kuehn}, {Kuropatkin}, {Lahav}, {Martini}, {Miquel}, {Nord}, {Ogando}, {Plazas}, {Reil}, {Roodman}, {Sako}, {Sanchez}, {Scarpine}, {Schubnell}, {Sevilla-Noarbe}, {Smith}, {Soares-Santos}, {Sobreira}, {Suchyta}, {Swanson}, {Tarle}, {Tucker}, {Vikram}, {Wester}, {Zhang}, {Zuntz}, \& {DES Collaboration}}]{Drlica_2015}
{Drlica-Wagner}, A., {Bechtol}, K., {Rykoff}, E.~S., {et~al.} 2015, \apj, 813, 109

\bibitem[{Fellhauer {et~al.}(2007)Fellhauer, Evans, Belokurov, Zucker, Yanny, Wilkinson, Gilmore, Irwin, Bramich, Vidrih, Hewett, \& Beers}]{Fellhauer_2007}
Fellhauer, M., Evans, N.~W., Belokurov, V., {et~al.} 2007, Monthly Notices of the Royal Astronomical Society, 375, 1171

\bibitem[{{Forbes}(2020)}]{Forbes_2020}
{Forbes}, D.~A. 2020, \mnras, 493, 847

\bibitem[{{Forbes} \& {Bridges}(2010)}]{Forbes_2010}
{Forbes}, D.~A. \& {Bridges}, T. 2010, \mnras, 404, 1203

\bibitem[{{Freeman} \& {Bland-Hawthorn}(2002)}]{Freeman_2002}
{Freeman}, K. \& {Bland-Hawthorn}, J. 2002, \araa, 40, 487

\bibitem[{{Gaia Collaboration} {et~al.}(2021){Gaia Collaboration}, {Brown}, {Vallenari}, {Prusti}, {de Bruijne}, {Babusiaux}, {Biermann}, {Creevey}, {Evans}, {Eyer}, {Hutton}, {Jansen}, {Jordi}, {Klioner}, {Lammers}, {Lindegren}, {Luri}, {Mignard}, {Panem}, {Pourbaix}, {Randich}, {Sartoretti}, {Soubiran}, {Walton}, {Arenou}, {Bailer-Jones}, {Bastian}, {Cropper}, {Drimmel}, {Katz}, {Lattanzi}, {van Leeuwen}, {Bakker}, {Cacciari}, {Casta{\~n}eda}, {De Angeli}, {Ducourant}, {Fabricius}, {Fouesneau}, {Fr{\'e}mat}, {Guerra}, {Guerrier}, {Guiraud}, {Jean-Antoine Piccolo}, {Masana}, {Messineo}, {Mowlavi}, {Nicolas}, {Nienartowicz}, {Pailler}, {Panuzzo}, {Riclet}, {Roux}, {Seabroke}, {Sordo}, {Tanga}, {Th{\'e}venin}, {Gracia-Abril}, {Portell}, {Teyssier}, {Altmann}, {Andrae}, {Bellas-Velidis}, {Benson}, {Berthier}, {Blomme}, {Brugaletta}, {Burgess}, {Busso}, {Carry}, {Cellino}, {Cheek}, {Clementini}, {Damerdji}, {Davidson}, {Delchambre}, {Dell'Oro}, {Fern{\'a}ndez-Hern{\'a}ndez}, {Galluccio}, {Garc{\'\i}a-Lario},
  {Garcia-Reinaldos}, {Gonz{\'a}lez-N{\'u}{\~n}ez}, {Gosset}, {Haigron}, {Halbwachs}, {Hambly}, {Harrison}, {Hatzidimitriou}, {Heiter}, {Hern{\'a}ndez}, {Hestroffer}, {Hodgkin}, {Holl}, {Jan{\ss}en}, {Jevardat de Fombelle}, {Jordan}, {Krone-Martins}, {Lanzafame}, {L{\"o}ffler}, {Lorca}, {Manteiga}, {Marchal}, {Marrese}, {Moitinho}, {Mora}, {Muinonen}, {Osborne}, {Pancino}, {Pauwels}, {Petit}, {Recio-Blanco}, {Richards}, {Riello}, {Rimoldini}, {Robin}, {Roegiers}, {Rybizki}, {Sarro}, {Siopis}, {Smith}, {Sozzetti}, {Ulla}, {Utrilla}, {van Leeuwen}, {van Reeven}, {Abbas}, {Abreu Aramburu}, {Accart}, {Aerts}, {Aguado}, {Ajaj}, {Altavilla}, {{\'A}lvarez}, {{\'A}lvarez Cid-Fuentes}, {Alves}, {Anderson}, {Anglada Varela}, {Antoja}, {Audard}, {Baines}, {Baker}, {Balaguer-N{\'u}{\~n}ez}, {Balbinot}, {Balog}, {Barache}, {Barbato}, {Barros}, {Barstow}, {Bartolom{\'e}}, {Bassilana}, {Bauchet}, {Baudesson-Stella}, {Becciani}, {Bellazzini}, {Bernet}, {Bertone}, {Bianchi}, {Blanco-Cuaresma}, {Boch}, {Bombrun}, {Bossini},
  {Bouquillon}, {Bragaglia}, {Bramante}, {Breedt}, {Bressan}, {Brouillet}, {Bucciarelli}, {Burlacu}, {Busonero}, {Butkevich}, {Buzzi}, {Caffau}, {Cancelliere}, {C{\'a}novas}, {Cantat-Gaudin}, {Carballo}, {Carlucci}, {Carnerero}, {Carrasco}, {Casamiquela}, {Castellani}, {Castro-Ginard}, {Castro Sampol}, {Chaoul}, {Charlot}, {Chemin}, {Chiavassa}, {Cioni}, {Comoretto}, {Cooper}, {Cornez}, {Cowell}, {Crifo}, {Crosta}, {Crowley}, {Dafonte}, {Dapergolas}, {David}, {David}, {de Laverny}, {De Luise}, {De March}, {De Ridder}, {de Souza}, {de Teodoro}, {de Torres}, {del Peloso}, {del Pozo}, {Delbo}, {Delgado}, {Delgado}, {Delisle}, {Di Matteo}, {Diakite}, {Diener}, {Distefano}, {Dolding}, {Eappachen}, {Edvardsson}, {Enke}, {Esquej}, {Fabre}, {Fabrizio}, {Faigler}, {Fedorets}, {Fernique}, {Fienga}, {Figueras}, {Fouron}, {Fragkoudi}, {Fraile}, {Franke}, {Gai}, {Garabato}, {Garcia-Gutierrez}, {Garc{\'\i}a-Torres}, {Garofalo}, {Gavras}, {Gerlach}, {Geyer}, {Giacobbe}, {Gilmore}, {Girona}, {Giuffrida}, {Gomel}, {Gomez},
  {Gonzalez-Santamaria}, {Gonz{\'a}lez-Vidal}, {Granvik}, {Guti{\'e}rrez-S{\'a}nchez}, {Guy}, {Hauser}, {Haywood}, {Helmi}, {Hidalgo}, {Hilger}, {H{\l}adczuk}, {Hobbs}, {Holland}, {Huckle}, {Jasniewicz}, {Jonker}, {Juaristi Campillo}, {Julbe}, {Karbevska}, {Kervella}, {Khanna}, {Kochoska}, {Kontizas}, {Kordopatis}, {Korn}, {Kostrzewa-Rutkowska}, {Kruszy{\'n}ska}, {Lambert}, {Lanza}, {Lasne}, {Le Campion}, {Le Fustec}, {Lebreton}, {Lebzelter}, {Leccia}, {Leclerc}, {Lecoeur-Taibi}, {Liao}, {Licata}, {Lindstr{\o}m}, {Lister}, {Livanou}, {Lobel}, {Madrero Pardo}, {Managau}, {Mann}, {Marchant}, {Marconi}, {Marcos Santos}, {Marinoni}, {Marocco}, {Marshall}, {Martin Polo}, {Mart{\'\i}n-Fleitas}, {Masip}, {Massari}, {Mastrobuono-Battisti}, {Mazeh}, {McMillan}, {Messina}, {Michalik}, {Millar}, {Mints}, {Molina}, {Molinaro}, {Moln{\'a}r}, {Montegriffo}, {Mor}, {Morbidelli}, {Morel}, {Morris}, {Mulone}, {Munoz}, {Muraveva}, {Murphy}, {Musella}, {Noval}, {Ord{\'e}novic}, {Orr{\`u}}, {Osinde}, {Pagani}, {Pagano},
  {Palaversa}, {Palicio}, {Panahi}, {Pawlak}, {Pe{\~n}alosa Esteller}, {Penttil{\"a}}, {Piersimoni}, {Pineau}, {Plachy}, {Plum}, {Poggio}, {Poretti}, {Poujoulet}, {Pr{\v{s}}a}, {Pulone}, {Racero}, {Ragaini}, {Rainer}, {Raiteri}, {Rambaux}, {Ramos}, {Ramos-Lerate}, {Re Fiorentin}, {Regibo}, {Reyl{\'e}}, {Ripepi}, {Riva}, {Rixon}, {Robichon}, {Robin}, {Roelens}, {Rohrbasser}, {Romero-G{\'o}mez}, {Rowell}, {Royer}, {Rybicki}, {Sadowski}, {Sagrist{\`a} Sell{\'e}s}, {Sahlmann}, {Salgado}, {Salguero}, {Samaras}, {Sanchez Gimenez}, {Sanna}, {Santove{\~n}a}, {Sarasso}, {Schultheis}, {Sciacca}, {Segol}, {Segovia}, {S{\'e}gransan}, {Semeux}, {Shahaf}, {Siddiqui}, {Siebert}, {Siltala}, {Slezak}, {Smart}, {Solano}, {Solitro}, {Souami}, {Souchay}, {Spagna}, {Spoto}, {Steele}, {Steidelm{\"u}ller}, {Stephenson}, {S{\"u}veges}, {Szabados}, {Szegedi-Elek}, {Taris}, {Tauran}, {Taylor}, {Teixeira}, {Thuillot}, {Tonello}, {Torra}, {Torra}, {Turon}, {Unger}, {Vaillant}, {van Dillen}, {Vanel}, {Vecchiato}, {Viala}, {Vicente},
  {Voutsinas}, {Weiler}, {Wevers}, {Wyrzykowski}, {Yoldas}, {Yvard}, {Zhao}, {Zorec}, {Zucker}, {Zurbach}, \& {Zwitter}}]{Gaia2021}
{Gaia Collaboration}, {Brown}, A.~G.~A., {Vallenari}, A., {et~al.} 2021, \aap, 649, A1

\bibitem[{{Gaia Collaboration} {et~al.}(2016){Gaia Collaboration}, {Prusti}, {de Bruijne}, {Brown}, {Vallenari}, {Babusiaux}, {Bailer-Jones}, {Bastian}, {Biermann}, {Evans}, {Eyer}, {Jansen}, {Jordi}, {Klioner}, {Lammers}, {Lindegren}, {Luri}, {Mignard}, {Milligan}, {Panem}, {Poinsignon}, {Pourbaix}, {Randich}, {Sarri}, {Sartoretti}, {Siddiqui}, {Soubiran}, {Valette}, {van Leeuwen}, {Walton}, {Aerts}, {Arenou}, {Cropper}, {Drimmel}, {H{\o}g}, {Katz}, {Lattanzi}, {O'Mullane}, {Grebel}, {Holland}, {Huc}, {Passot}, {Bramante}, {Cacciari}, {Casta{\~n}eda}, {Chaoul}, {Cheek}, {De Angeli}, {Fabricius}, {Guerra}, {Hern{\'a}ndez}, {Jean-Antoine-Piccolo}, {Masana}, {Messineo}, {Mowlavi}, {Nienartowicz}, {Ord{\'o}{\~n}ez-Blanco}, {Panuzzo}, {Portell}, {Richards}, {Riello}, {Seabroke}, {Tanga}, {Th{\'e}venin}, {Torra}, {Els}, {Gracia-Abril}, {Comoretto}, {Garcia-Reinaldos}, {Lock}, {Mercier}, {Altmann}, {Andrae}, {Astraatmadja}, {Bellas-Velidis}, {Benson}, {Berthier}, {Blomme}, {Busso}, {Carry}, {Cellino}, {Clementini},
  {Cowell}, {Creevey}, {Cuypers}, {Davidson}, {De Ridder}, {de Torres}, {Delchambre}, {Dell'Oro}, {Ducourant}, {Fr{\'e}mat}, {Garc{\'\i}a-Torres}, {Gosset}, {Halbwachs}, {Hambly}, {Harrison}, {Hauser}, {Hestroffer}, {Hodgkin}, {Huckle}, {Hutton}, {Jasniewicz}, {Jordan}, {Kontizas}, {Korn}, {Lanzafame}, {Manteiga}, {Moitinho}, {Muinonen}, {Osinde}, {Pancino}, {Pauwels}, {Petit}, {Recio-Blanco}, {Robin}, {Sarro}, {Siopis}, {Smith}, {Smith}, {Sozzetti}, {Thuillot}, {van Reeven}, {Viala}, {Abbas}, {Abreu Aramburu}, {Accart}, {Aguado}, {Allan}, {Allasia}, {Altavilla}, {{\'A}lvarez}, {Alves}, {Anderson}, {Andrei}, {Anglada Varela}, {Antiche}, {Antoja}, {Ant{\'o}n}, {Arcay}, {Atzei}, {Ayache}, {Bach}, {Baker}, {Balaguer-N{\'u}{\~n}ez}, {Barache}, {Barata}, {Barbier}, {Barblan}, {Baroni}, {Barrado y Navascu{\'e}s}, {Barros}, {Barstow}, {Becciani}, {Bellazzini}, {Bellei}, {Bello Garc{\'\i}a}, {Belokurov}, {Bendjoya}, {Berihuete}, {Bianchi}, {Bienaym{\'e}}, {Billebaud}, {Blagorodnova}, {Blanco-Cuaresma}, {Boch},
  {Bombrun}, {Borrachero}, {Bouquillon}, {Bourda}, {Bouy}, {Bragaglia}, {Breddels}, {Brouillet}, {Br{\"u}semeister}, {Bucciarelli}, {Budnik}, {Burgess}, {Burgon}, {Burlacu}, {Busonero}, {Buzzi}, {Caffau}, {Cambras}, {Campbell}, {Cancelliere}, {Cantat-Gaudin}, {Carlucci}, {Carrasco}, {Castellani}, {Charlot}, {Charnas}, {Charvet}, {Chassat}, {Chiavassa}, {Clotet}, {Cocozza}, {Collins}, {Collins}, {Costigan}, {Crifo}, {Cross}, {Crosta}, {Crowley}, {Dafonte}, {Damerdji}, {Dapergolas}, {David}, {David}, {De Cat}, {de Felice}, {de Laverny}, {De Luise}, {De March}, {de Martino}, {de Souza}, {Debosscher}, {del Pozo}, {Delbo}, {Delgado}, {Delgado}, {di Marco}, {Di Matteo}, {Diakite}, {Distefano}, {Dolding}, {Dos Anjos}, {Drazinos}, {Dur{\'a}n}, {Dzigan}, {Ecale}, {Edvardsson}, {Enke}, {Erdmann}, {Escolar}, {Espina}, {Evans}, {Eynard Bontemps}, {Fabre}, {Fabrizio}, {Faigler}, {Falc{\~a}o}, {Farr{\`a}s Casas}, {Faye}, {Federici}, {Fedorets}, {Fern{\'a}ndez-Hern{\'a}ndez}, {Fernique}, {Fienga}, {Figueras}, {Filippi},
  {Findeisen}, {Fonti}, {Fouesneau}, {Fraile}, {Fraser}, {Fuchs}, {Furnell}, {Gai}, {Galleti}, {Galluccio}, {Garabato}, {Garc{\'\i}a-Sedano}, {Gar{\'e}}, {Garofalo}, {Garralda}, {Gavras}, {Gerssen}, {Geyer}, {Gilmore}, {Girona}, {Giuffrida}, {Gomes}, {Gonz{\'a}lez-Marcos}, {Gonz{\'a}lez-N{\'u}{\~n}ez}, {Gonz{\'a}lez-Vidal}, {Granvik}, {Guerrier}, {Guillout}, {Guiraud}, {G{\'u}rpide}, {Guti{\'e}rrez-S{\'a}nchez}, {Guy}, {Haigron}, {Hatzidimitriou}, {Haywood}, {Heiter}, {Helmi}, {Hobbs}, {Hofmann}, {Holl}, {Holland}, {Hunt}, {Hypki}, {Icardi}, {Irwin}, {Jevardat de Fombelle}, {Jofr{\'e}}, {Jonker}, {Jorissen}, {Julbe}, {Karampelas}, {Kochoska}, {Kohley}, {Kolenberg}, {Kontizas}, {Koposov}, {Kordopatis}, {Koubsky}, {Kowalczyk}, {Krone-Martins}, {Kudryashova}, {Kull}, {Bachchan}, {Lacoste-Seris}, {Lanza}, {Lavigne}, {Le Poncin-Lafitte}, {Lebreton}, {Lebzelter}, {Leccia}, {Leclerc}, {Lecoeur-Taibi}, {Lemaitre}, {Lenhardt}, {Leroux}, {Liao}, {Licata}, {Lindstr{\o}m}, {Lister}, {Livanou}, {Lobel}, {L{\"o}ffler},
  {L{\'o}pez}, {Lopez-Lozano}, {Lorenz}, {Loureiro}, {MacDonald}, {Magalh{\~a}es Fernandes}, {Managau}, {Mann}, {Mantelet}, {Marchal}, {Marchant}, {Marconi}, {Marie}, {Marinoni}, {Marrese}, {Marschalk{\'o}}, {Marshall}, {Mart{\'\i}n-Fleitas}, {Martino}, {Mary}, {Matijevi{\v{c}}}, {Mazeh}, {McMillan}, {Messina}, {Mestre}, {Michalik}, {Millar}, {Miranda}, {Molina}, {Molinaro}, {Molinaro}, {Moln{\'a}r}, {Moniez}, {Montegriffo}, {Monteiro}, {Mor}, {Mora}, {Morbidelli}, {Morel}, {Morgenthaler}, {Morley}, {Morris}, {Mulone}, {Muraveva}, {Musella}, {Narbonne}, {Nelemans}, {Nicastro}, {Noval}, {Ord{\'e}novic}, {Ordieres-Mer{\'e}}, {Osborne}, {Pagani}, {Pagano}, {Pailler}, {Palacin}, {Palaversa}, {Parsons}, {Paulsen}, {Pecoraro}, {Pedrosa}, {Pentik{\"a}inen}, {Pereira}, {Pichon}, {Piersimoni}, {Pineau}, {Plachy}, {Plum}, {Poujoulet}, {Pr{\v{s}}a}, {Pulone}, {Ragaini}, {Rago}, {Rambaux}, {Ramos-Lerate}, {Ranalli}, {Rauw}, {Read}, {Regibo}, {Renk}, {Reyl{\'e}}, {Ribeiro}, {Rimoldini}, {Ripepi}, {Riva}, {Rixon},
  {Roelens}, {Romero-G{\'o}mez}, {Rowell}, {Royer}, {Rudolph}, {Ruiz-Dern}, {Sadowski}, {Sagrist{\`a} Sell{\'e}s}, {Sahlmann}, {Salgado}, {Salguero}, {Sarasso}, {Savietto}, {Schnorhk}, {Schultheis}, {Sciacca}, {Segol}, {Segovia}, {Segransan}, {Serpell}, {Shih}, {Smareglia}, {Smart}, {Smith}, {Solano}, {Solitro}, {Sordo}, {Soria Nieto}, {Souchay}, {Spagna}, {Spoto}, {Stampa}, {Steele}, {Steidelm{\"u}ller}, {Stephenson}, {Stoev}, {Suess}, {S{\"u}veges}, {Surdej}, {Szabados}, {Szegedi-Elek}, {Tapiador}, {Taris}, {Tauran}, {Taylor}, {Teixeira}, {Terrett}, {Tingley}, {Trager}, {Turon}, {Ulla}, {Utrilla}, {Valentini}, {van Elteren}, {Van Hemelryck}, {van Leeuwen}, {Varadi}, {Vecchiato}, {Veljanoski}, {Via}, {Vicente}, {Vogt}, {Voss}, {Votruba}, {Voutsinas}, {Walmsley}, {Weiler}, {Weingrill}, {Werner}, {Wevers}, {Whitehead}, {Wyrzykowski}, {Yoldas}, {{\v{Z}}erjal}, {Zucker}, {Zurbach}, {Zwitter}, {Alecu}, {Allen}, {Allende Prieto}, {Amorim}, {Anglada-Escud{\'e}}, {Arsenijevic}, {Azaz}, {Balm}, {Beck}, {Bernstein},
  {Bigot}, {Bijaoui}, {Blasco}, {Bonfigli}, {Bono}, {Boudreault}, {Bressan}, {Brown}, {Brunet}, {Bunclark}, {Buonanno}, {Butkevich}, {Carret}, {Carrion}, {Chemin}, {Ch{\'e}reau}, {Corcione}, {Darmigny}, {de Boer}, {de Teodoro}, {de Zeeuw}, {Delle Luche}, {Domingues}, {Dubath}, {Fodor}, {Fr{\'e}zouls}, {Fries}, {Fustes}, {Fyfe}, {Gallardo}, {Gallegos}, {Gardiol}, {Gebran}, {Gomboc}, {G{\'o}mez}, {Grux}, {Gueguen}, {Heyrovsky}, {Hoar}, {Iannicola}, {Isasi Parache}, {Janotto}, {Joliet}, {Jonckheere}, {Keil}, {Kim}, {Klagyivik}, {Klar}, {Knude}, {Kochukhov}, {Kolka}, {Kos}, {Kutka}, {Lainey}, {LeBouquin}, {Liu}, {Loreggia}, {Makarov}, {Marseille}, {Martayan}, {Martinez-Rubi}, {Massart}, {Meynadier}, {Mignot}, {Munari}, {Nguyen}, {Nordlander}, {Ocvirk}, {O'Flaherty}, {Olias Sanz}, {Ortiz}, {Osorio}, {Oszkiewicz}, {Ouzounis}, {Palmer}, {Park}, {Pasquato}, {Peltzer}, {Peralta}, {P{\'e}turaud}, {Pieniluoma}, {Pigozzi}, {Poels}, {Prat}, {Prod'homme}, {Raison}, {Rebordao}, {Risquez}, {Rocca-Volmerange}, {Rosen},
  {Ruiz-Fuertes}, {Russo}, {Sembay}, {Serraller Vizcaino}, {Short}, {Siebert}, {Silva}, {Sinachopoulos}, {Slezak}, {Soffel}, {Sosnowska}, {Strai{\v{z}}ys}, {ter Linden}, {Terrell}, {Theil}, {Tiede}, {Troisi}, {Tsalmantza}, {Tur}, {Vaccari}, {Vachier}, {Valles}, {Van Hamme}, {Veltz}, {Virtanen}, {Wallut}, {Wichmann}, {Wilkinson}, {Ziaeepour}, \& {Zschocke}}]{Gaia2016}
{Gaia Collaboration}, {Prusti}, T., {de Bruijne}, J.~H.~J., {et~al.} 2016, \aap, 595, A1

\bibitem[{{Gnedin} \& {Ostriker}(1997)}]{Gnedin_1997}
{Gnedin}, O.~Y. \& {Ostriker}, J.~P. 1997, \apj, 474, 223

\bibitem[{{Grillmair}(2006)}]{Grillmair_2006}
{Grillmair}, C.~J. 2006, \apjl, 645, L37

\bibitem[{{Grossi} {et~al.}(2020){Grossi}, {Garc{\'\i}a-Benito}, {Cortesi}, {Gon{\c{c}}alves}, {Gon{\c{c}}alves}, {Lopes}, {Men{\'e}ndez-Delmestre}, \& {Telles}}]{Grossi_2020}
{Grossi}, M., {Garc{\'\i}a-Benito}, R., {Cortesi}, A., {et~al.} 2020, \mnras, 498, 1939

\bibitem[{{Hansen} {et~al.}(2020{\natexlab{a}}){Hansen}, {Marshall}, {Simon}, {Li}, {Bernstein}, {Pace}, {Ferguson}, {Nagasawa}, {Kuehn}, {Carollo}, {Geha}, {James}, {Walker}, {Diehl}, {Aguena}, {Allam}, {Avila}, {Bertin}, {Brooks}, {Buckley-Geer}, {Burke}, {Rosell}, {Kind}, {Carretero}, {Costanzi}, {Da Costa}, {Desai}, {De Vicente}, {Doel}, {Eckert}, {Eifler}, {Everett}, {Ferrero}, {Frieman}, {Garc{\'\i}a-Bellido}, {Gaztanaga}, {Gerdes}, {Gruen}, {Gruendl}, {Gschwend}, {Gutierrez}, {Hinton}, {Hollowood}, {Honscheid}, {Kuropatkin}, {Maia}, {March}, {Miquel}, {Palmese}, {Paz-Chinch{\'o}n}, {Plazas}, {Sanchez}, {Santiago}, {Scarpine}, {Serrano}, {Smith}, {Soares-Santos}, {Suchyta}, {Swanson}, {Tarle}, {Varga}, {Wilkinson}, \& {DES Collaboration}}]{Hansen_2020b}
{Hansen}, T.~T., {Marshall}, J.~L., {Simon}, J.~D., {et~al.} 2020{\natexlab{a}}, \apj, 897, 183

\bibitem[{{Hansen} {et~al.}(2020{\natexlab{b}}){Hansen}, {Riley}, {Strigari}, {Marshall}, {Ferguson}, {Zepeda}, \& {Sneden}}]{Hansen_2020a}
{Hansen}, T.~T., {Riley}, A.~H., {Strigari}, L.~E., {et~al.} 2020{\natexlab{b}}, \apj, 901, 23

\bibitem[{Harris {et~al.}(2020)Harris, Millman, van~der Walt, Gommers, Virtanen, Cournapeau, Wieser, Taylor, Berg, Smith, Kern, Picus, Hoyer, van Kerkwijk, Brett, Haldane, del R{\'{i}}o, Wiebe, Peterson, G{\'{e}}rard-Marchant, Sheppard, Reddy, Weckesser, Abbasi, Gohlke, \& Oliphant}]{Harris2020}
Harris, C.~R., Millman, K.~J., van~der Walt, S.~J., {et~al.} 2020, Nature, 585, 357

\bibitem[{{Harris}(1996)}]{Harris_1996}
{Harris}, W.~E. 1996, \aj, 112, 1487

\bibitem[{{Hawkins} {et~al.}(2020){Hawkins}, {Lucey}, {Ting}, {Ji}, {Katzberg}, {Thompson}, {El-Badry}, {Teske}, {Nelson}, \& {Carrillo}}]{Hawkins_2020}
{Hawkins}, K., {Lucey}, M., {Ting}, Y.-S., {et~al.} 2020, \mnras, 492, 1164

\bibitem[{Hawkins {et~al.}(2023)Hawkins, Price-Whelan, Sheffield, Subrahimovic, Beaton, Belokurov, Erkal, Koposov, Lane, Laporte, \& Nitschelm}]{Hawkins_2023}
Hawkins, K., Price-Whelan, A.~M., Sheffield, A.~A., {et~al.} 2023, The Astrophysical Journal, 948, 123

\bibitem[{{Haywood} {et~al.}(2018){Haywood}, {Di Matteo}, {Lehnert}, {Snaith}, {Khoperskov}, \& {G{\'o}mez}}]{Haywood_2018}
{Haywood}, M., {Di Matteo}, P., {Lehnert}, M.~D., {et~al.} 2018, \apj, 863, 113

\bibitem[{{Helmi} {et~al.}(2018){Helmi}, {Babusiaux}, {Koppelman}, {Massari}, {Veljanoski}, \& {Brown}}]{Helmi_2018}
{Helmi}, A., {Babusiaux}, C., {Koppelman}, H.~H., {et~al.} 2018, \nat, 563, 85

\bibitem[{{Helmi} {et~al.}(1999){Helmi}, {White}, {de Zeeuw}, \& {Zhao}}]{Helmi_1999}
{Helmi}, A., {White}, S. D.~M., {de Zeeuw}, P.~T., \& {Zhao}, H. 1999, \nat, 402, 53

\bibitem[{{Hinton} \& {Roweis}(2002)}]{Hinton2002}
{Hinton}, G. \& {Roweis}, S. 2002, Advances in Neural Processing Systems, 15, 833

\bibitem[{{Horta} {et~al.}(2020){Horta}, {Schiavon}, {Mackereth}, {Beers}, {Fern{\'a}ndez-Trincado}, {Frinchaboy}, {Garc{\'\i}a-Hern{\'a}ndez}, {Geisler}, {Hasselquist}, {J{\"o}nsson}, {Lane}, {Majewski}, {M{\'e}sz{\'a}ros}, {Bidin}, {Nataf}, {Roman-Lopes}, {Nitschelm}, {Vargas-Gonz{\'a}lez}, \& {Zasowski}}]{Horta_2020}
{Horta}, D., {Schiavon}, R.~P., {Mackereth}, J.~T., {et~al.} 2020, \mnras, 493, 3363

\bibitem[{{Hughes} {et~al.}(2022){Hughes}, {Spitler}, {Zucker}, {Nordlander}, {Simpson}, {da Costa}, {Ting}, {Li}, {Bland-Hawthorn}, {Buder}, {Casey}, {de Silva}, {D'Orazi}, {Freeman}, {Hayden}, {Kos}, {Lewis}, {Lin}, {Lind}, {Martell}, {Schlesinger}, {Sharma}, {Zwitter}, \& {GALAH Collaboration}}]{Hughes_2022}
{Hughes}, A. C.~N., {Spitler}, L.~R., {Zucker}, D.~B., {et~al.} 2022, \apj, 930, 47

\bibitem[{{Hunter}(2007)}]{Hunter2007}
{Hunter}, J.~D. 2007, Computing in Science and Engineering, 9, 90

\bibitem[{{Ibata} {et~al.}(2021){Ibata}, {Malhan}, {Martin}, {Aubert}, {Famaey}, {Bianchini}, {Monari}, {Siebert}, {Thomas}, {Bellazzini}, {Bonifacio}, {Caffau}, \& {Renaud}}]{Ibata_2021}
{Ibata}, R., {Malhan}, K., {Martin}, N., {et~al.} 2021, \apj, 914, 123

\bibitem[{{Ibata} {et~al.}(2022){Ibata}, {Malhan}, {Martin}, {Aubert}, {Famaey}, {Bianchini}, {Monari}, {Siebert}, {Thomas}, {Bellazzini}, {Bonifacio}, {Caffau}, \& {Renaud}}]{Ibata_2022_cat}
{Ibata}, R., {Malhan}, K., {Martin}, N., {et~al.} 2022, {VizieR Online Data Catalog: Gaia DR2 and EDR3 stars with sp. follow-up (Ibata+, 2021)}, VizieR On-line Data Catalog: J/ApJ/914/123. Originally published in: 2021ApJ...914..123I

\bibitem[{{Ibata} {et~al.}(2024){Ibata}, {Malhan}, {Tenachi}, {Ardern-Arentsen}, {Bellazzini}, {Bianchini}, {Bonifacio}, {Caffau}, {Diakogiannis}, {Errani}, {Famaey}, {Ferrone}, {Martin}, {di Matteo}, {Monari}, {Renaud}, {Starkenburg}, {Thomas}, {Viswanathan}, \& {Yuan}}]{Ibata_2024}
{Ibata}, R., {Malhan}, K., {Tenachi}, W., {et~al.} 2024, \apj, 967, 89

\bibitem[{{Ibata} {et~al.}(2019){Ibata}, {Bellazzini}, {Malhan}, {Martin}, \& {Bianchini}}]{Ibata_2019}
{Ibata}, R.~A., {Bellazzini}, M., {Malhan}, K., {Martin}, N., \& {Bianchini}, P. 2019, Nature Astronomy, 3, 667

\bibitem[{{Ibata} {et~al.}(1994){Ibata}, {Gilmore}, \& {Irwin}}]{Ibata_1994}
{Ibata}, R.~A., {Gilmore}, G., \& {Irwin}, M.~J. 1994, \nat, 370, 194

\bibitem[{{Ji} {et~al.}(2020){Ji}, {Li}, {Hansen}, {Casey}, {Koposov}, {Pace}, {Mackey}, {Lewis}, {Simpson}, {Bland-Hawthorn}, {Cullinane}, {Da Costa}, {Hattori}, {Martell}, {Kuehn}, {Erkal}, {Shipp}, {Wan}, \& {Zucker}}]{Ji_2020}
{Ji}, A.~P., {Li}, T.~S., {Hansen}, T.~T., {et~al.} 2020, \aj, 160, 181

\bibitem[{{Koposov} {et~al.}(2019){Koposov}, {Belokurov}, {Li}, {Mateu}, {Erkal}, {Grillmair}, {Hendel}, {Price-Whelan}, {Laporte}, {Hawkins}, {Sohn}, {del Pino}, {Evans}, {Slater}, {Kallivayalil}, {Navarro}, \& {Orphan Aspen Treasury Collaboration}}]{Koposov_2019}
{Koposov}, S.~E., {Belokurov}, V., {Li}, T.~S., {et~al.} 2019, \mnras, 485, 4726

\bibitem[{Koposov {et~al.}(2023)Koposov, Erkal, Li, Da Costa, Cullinane, Ji, Kuehn, Lewis, Pace, Shipp, Zucker, Bland-Hawthorn, Lilleengen, Martell, \& Collaboration)}]{Koposov_2023}
Koposov, S.~E., Erkal, D., Li, T.~S., {et~al.} 2023, Monthly Notices of the Royal Astronomical Society, 521, 4936

\bibitem[{{Koppelman} {et~al.}(2020){Koppelman}, {Bos}, \& {Helmi}}]{Koppelman_2020}
{Koppelman}, H.~H., {Bos}, R. O.~Y., \& {Helmi}, A. 2020, \aap, 642, L18

\bibitem[{{Koppelman} {et~al.}(2019{\natexlab{a}}){Koppelman}, {Helmi}, {Massari}, {Price-Whelan}, \& {Starkenburg}}]{Koppelman_2019b}
{Koppelman}, H.~H., {Helmi}, A., {Massari}, D., {Price-Whelan}, A.~M., \& {Starkenburg}, T.~K. 2019{\natexlab{a}}, \aap, 631, L9

\bibitem[{{Koppelman} {et~al.}(2019{\natexlab{b}}){Koppelman}, {Helmi}, {Massari}, {Roelenga}, \& {Bastian}}]{Koppelman_2019a}
{Koppelman}, H.~H., {Helmi}, A., {Massari}, D., {Roelenga}, S., \& {Bastian}, U. 2019{\natexlab{b}}, \aap, 625, A5

\bibitem[{{Kos} {et~al.}(2018){Kos}, {Bland-Hawthorn}, {Freeman}, {Buder}, {Traven}, {De Silva}, {Sharma}, {Asplund}, {Duong}, {Lin}, {Lind}, {Martell}, {Simpson}, {Stello}, {Zucker}, {Zwitter}, {Anguiano}, {Da Costa}, {D'Orazi}, {Horner}, {Kafle}, {Lewis}, {Munari}, {Nataf}, {Ness}, {Reid}, {Schlesinger}, {Ting}, \& {Wyse}}]{Kos_2018}
{Kos}, J., {Bland-Hawthorn}, J., {Freeman}, K., {et~al.} 2018, \mnras, 473, 4612

\bibitem[{Kruijssen {et~al.}(2020)Kruijssen, Pfeffer, Chevance, Bonaca, Trujillo-Gomez, Bastian, Reina-Campos, Crain, \& Hughes}]{Kruijssen_2020}
Kruijssen, J. M.~D., Pfeffer, J.~L., Chevance, M., {et~al.} 2020, Monthly Notices of the Royal Astronomical Society, 498, 2472

\bibitem[{{Kundu} {et~al.}(2021){Kundu}, {Navarrete}, {Fern{\'a}ndez-Trincado}, {Minniti}, {Singh}, {Sbordone}, {Piatti}, \& {Reyl{\'e}}}]{Kundu_2021}
{Kundu}, R., {Navarrete}, C., {Fern{\'a}ndez-Trincado}, J.~G., {et~al.} 2021, \aap, 645, A116

\bibitem[{{Kushniruk} {et~al.}(2025){Kushniruk}, {Youakim}, \& {Lind}}]{Kushniruk_2025}
{Kushniruk}, I., {Youakim}, K., \& {Lind}, K. 2025, A\&A, Submitted

\bibitem[{{Lane} {et~al.}(2020){Lane}, {Navarro}, {Fattahi}, {Oman}, \& {Bovy}}]{Lane_2020}
{Lane}, J. M.~M., {Navarro}, J.~F., {Fattahi}, A., {Oman}, K.~A., \& {Bovy}, J. 2020, \mnras, 492, 4164

\bibitem[{{Leaman} {et~al.}(2013){Leaman}, {VandenBerg}, \& {Mendel}}]{Leaman_2013}
{Leaman}, R., {VandenBerg}, D.~A., \& {Mendel}, J.~T. 2013, \mnras, 436, 122

\bibitem[{{Li} {et~al.}(2022){Li}, {Ji}, {Pace}, {Erkal}, {Koposov}, {Shipp}, {Da Costa}, {Cullinane}, {Kuehn}, {Lewis}, {Mackey}, {Simpson}, {Zucker}, {Ferguson}, {Martell}, {Bland-Hawthorn}, {Balbinot}, {Tavangar}, {Drlica-Wagner}, {De Silva}, \& {Simon}}]{Li_2022}
{Li}, T.~S., {Ji}, A.~P., {Pace}, A.~B., {et~al.} 2022, \apj, 928, 30

\bibitem[{Li {et~al.}(2019)Li, Koposov, Zucker, Lewis, Kuehn, Simpson, Ji, Shipp, Mao, Geha, Pace, Mackey, Allam, Tucker, Da~Costa, Erkal, Simon, Mould, Martell, Wan, De~Silva, Bechtol, Balbinot, Belokurov, Bland-Hawthorn, Casey, Cullinane, Drlica-Wagner, Sharma, Vivas, Wechsler, Yanny, \& Collaboration)}]{Li_2019}
Li, T.~S., Koposov, S.~E., Zucker, D.~B., {et~al.} 2019, Monthly Notices of the Royal Astronomical Society, 490, 3508

\bibitem[{{Limberg} {et~al.}(2024){Limberg}, {Ji}, {Naidu}, {Chiti}, {Rossi}, {Usman}, {Ting}, {Zaritsky}, {Bonaca}, {Borbolato}, {Speagle}, {Chandra}, \& {Conroy}}]{Limberg_2024}
{Limberg}, G., {Ji}, A.~P., {Naidu}, R.~P., {et~al.} 2024, \mnras, 530, 2512

\bibitem[{{Lindegren} {et~al.}(2021){Lindegren}, {Klioner}, {Hern{\'a}ndez}, {Bombrun}, {Ramos-Lerate}, {Steidelm{\"u}ller}, {Bastian}, {Biermann}, {de Torres}, {Gerlach}, {Geyer}, {Hilger}, {Hobbs}, {Lammers}, {McMillan}, {Stephenson}, {Casta{\~n}eda}, {Davidson}, {Fabricius}, {Gracia-Abril}, {Portell}, {Rowell}, {Teyssier}, {Torra}, {Bartolom{\'e}}, {Clotet}, {Garralda}, {Gonz{\'a}lez-Vidal}, {Torra}, {Abbas}, {Altmann}, {Anglada Varela}, {Balaguer-N{\'u}{\~n}ez}, {Balog}, {Barache}, {Becciani}, {Bernet}, {Bertone}, {Bianchi}, {Bouquillon}, {Brown}, {Bucciarelli}, {Busonero}, {Butkevich}, {Buzzi}, {Cancelliere}, {Carlucci}, {Charlot}, {Cioni}, {Crosta}, {Crowley}, {del Peloso}, {del Pozo}, {Drimmel}, {Esquej}, {Fienga}, {Fraile}, {Gai}, {Garcia-Reinaldos}, {Guerra}, {Hambly}, {Hauser}, {Jan{\ss}en}, {Jordan}, {Kostrzewa-Rutkowska}, {Lattanzi}, {Liao}, {Licata}, {Lister}, {L{\"o}ffler}, {Marchant}, {Masip}, {Mignard}, {Mints}, {Molina}, {Mora}, {Morbidelli}, {Murphy}, {Pagani}, {Panuzzo}, {Pe{\~n}alosa
  Esteller}, {Poggio}, {Re Fiorentin}, {Riva}, {Sagrist{\`a} Sell{\'e}s}, {Sanchez Gimenez}, {Sarasso}, {Sciacca}, {Siddiqui}, {Smart}, {Souami}, {Spagna}, {Steele}, {Taris}, {Utrilla}, {van Reeven}, \& {Vecchiato}}]{Lindegren2021}
{Lindegren}, L., {Klioner}, S.~A., {Hern{\'a}ndez}, J., {et~al.} 2021, \aap, 649, A2

\bibitem[{{Malhan}(2022)}]{Malhan_2022b}
{Malhan}, K. 2022, \apjl, 930, L9

\bibitem[{{Malhan} \& {Ibata}(2018)}]{Malhan_2018a}
{Malhan}, K. \& {Ibata}, R.~A. 2018, \mnras, 477, 4063

\bibitem[{{Malhan} {et~al.}(2019){Malhan}, {Ibata}, {Carlberg}, {Bellazzini}, {Famaey}, \& {Martin}}]{Malhan_2019}
{Malhan}, K., {Ibata}, R.~A., {Carlberg}, R.~G., {et~al.} 2019, \apjl, 886, L7

\bibitem[{{Malhan} {et~al.}(2018){Malhan}, {Ibata}, \& {Martin}}]{Malhan_2018b}
{Malhan}, K., {Ibata}, R.~A., \& {Martin}, N.~F. 2018, \mnras, 481, 3442

\bibitem[{{Malhan} {et~al.}(2022){Malhan}, {Ibata}, {Sharma}, {Famaey}, {Bellazzini}, {Carlberg}, {D'Souza}, {Yuan}, {Martin}, \& {Thomas}}]{Malhan_2022a}
{Malhan}, K., {Ibata}, R.~A., {Sharma}, S., {et~al.} 2022, \apj, 926, 107

\bibitem[{{Malhan} {et~al.}(2021){Malhan}, {Yuan}, {Ibata}, {Arentsen}, {Bellazzini}, \& {Martin}}]{Malhan_2021}
{Malhan}, K., {Yuan}, Z., {Ibata}, R.~A., {et~al.} 2021, \apj, 920, 51

\bibitem[{{Martell} {et~al.}(2016){Martell}, {Shetrone}, {Lucatello}, {Schiavon}, {M{\'e}sz{\'a}ros}, {Allende Prieto}, {Garc{\'\i}a-Hern{\'a}ndez}, {Beers}, \& {Nidever}}]{Martell_2016}
{Martell}, S.~L., {Shetrone}, M.~D., {Lucatello}, S., {et~al.} 2016, \apj, 825, 146

\bibitem[{{Martin} {et~al.}(2022){Martin}, {Venn}, {Aguado}, {Starkenburg}, {Gonz{\'a}lez Hern{\'a}ndez}, {Ibata}, {Bonifacio}, {Caffau}, {Sestito}, {Arentsen}, {Allende Prieto}, {Carlberg}, {Fabbro}, {Fouesneau}, {Hill}, {Jablonka}, {Kordopatis}, {Lardo}, {Malhan}, {Mashonkina}, {McConnachie}, {Navarro}, {S{\'a}nchez-Janssen}, {Thomas}, {Yuan}, \& {Mucciarelli}}]{Martin_2022}
{Martin}, N.~F., {Venn}, K.~A., {Aguado}, D.~S., {et~al.} 2022, \nat, 601, 45

\bibitem[{{Massari} {et~al.}(2019){Massari}, {Koppelman}, \& {Helmi}}]{Massari_2019}
{Massari}, D., {Koppelman}, H.~H., \& {Helmi}, A. 2019, \aap, 630, L4

\bibitem[{{Mateu}(2023)}]{Mateu_2023}
{Mateu}, C. 2023, \mnras, 520, 5225

\bibitem[{{McConnachie} \& {Venn}(2020{\natexlab{a}})}]{McConnachie_2020a}
{McConnachie}, A.~W. \& {Venn}, K.~A. 2020{\natexlab{a}}, \aj, 160, 124

\bibitem[{{McConnachie} \& {Venn}(2020{\natexlab{b}})}]{McConnachie_2020b}
{McConnachie}, A.~W. \& {Venn}, K.~A. 2020{\natexlab{b}}, Research Notes of the American Astronomical Society, 4, 229

\bibitem[{{Meza} {et~al.}(2005){Meza}, {Navarro}, {Abadi}, \& {Steinmetz}}]{Meza_2005}
{Meza}, A., {Navarro}, J.~F., {Abadi}, M.~G., \& {Steinmetz}, M. 2005, \mnras, 359, 93

\bibitem[{{Myeong} {et~al.}(2018{\natexlab{a}}){Myeong}, {Evans}, {Belokurov}, {Sanders}, \& {Koposov}}]{Myeong_2018b}
{Myeong}, G.~C., {Evans}, N.~W., {Belokurov}, V., {Sanders}, J.~L., \& {Koposov}, S.~E. 2018{\natexlab{a}}, \mnras, 478, 5449

\bibitem[{{Myeong} {et~al.}(2018{\natexlab{b}}){Myeong}, {Evans}, {Belokurov}, {Sanders}, \& {Koposov}}]{Myeong_2018a}
{Myeong}, G.~C., {Evans}, N.~W., {Belokurov}, V., {Sanders}, J.~L., \& {Koposov}, S.~E. 2018{\natexlab{b}}, \apjl, 863, L28

\bibitem[{Myeong {et~al.}(2019)Myeong, Vasiliev, Iorio, Evans, \& Belokurov}]{Myeong_2019}
Myeong, G.~C., Vasiliev, E., Iorio, G., Evans, N.~W., \& Belokurov, V. 2019, Monthly Notices of the Royal Astronomical Society, 488, 1235

\bibitem[{{Naidu} {et~al.}(2020){Naidu}, {Conroy}, {Bonaca}, {Johnson}, {Ting}, {Caldwell}, {Zaritsky}, \& {Cargile}}]{Naidu_2020}
{Naidu}, R.~P., {Conroy}, C., {Bonaca}, A., {et~al.} 2020, \apj, 901, 48

\bibitem[{Newberg {et~al.}(2010)Newberg, Willett, Yanny, \& Xu}]{Newberg_2010}
Newberg, H.~J., Willett, B.~A., Yanny, B., \& Xu, Y. 2010, The Astrophysical Journal, 711, 32

\bibitem[{{Ortigoza-Urdaneta} {et~al.}(2023){Ortigoza-Urdaneta}, {Vieira}, {Fern{\'a}ndez-Trincado}, {Queiroz}, {Barbuy}, {Beers}, {Chiappini}, {Anders}, {Minniti}, \& {Tang}}]{Ortigoza-Urdaneta_2023}
{Ortigoza-Urdaneta}, M., {Vieira}, K., {Fern{\'a}ndez-Trincado}, J.~G., {et~al.} 2023, \aap, 676, A140

\bibitem[{{Palau} \& {Miralda-Escud{\'e}}(2019)}]{Palau_2019}
{Palau}, C.~G. \& {Miralda-Escud{\'e}}, J. 2019, \mnras, 488, 1535

\bibitem[{{Price-Jones} \& {Bovy}(2019)}]{Price-Jones_2019}
{Price-Jones}, N. \& {Bovy}, J. 2019, \mnras, 487, 871

\bibitem[{{Prudil} {et~al.}(2021){Prudil}, {Hanke}, {Lemasle}, {Crestani}, {Braga}, {Fabrizio}, {Koch-Hansen}, {Bono}, {Grebel}, {Matsunaga}, {Marengo}, {da Silva}, {Dall'Ora}, {Mart{\'\i}nez-V{\'a}zquez}, {Altavilla}, {Lala}, {Chaboyer}, {Ferraro}, {Fiorentino}, {Gilligan}, {Nonino}, \& {Th{\'e}venin}}]{Prudil_2021}
{Prudil}, Z., {Hanke}, M., {Lemasle}, B., {et~al.} 2021, \aap, 648, A78

\bibitem[{{Python Core Team}(2019)}]{python}
{Python Core Team}. 2019, {Python: A dynamic, open source programming language}, {Python Software Foundation}

\bibitem[{{Queiroz} {et~al.}(2018){Queiroz}, {Anders}, {Santiago}, {Chiappini}, {Steinmetz}, {Dal Ponte}, {Stassun}, {da Costa}, {Maia}, {Crestani}, {Beers}, {Fern{\'a}ndez-Trincado}, {Garc{\'\i}a-Hern{\'a}ndez}, {Roman-Lopes}, \& {Zamora}}]{Queiroz_2018}
{Queiroz}, A.~B.~A., {Anders}, F., {Santiago}, B.~X., {et~al.} 2018, \mnras, 476, 2556

\bibitem[{{Riley} \& {Strigari}(2020)}]{Riley_2020}
{Riley}, A.~H. \& {Strigari}, L.~E. 2020, \mnras, 494, 983

\bibitem[{{Santiago} {et~al.}(2022){Santiago}, {Chaushev}, \& {Sallum}}]{Santiago_2022}
{Santiago}, C., {Chaushev}, A., \& {Sallum}, S. 2022, in American Astronomical Society Meeting Abstracts, Vol. 240, American Astronomical Society Meeting \#240, 418.08

\bibitem[{{Sesar} {et~al.}(2015){Sesar}, {Bovy}, {Bernard}, {Caldwell}, {Cohen}, {Fouesneau}, {Johnson}, {Ness}, {Ferguson}, {Martin}, {Price-Whelan}, {Rix}, {Schlafly}, {Burgett}, {Chambers}, {Flewelling}, {Hodapp}, {Kaiser}, {Magnier}, {Platais}, {Tonry}, {Waters}, \& {Wyse}}]{Sesar_2015}
{Sesar}, B., {Bovy}, J., {Bernard}, E.~J., {et~al.} 2015, \apj, 809, 59

\bibitem[{{Sestito} {et~al.}(2020){Sestito}, {Martin}, {Starkenburg}, {Arentsen}, {Ibata}, {Longeard}, {Kielty}, {Youakim}, {Venn}, {Aguado}, {Carlberg}, {Gonz{\'a}lez Hern{\'a}ndez}, {Hill}, {Jablonka}, {Kordopatis}, {Malhan}, {Navarro}, {S{\'a}nchez-Janssen}, {Thomas}, {Tolstoy}, {Wilson}, {Palicio}, {Bialek}, {Garcia-Dias}, {Lucchesi}, {North}, {Osorio}, {Patrick}, \& {Peralta de Arriba}}]{Sestito_2020}
{Sestito}, F., {Martin}, N.~F., {Starkenburg}, E., {et~al.} 2020, \mnras, 497, L7

\bibitem[{Shipp {et~al.}(2018)Shipp, Drlica-Wagner, Balbinot, Ferguson, Erkal, Li, Bechtol, Belokurov, Buncher, Carollo, Kind, Kuehn, Marshall, Pace, Rykoff, Sevilla-Noarbe, Sheldon, Strigari, Vivas, Yanny, Zenteno, Abbott, Abdalla, Allam, Avila, Bertin, Brooks, Burke, Carretero, Castander, Cawthon, Crocce, Cunha, D’Andrea, da~Costa, Davis, Vicente, Desai, Diehl, Doel, Evrard, Flaugher, Fosalba, Frieman, García-Bellido, Gaztanaga, Gerdes, Gruen, Gruendl, Gschwend, Gutierrez, Hartley, Honscheid, Hoyle, James, Johnson, Krause, Kuropatkin, Lahav, Lin, Maia, March, Martini, Menanteau, Miller, Miquel, Nichol, Plazas, Romer, Sako, Sanchez, Santiago, Scarpine, Schindler, Schubnell, Smith, Smith, Sobreira, Suchyta, Swanson, Tarle, Thomas, Tucker, Walker, Wechsler, \& (DES Collaboration)}]{Shipp_2018}
Shipp, N., Drlica-Wagner, A., Balbinot, E., {et~al.} 2018, The Astrophysical Journal, 862, 114

\bibitem[{{Simon} {et~al.}(2020){Simon}, {Li}, {Erkal}, {Pace}, {Drlica-Wagner}, {James}, {Marshall}, {Bechtol}, {Hansen}, {Kuehn}, {Lidman}, {Allam}, {Annis}, {Avila}, {Bertin}, {Brooks}, {Burke}, {Rosell}, {Carrasco Kind}, {Carretero}, {da Costa}, {De Vicente}, {Desai}, {Doel}, {Eifler}, {Everett}, {Fosalba}, {Frieman}, {Garc{\'\i}a-Bellido}, {Gaztanaga}, {Gerdes}, {Gruen}, {Gruendl}, {Gschwend}, {Gutierrez}, {Hollowood}, {Honscheid}, {Krause}, {Kuropatkin}, {MacCrann}, {Maia}, {March}, {Miquel}, {Palmese}, {Paz-Chinch{\'o}n}, {Plazas}, {Reil}, {Roodman}, {Sanchez}, {Santiago}, {Scarpine}, {Schubnell}, {Serrano}, {Smith}, {Suchyta}, {Tarle}, {Walker}, \& {DES Collaboration}}]{Simon_2020}
{Simon}, J.~D., {Li}, T.~S., {Erkal}, D., {et~al.} 2020, \apj, 892, 137

\bibitem[{Sollima(2020)}]{Sollima_2020}
Sollima, A. 2020, Monthly Notices of the Royal Astronomical Society, 495, 2222

\bibitem[{{Sun} {et~al.}(2023){Sun}, {Wang}, {Liu}, {Long}, {Chen}, \& {Gao}}]{Sun_2023}
{Sun}, G., {Wang}, Y., {Liu}, C., {et~al.} 2023, Research in Astronomy and Astrophysics, 23, 015013

\bibitem[{{Taylor}(2005)}]{Taylor2005}
{Taylor}, M.~B. 2005, in Astronomical Society of the Pacific Conference Series, Vol. 347, Astronomical Data Analysis Software and Systems XIV, ed. P.~{Shopbell}, M.~{Britton}, \& R.~{Ebert}, 29

\bibitem[{Thomas {et~al.}(2020)Thomas, Jensen, McConnachie, Côté, Venn, Longeard, Carlberg, Chapman, Cuillandre, Famaey, Ferrarese, Gwyn, Hammer, Ibata, Malhan, Martin, Mei, Navarro, Reylé, \& Starkenburg}]{Thomas_2020}
Thomas, G.~F., Jensen, J., McConnachie, A., {et~al.} 2020, The Astrophysical Journal, 902, 89

\bibitem[{{Traven} {et~al.}(2020){Traven}, {Feltzing}, {Merle}, {Van der Swaelmen}, {{\v{C}}otar}, {Church}, {Zwitter}, {Ting}, {Sahlholdt}, {Asplund}, {Bland-Hawthorn}, {De Silva}, {Freeman}, {Martell}, {Sharma}, {Zucker}, {Buder}, {Casey}, {D'Orazi}, {Kos}, {Lewis}, {Lin}, {Lind}, {Simpson}, {Stello}, {Munari}, \& {Wittenmyer}}]{Traven_2020}
{Traven}, G., {Feltzing}, S., {Merle}, T., {et~al.} 2020, \aap, 638, A145

\bibitem[{{Traven} {et~al.}(2017){Traven}, {Matijevi{\v{c}}}, {Zwitter}, {{\v{Z}}erjal}, {Kos}, {Asplund}, {Bland-Hawthorn}, {Casey}, {De Silva}, {Freeman}, {Lin}, {Martell}, {Schlesinger}, {Sharma}, {Simpson}, {Zucker}, {Anguiano}, {Da Costa}, {Duong}, {Horner}, {Hyde}, {Kafle}, {Munari}, {Nataf}, {Navin}, {Reid}, \& {Ting}}]{Traven_2017}
{Traven}, G., {Matijevi{\v{c}}}, G., {Zwitter}, T., {et~al.} 2017, \apjs, 228, 24

\bibitem[{Ulyanov(2016)}]{Ulyanov_2016}
Ulyanov, D. 2016, Multicore-TSNE, \url{https://github.com/DmitryUlyanov/Multicore-TSNE}

\bibitem[{{van der Maaten}(2013)}]{vanderMaaten2013}
{van der Maaten}, L. 2013, arXiv e-prints, arXiv:1301.3342

\bibitem[{{van der Maaten} \& {Hinton}(2008)}]{vanderMaaten2008}
{van der Maaten}, L. \& {Hinton}, G. 2008, Journal of Machine Learning Research, 1, 1

\bibitem[{{Vasiliev}(2019)}]{Vasiliev_2019}
{Vasiliev}, E. 2019, \mnras, 484, 2832

\bibitem[{{Virtanen} {et~al.}(2020){Virtanen}, {Gommers}, {Oliphant}, {Haberland}, {Reddy}, {Cournapeau}, {Burovski}, {Peterson}, {Weckesser}, {Bright}, {van der Walt}, {Brett}, {Wilson}, {Jarrod Millman}, {Mayorov}, {Nelson}, {Jones}, {Kern}, {Larson}, {Carey}, {Polat}, {Feng}, {Moore}, {VanderPlas}, {Laxalde}, {Perktold}, {Cimrman}, {Henriksen}, {Quintero}, {Harris}, {Archibald}, {Ribeiro}, {Pedregosa}, {van Mulbregt}, \& {Contributors}}]{scipy}
{Virtanen}, P., {Gommers}, R., {Oliphant}, T.~E., {et~al.} 2020, Nature Methods

\bibitem[{{W}es {M}c{K}inney(2010)}]{Mckinney2010}
{W}es {M}c{K}inney. 2010, in {P}roceedings of the 9th {P}ython in {S}cience {C}onference, ed. {S}t\'efan van~der {W}alt \& {J}arrod {M}illman, 56 -- 61

\bibitem[{Yang {et~al.}(2023)Yang, Zhao, Tang, Ye, \& Zhao}]{Yang_2023}
Yang, Y., Zhao, J.-K., Tang, X.-Z., Ye, X.-H., \& Zhao, G. 2023, The Astrophysical Journal, 953, 130

\bibitem[{{Youakim} {et~al.}(2023){Youakim}, {Lind}, \& {Kushniruk}}]{Youakim_2023}
{Youakim}, K., {Lind}, K., \& {Kushniruk}, I. 2023, \mnras, 524, 2630

\bibitem[{{Yuan} {et~al.}(2020){Yuan}, {Chang}, {Beers}, \& {Huang}}]{Yuan_2020}
{Yuan}, Z., {Chang}, J., {Beers}, T.~C., \& {Huang}, Y. 2020, \apjl, 898, L37

\bibitem[{{Yuan} {et~al.}(2022){Yuan}, {Martin}, {Ibata}, {Caffau}, {Bonifacio}, {Mashonkina}, {Errani}, {Doliva-Dolinsky}, {Starkenburg}, {Venn}, {Arentsen}, {Aguado}, {Bellazzini}, {Famaey}, {Fouesneau}, {Gonz{\'a}lez Hern{\'a}ndez}, {Jablonka}, {Lardo}, {Malhan}, {Navarro}, {S{\'a}nchez Janssen}, {Sestito}, {Thomas}, {Viswanathan}, \& {Vitali}}]{Yuan_2022}
{Yuan}, Z., {Martin}, N.~F., {Ibata}, R.~A., {et~al.} 2022, \mnras, 514, 1664

\bibitem[{{Zinn}(1993)}]{Zinn_1993}
{Zinn}, R. 1993, in Astronomical Society of the Pacific Conference Series, Vol.~48, The Globular Cluster-Galaxy Connection, ed. G.~H. {Smith} \& J.~P. {Brodie}, 38

\end{thebibliography}
%

\begin{appendix}

\onecolumn
\section{Different clustering scales with changing perplexity}

\label{app:choosing_perplexity}
\FloatBarrier
   \begin{figure}[h!]
   \includegraphics[width=\textwidth]{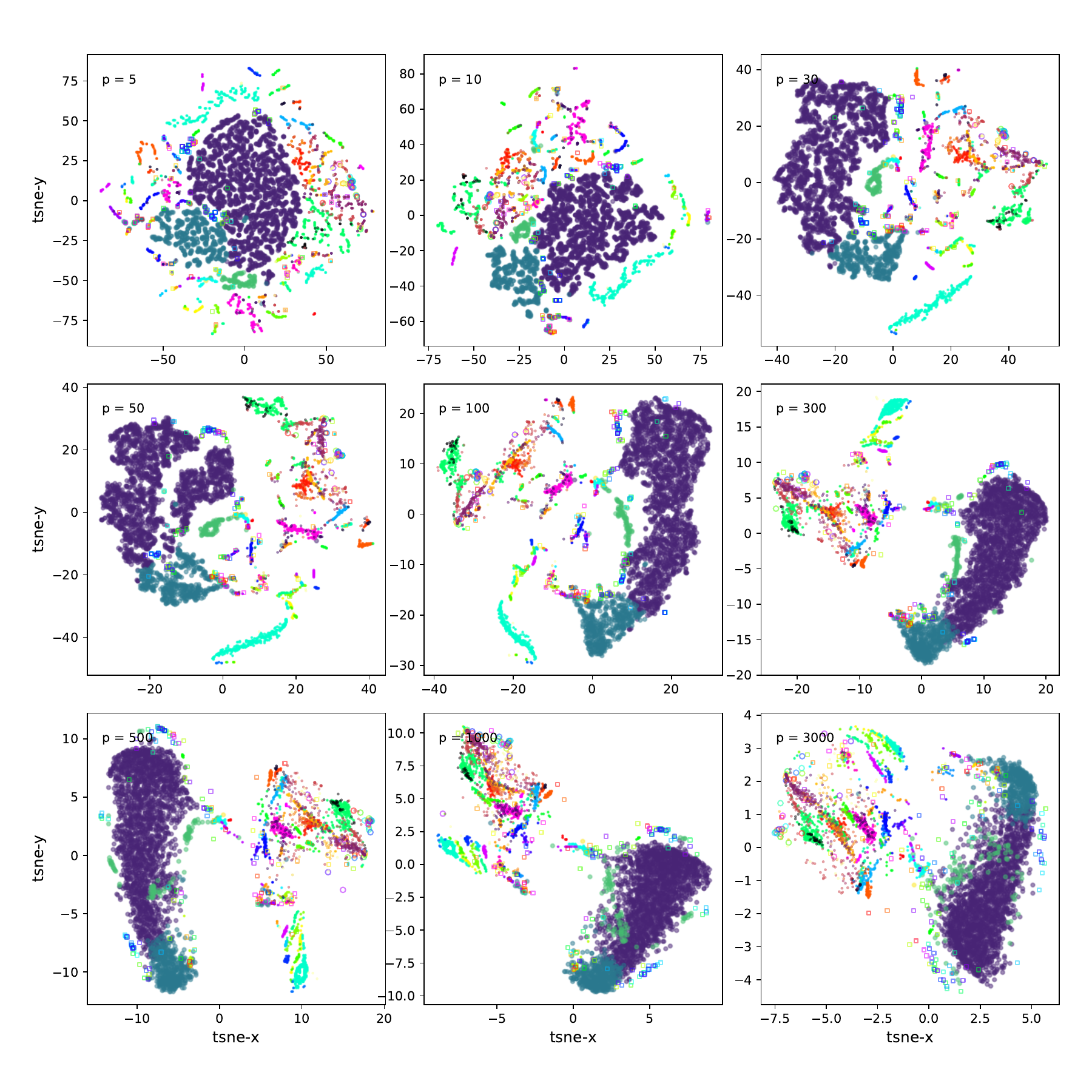}
   \label{fig_app:perp_grid}
   \caption{t-SNE latent space projections for different values of perplexity showing the different scales of clustering used in the selection of groups. Markers are the same as described in Figure 1.}
    \end{figure}
    
\FloatBarrier
\newpage

\section{Groups of globular clusters}
\label{app:gc_analyis}
\FloatBarrier
    \begin{figure}[h!]
    \centering
     \includegraphics[width=\textwidth]{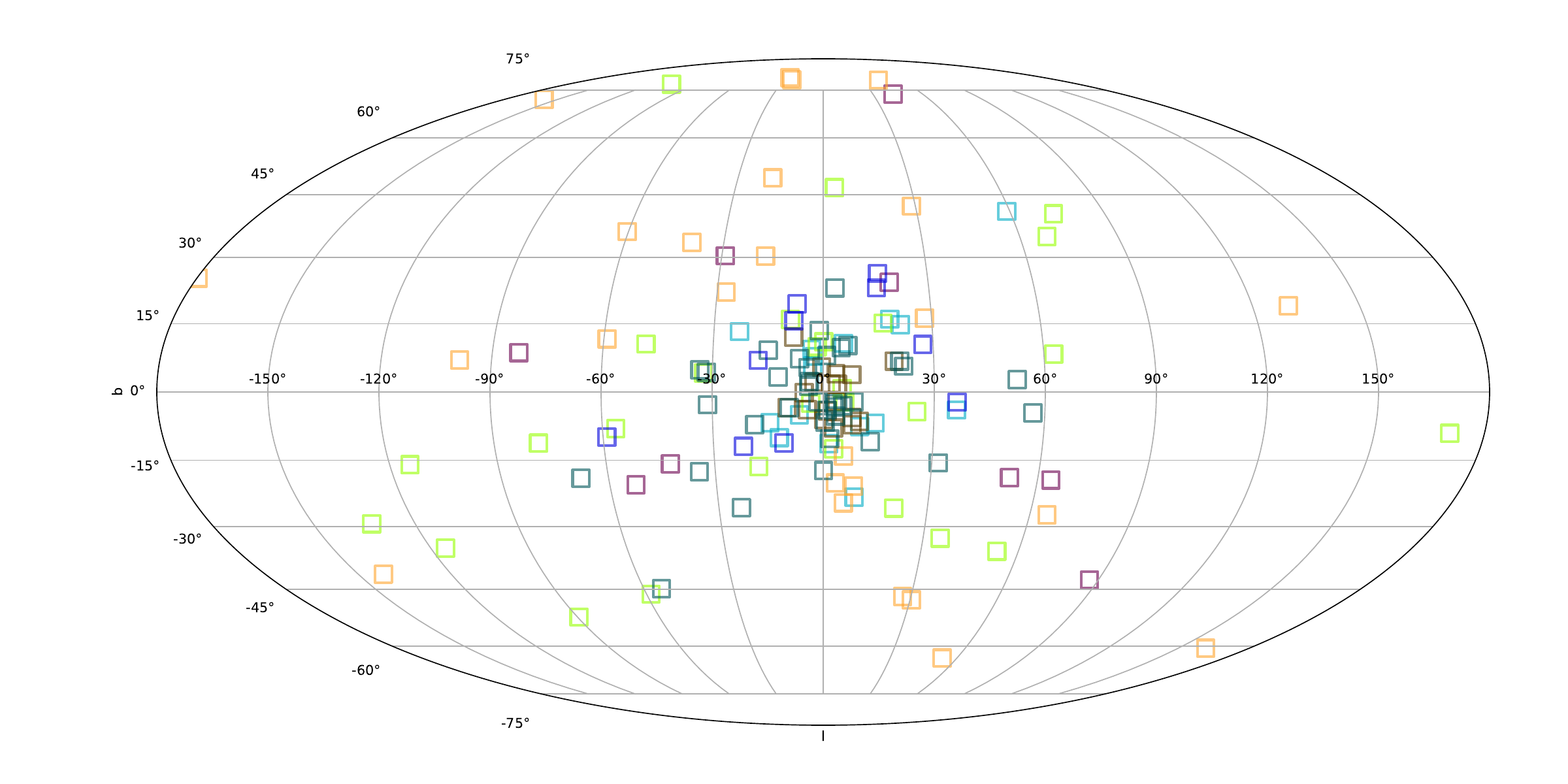}
     \includegraphics[width=\textwidth]{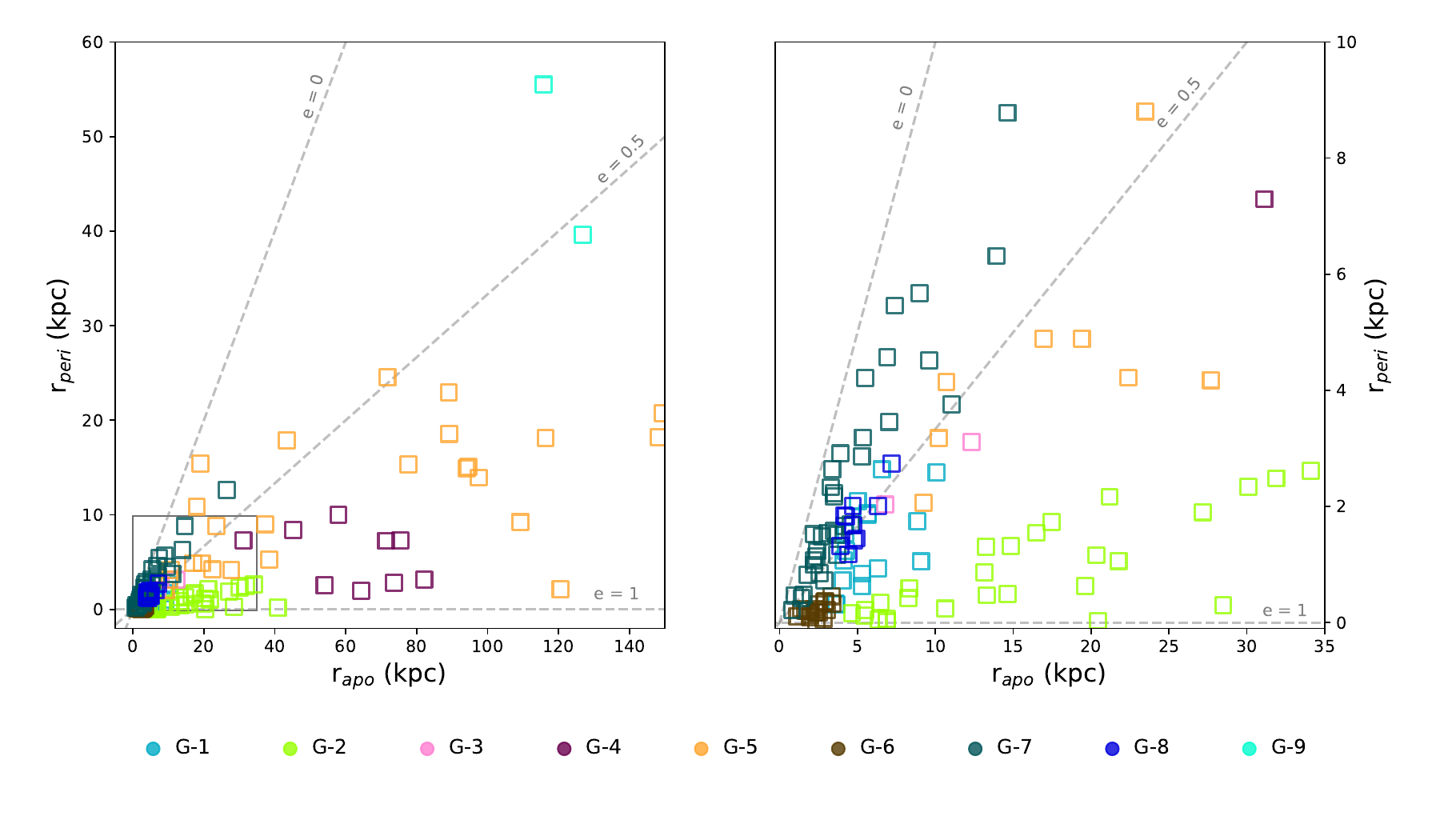}
   \caption{Top: Sky plot showing only the GCs from the selected groups plotted in Galactic coordinates. Bottom left: Galactic pericentre vs apocentre distances for the GCs. Bottom right: Zoomed-in view of the left panel, with axes according to the black box in the left panel. The grey dashed lines show lines of constant eccentricity.}
   \label{fig:all_gcs_validation_plot}
    \end{figure}
    
\FloatBarrier
\newpage
\section{Subgroups of GSE globular clusters}
\label{app:gse_subgroup_analyis}
\FloatBarrier
    \begin{figure}[h!]
    \label{fig:gse_validation_plot}
    \centering
     \includegraphics[width=\textwidth]{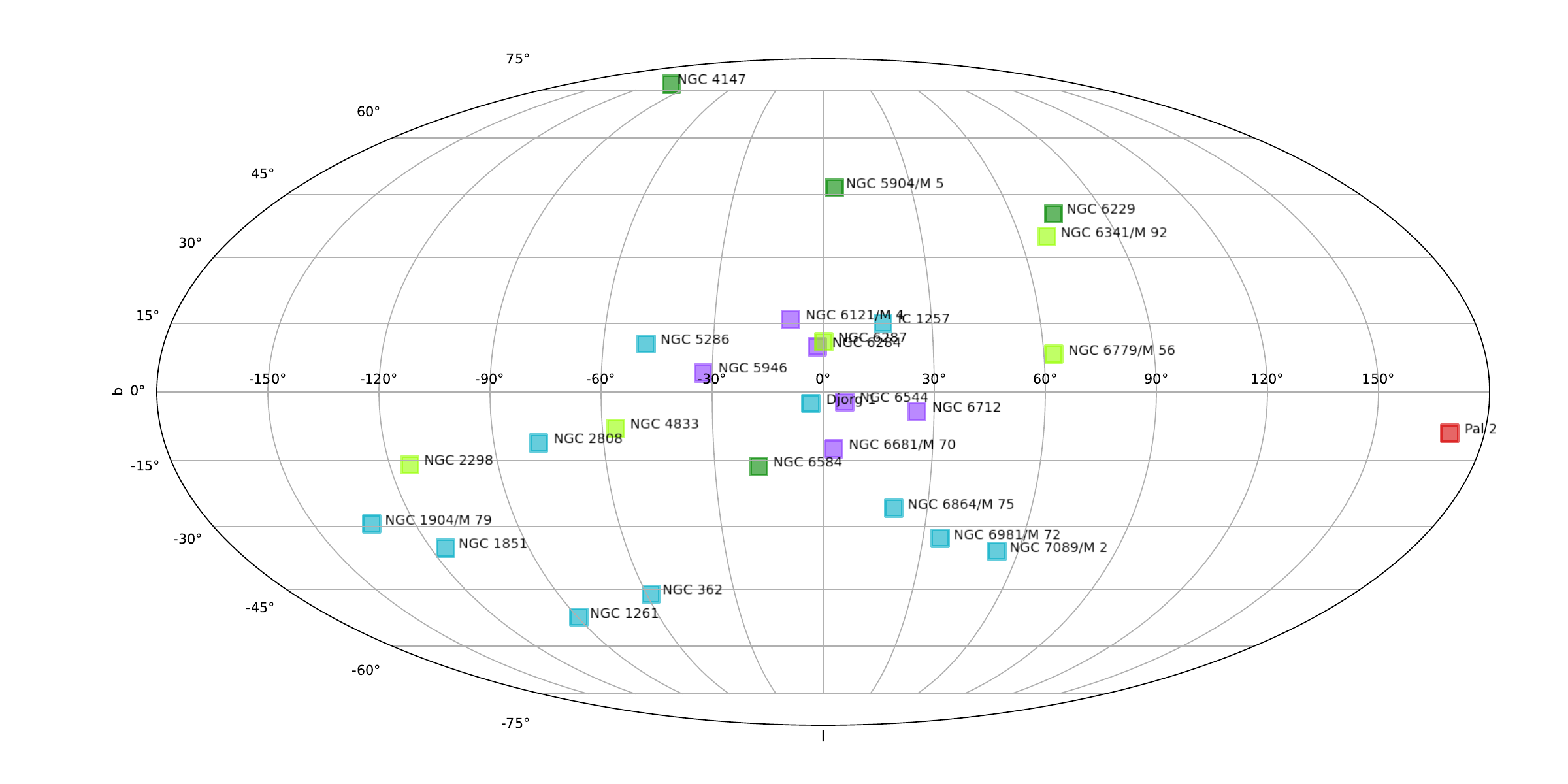}
     \vspace{-0.5cm}
     \includegraphics[width=\textwidth]{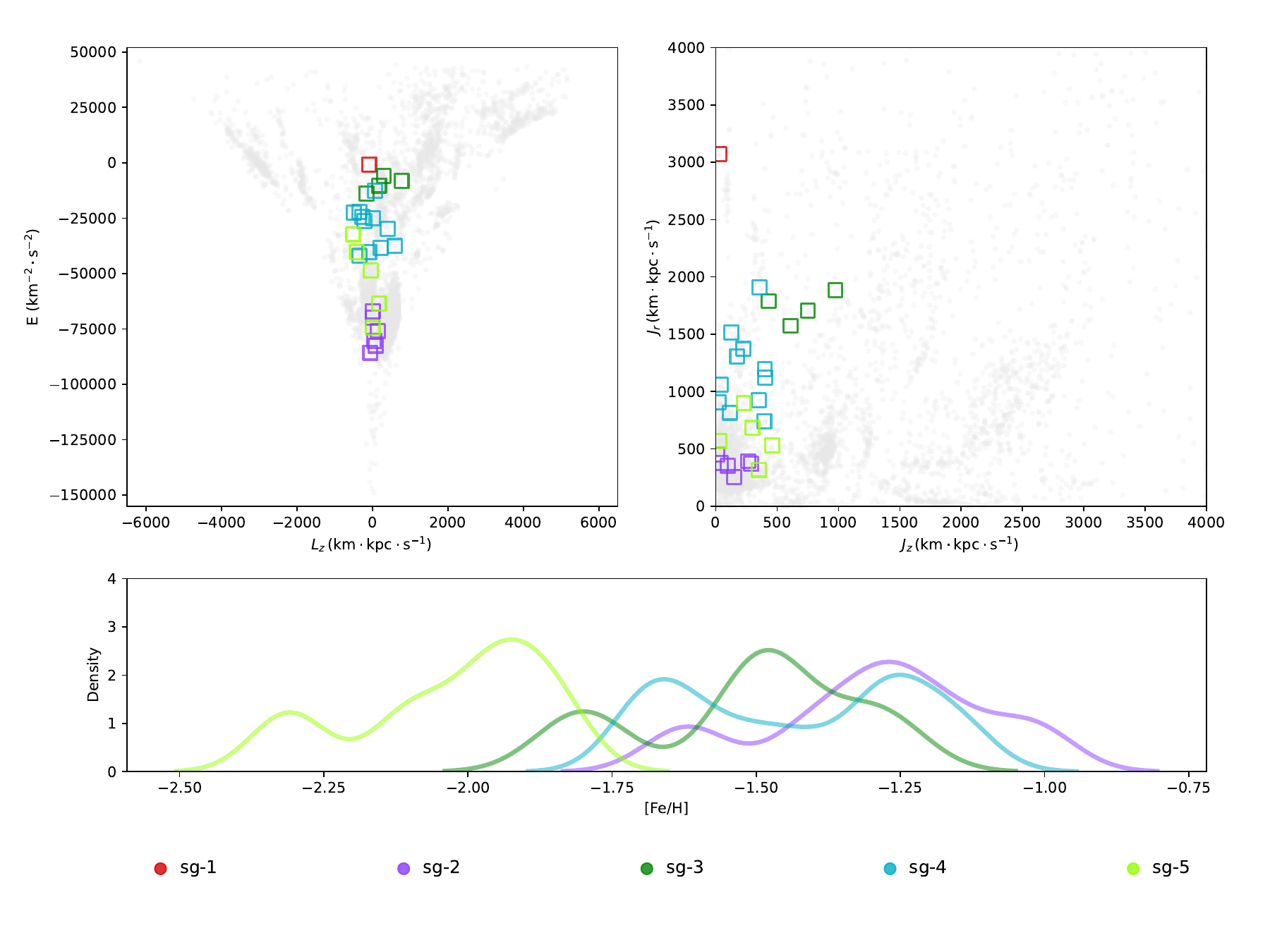}
     \vspace{-0.6cm}
   \caption{Top: Globular clusters from the GSE selected subgroups plotted in Galactic coordinates. Middle: Kinematic parameter spaces E-L$_z$ and J$_r$-J$_z$ are shown in the left and right panels, respectively. Each square symbol represents one GC and the grey points show the rest of the dataset to give context to where these GC populations lie. Bottom: Kernel density estimation of [Fe/H] distributions for each GSE subgroup.}
    \end{figure}
\FloatBarrier

\section{Full table of selected groups}
\label{app:full_groups_table}
\renewcommand{\arraystretch}{1.1}

\begin{longtable}{|c|c|l|l|l|}
\caption{Table summarising all of the identified groups and subgroups from the t-SNE latent space selection.} \\ \hline
\label{tab:large_associations}
Gr. & \begin{tabular}[c]{@{}c@{}}sub Gr.\\ (\#)\end{tabular} & \multicolumn{1}{c|}{Member GCs} & \multicolumn{1}{c|}{Member stellar streams} & \multicolumn{1}{c|}{Member satellite galaxies} \\ \hline \hline
\endfirsthead
\caption{continued.} \\
\hline
Gr. & \begin{tabular}[c]{@{}c@{}}sub Gr.\\ (\#)\end{tabular} & \multicolumn{1}{c|}{Member GCs} & \multicolumn{1}{c|}{Member stellar streams} & \multicolumn{1}{c|}{Member satellite galaxies} \\ \hline \hline
\endhead
\hline
\endfoot
\multirow{2}{*}{\begin{tabular}[c]{@{}c@{}}G-1\\ Splash\end{tabular}} & \begin{tabular}[c]{@{}c@{}}-\\ (11)\end{tabular} & \begin{tabular}[c]{@{}l@{}}NGC 4372, NGC 6093/M 80,\\ NGC 6144, NGC 6139,\\ NGC 6749, NGC 6218/M 12, \\ NGC 6254/M 10, NGC 6397, \\ NGC 6535, NGC 6541\end{tabular} & NGC 6397 & \\ \hline
- & \begin{tabular}[c]{@{}c@{}}sg-1\\ (2)\end{tabular} & Pal 2 & Hrid &  \\ \hline 
- & \begin{tabular}[c]{@{}c@{}}Kraken/Koala\\sg-2\\ (6)\end{tabular} & \begin{tabular}[c]{@{}l@{}}NGC 5946, NGC 6121/M 4,\\ NGC 6284, NGC 6544,\\ NGC 6681/M 70, NGC 6712\end{tabular} &  &  \\ \hline
 \multirow{2}{*}{\begin{tabular}[c]{@{}c@{}}\\ \\ \\ G-2\\ GSE\end{tabular}} & 
 \begin{tabular}[c]{@{}c@{}}sg-3\\ (5)\end{tabular} & \begin{tabular}[c]{@{}l@{}}NGC 4147, NGC 5904/M 5,\\ NGC 6229, NGC 6584\end{tabular} & M5 &  \\ \cline{2-5} 
 & \begin{tabular}[c]{@{}c@{}}sg-4\\ (13)\end{tabular} & \begin{tabular}[c]{@{}l@{}}NGC 362, NGC 1261, \\ NGC 1851, NGC 1904/M 79, \\ NGC 2808, NGC 5286, \\ IC 1257, Djorg 1, \\ NGC 6864/M 75, \\ NGC 6981/M 72, \\ NGC 7089/M 2\end{tabular} & \begin{tabular}[c]{@{}l@{}}NGC 1851, \\ NGC 2808\end{tabular} &  \\ \hline 
 - & \begin{tabular}[c]{@{}c@{}}Pontus\\sg-5\\ (9)\end{tabular} & \begin{tabular}[c]{@{}l@{}}NGC 2298, NGC 6341/M 92,\\ NGC 6779/M 56, NGC 6287\\ NGC 7099/M 30, NGC 4833, \\ ESO 280/SC 06 \end{tabular} & \begin{tabular}[c]{@{}l@{}} M 92, \\ C-19 \citep{Yuan_2022} \end{tabular} &  \\ \hline
\begin{tabular}[c]{@{}c@{}}G-3\\ Thamnos\end{tabular} & \begin{tabular}[c]{@{}c@{}}sg-6\\ (5)\end{tabular} & NGC 288, NGC 5139/\ocen & \begin{tabular}[c]{@{}l@{}}NGC 288 \\ Fimbulthul, C-7\end{tabular} &  \\ \hline
- & \begin{tabular}[c]{@{}c@{}}Candidate Merger \\sg-7\\ (9)\end{tabular} & \begin{tabular}[c]{@{}l@{}}NGC 5466, NGC 5694, \\ Pal 15, NGC 6934, Pal 13\end{tabular} & \begin{tabular}[c]{@{}l@{}}Gaia-10, NGC 5466,\\ Tuc-III \end{tabular} & Tucana III \\ \hline 
 \multirow{3}{*}{\begin{tabular}[c]{@{}c@{}}  G-4\\ Sequoia/\\Arjuna/\\I'itoi \end{tabular}}
 & \begin{tabular}[c]{@{}c@{}}Sequoia\\sg-8\\ (14)\end{tabular} & \begin{tabular}[c]{@{}l@{}}NGC 3201, IC 4499, \\ NGC 6101\end{tabular} & \begin{tabular}[c]{@{}l@{}}Gaia-12, NGC 1261, \\ Leiptr, Gj\"oll, Ylgr, \\ Gaia-1, Phlegethon \\ Gaia-6, Gaia-9,\\ NGC 6101, Gaia-11 \\ \end{tabular} &  \\ \cline{2-5} 
 & \begin{tabular}[c]{@{}c@{}}I'itoi\\sg-9\\ (2)\end{tabular} &  & GD-1, Kshir &  \\ \hline

\multirow{5}{*}{\begin{tabular}[c]{@{}c@{}} \\ \\ \\ \\ \\  Accreted\\ Structures\\ G-5\end{tabular}} & \begin{tabular}[c]{@{}c@{}}-\\ (13)\end{tabular} & \begin{tabular}[c]{@{}l@{}}Rup 106, AM 4, \\ NGC 5824\end{tabular} & \begin{tabular}[c]{@{}l@{}}Slidr, Sylgr, Elqui,\\ Chenab, ATLAS,\\ Aliqua\end{tabular} & \begin{tabular}[c]{@{}l@{}}Reticulum II, Antlia II,\\ Segue II, Willman I\end{tabular} \\ \cline{2-5} 
 & \begin{tabular}[c]{@{}c@{}}Helmi Streams\\sg-10 (8)\end{tabular} & \begin{tabular}[c]{@{}l@{}}NGC 5272/M 3, NGC 5634,\\ NGC 5897, NGC 6426, \\ NGC 7078/M 15, NGC 7492\end{tabular} & Svol, Ophiuchus, &  \\ \cline{2-5} 
 & \begin{tabular}[c]{@{}c@{}}sg-11\\ (4)\end{tabular} & NGC 5024/M 53 & \begin{tabular}[c]{@{}l@{}}NGC 1261, Gaia-8,\\ Jhelum\end{tabular} &  \\ \cline{2-5} 
 & \begin{tabular}[c]{@{}c@{}}LMS-1/Wukong\\sg-12\\ (5)\end{tabular} & NGC 5053 & \begin{tabular}[c]{@{}l@{}}Phoenix, Kwando, \\ LMS-1 \citep{Malhan_2021}, \\ Wukong \citep{Limberg_2024}\end{tabular} &  \\ \cline{2-5} 
 & \begin{tabular}[c]{@{}c@{}}sg-13\\ (2)\end{tabular} &  & Orphan & Grus II \\ \cline{2-5} 
 \multirow{3}{*}{\begin{tabular}[c]{@{}c@{}} \\   Accreted\\ Structures\\ G-5\end{tabular}} & \begin{tabular}[c]{@{}c@{}}Sagittarius\\sg-14\\ (22)\end{tabular} & \begin{tabular}[c]{@{}l@{}}Whiting 1, Eridanus, \\ NGC 2419, Pyxis, Pal 4, \\ Pal 14, NGC 6715/M 54, \\ Terzan 7, Arp 2, Pal 4,\\ Terzan 8, Pal 12\end{tabular} & \begin{tabular}[c]{@{}l@{}}NGC 5466, Indus, \\ Sagittarius\end{tabular} & \begin{tabular}[c]{@{}l@{}}Draco II, Pegasus III, \\ Triangulum II, \\ Canes Venatici II, \\ Leo IV, Pisces II, Segue I \end{tabular} \\ \cline{2-5} 
 & \begin{tabular}[c]{@{}c@{}}sg-15\\ (3)\end{tabular} & NGC 4590/M 68 & SGP-S, Fj\"orm &  \\ \cline{2-5} 
 & \begin{tabular}[c]{@{}c@{}}sg-16\\ (3)\end{tabular} & Pal 1 & Pal-5, Ravi &  \\ \hline
\begin{tabular}[c]{@{}c@{}}G-6\\ Bulge GCs \end{tabular} & \begin{tabular}[c]{@{}c@{}}-\\ (18)\end{tabular} & \begin{tabular}[c]{@{}l@{}}ESO 452/SC 11, NGC 6355,\\ HP 1/BH 229, Liller 1, \\ NGC 6380/Ton 1, NGC 6401, \\ Pal 6, Terzan 5/11, \\ NGC 6440, NGC 6453, \\ Terzan 9, NGC 6517, \\ NGC 6528, NGC 6558, \\ NGC 6624, NGC 6626/M 28, \\ NGC 6638, NGC 6642\end{tabular} &  &  \\ \hline

\hline
\begin{tabular}[c]{@{}c@{}}G-7\\ Post-disc \\ GCs\end{tabular} & \begin{tabular}[c]{@{}c@{}}-\\ (37)\end{tabular} & \begin{tabular}[c]{@{}l@{}}NGC 104/47 Tuc, E 3,\\ NGC 5927, BH 176, \\ Lynga 7/BH 184, Terzan 3,\\ NGC 6171/M 107, NGC 6235, \\ NGC 6256, NGC 6266/M 62,\\ NGC 6304, NGC 6325, \\ NGC 6352, Terzan 2/HP 3, \\ Terzan 4/HP 4, NGC 6362, \\ Ton 2/Pismis 26, NGC 6342,\\ Terzan 6/HP 5, NGC 6356,\\ Djorg 2/ESO 456, NGC 6522, \\ NGC 6539, NGC 6540/Djorg, \\ NGC 6553, IC 1276/Pal 7, \\ NGC 6569, BH 261/AL 3, \\ NGC 6637/M 69, Terzan 12,\\ NGC 6717/Pal 9, NGC 6723,\\ NGC 6752, Pal 10, Pal 11, \\ NGC 6838/M 71\end{tabular} & Gaia-7 &  \\ \hline
\begin{tabular}[c]{@{}c@{}}G-8\\ Pre-disc \\ GCs \end{tabular} &
\begin{tabular}[c]{@{}c@{}}-\\ (16)\end{tabular} & \begin{tabular}[c]{@{}l@{}}NGC 5986, NGC 6205/M 13,\\ NGC 6273/M 19, NGC 6293,\\ NGC 6316, NGC 6333/M 9,\\ NGC 6366, NGC 6388, \\ NGC 6402/M 14, NGC 6441,\\ NGC 6496, NGC 6652, \\ NGC 6656/M 22, Pal 8, \\ NGC 6760, NGC 6809/M 55\end{tabular} &  &  \\ \hline
\begin{tabular}[c]{@{}c@{}} G-9 \\ High-energy \\ Satellite \\ Galaxies \end{tabular} & \begin{tabular}[c]{@{}c@{}} - \\ (19)\end{tabular} & AM 1, Pal 3 &  & \begin{tabular}[c]{@{}l@{}}Horologium I, Draco,\\ Horologium II, Fornax,\\ Tucana IV, Tucana V, \\ Aquarius II, Bootes I, \\ Carina, Crater II,\\ Coma Berenices, \\ Leo II, Leo V, Sculptor, \\ Ursa Major I, \\ Ursa Major II, \\ Ursa Minor\end{tabular}
 \\ \hline
\end{longtable}

\end{appendix}
\label{LastPage}
\end{document}